\documentclass[fleqn,usenatbib]{mnras}

\usepackage[varg]{newtxmath}
\usepackage[T1]{fontenc}
\usepackage{ae,aecompl}

\usepackage{graphicx}	% Including figure files
\usepackage{color}
\usepackage{journals}
\usepackage{etoolbox}
\makeatletter
\patchcmd\@combinedblfloats{\box\@outputbox}{\unvbox\@outputbox}{}{%
  \errmessage{\noexpand\@combinedblfloats could not be patched}%
}%
\makeatother

\newcommand{\msun}{{\rm M_\odot}}

\newcommand{\lm}{\log\,(m/{\rm M}_\odot)}
\newcommand{\lM}{\log\,(M/{\rm M}_\odot)}
\definecolor{darkgreen}{rgb}{0.0,0.5,0.0}
\definecolor{darkred}{rgb}{0.5,0.0,0.0}
\definecolor{brown}{rgb}{0.65,.16,0.16}
\definecolor{grey}{rgb}{0.4,0.5,0.6}

\definecolor{cgcol}{rgb}{0.0,0.0,0.627}
\definecolor{mnwcol}{rgb}{0.0,0.565,0.0}
\definecolor{bwccol}{rgb}{0.561,0.0,0.561}
\definecolor{mgpcol}{rgb}{0.942,0.,0.}
\definecolor{guocol}{rgb}{0,0.56,0.56}
\newcommand{\CG}{C+G}
\newcommand{\MNW}{MNW}
\newcommand{\BWC}{BWC}
\newcommand{\MCP}{MCP}

\newcommand{\Hen}{Henriques}
\title[The predicted frequency of very young galaxies]{The frequency of very young
  galaxies in the local Universe:  
I. A test for galaxy formation and cosmological models}

\author[D. P. Tweed et al.]
{D. P. Tweed$^{1,2,3}$\thanks{E-mail:dylan.tweed@gmail.com}, 
G. A. Mamon$^2$\thanks{E-mail:gam@iap.fr}, 
T. X. Thuan$^{4,2}$,
A. Cattaneo$^{5,2}$,
A. Dekel$^{3,6}$,
\newauthor 
N. Menci$^7$,
F. Calura$^8$,
J. Silk$^{2,9,10}$
 \\ 
$^1$Center for Astronomy and Astrophysics, Department of Physics, Shanghai
Jiao Tong University, Shanghai 200240, China\\
$^2$Institut d'Astrophysique de Paris (UMR 7095: CNRS \& Sorbonne Universit\'e), 98 bis
Bd Arago, F-75014 Paris, France\\ 
$^3$Center for Astrophysics and Planetary Science, 
Racah Institute of Physics, The Hebrew University, Jerusalem, Israel\\
$^4$Astronomy Department, University of Virginia, P.O. Box 400325,
  Charlottesville, VA 22904-4325\\ 
$^5$GEPI (UMR 8111: Observatoire de Paris), 61 av. de l'Observatoire,
  75014 Paris, France\\
$^6$Santa Cruz Institute for Particle Physics, University of California, Santa Cruz CA 95064, USA\\  
$^7$INAF -- Osservatorio Astronomico di Roma, via di Frascati 33, I-00040 Monte Porzio Catone, Italy\\
$^8$INAF - Osservatorio di Astrofisica e Scienza dello Spazio di Bologna, Via
Gobetti 93/3, I-40129 Bologna, Italy\\
$^9$Dept. of Physics \& Astronomy, The Johns Hopkins University, Baltimore
MD 21218, USA\\
$^{10}$Beecroft Institute for Particle Astrophysics and Cosmology, Univ. of
  Oxford, Keble Road, Oxford OX1 3RH, UK
}

\date{Accepted 2018 February 21. Received 2018 January 23; in original form 2017 May 15}

\pubyear{2018}
\begin{document}

\label{firstpage}
\pagerange{\pageref{firstpage}--\pageref{lastpage}} \pubyear{2017}

\maketitle

\begin{abstract}
In the local Universe, the existence of very young galaxies (VYGs), having
formed at least half their stellar mass in the last 1 Gyr, is debated. We
predict the present-day fraction of VYGs among central galaxies as a function
of galaxy stellar mass. For this, we apply to high mass resolution
Monte-Carlo halo merger trees (MCHMTs) three (one) analytical models of galaxy formation,
where the ratio of stellar to halo mass (mass growth rate) is a function of halo mass and redshift. 
Galaxy merging is delayed until orbital decay by
dynamical friction. With starbursts associated with halo mergers, our models
predict typically one percent of VYGs up to galaxy masses of $10^{10}\,\msun$,
falling rapidly at higher masses, and VYGs are usually associated with
recent major mergers of their haloes. Without these
starbursts, two of the models have VYG fractions reduced by 1 or 2 dex at low
or intermediate stellar masses, and VYGs are rarely associated with major
halo mergers. In comparison, the state-of-the-art semi-analytical model (SAM)
of Henriques et al. produces only 0.01 per cent of VYGs at intermediate
masses. Finally, the Menci et al.  SAM run on MCMHTs with Warm Dark Matter
cosmology generates 10 times more VYGs at masses below $10^8\,\msun$ than when run with Cold
Dark Matter. The wide range in these VYG fractions illustrates the usefulness
of VYGs to constrain both galaxy formation and cosmological models.

\end{abstract}


\section{Introduction}

In the standard $\Lambda$ Cold Dark Matter ($\Lambda$CDM) paradigm,
galaxies form by dissipative collapse inside dark matter haloes: the smaller
haloes are the first to detach from the Hubble expansion and collapse. The larger
most massive structures, clusters, groups and massive ellipticals, form later
through mergers.

The mass function and growth of dark matter haloes is
 well understood thanks to  the  
\cite{Press&Schechter74} theory and its extensions  
\citep{Bond+91,Bower91}, confirmed by large-scale cosmological $N$-body
simulations \citep{Efstathiou+88,Carlberg&Couchman89,Springel+05,Warren+06,Tinker+08}.
However,
it has been a considerable challenge to  understand how galaxies form stars within
these haloes, because of the numerous physical processes involved. We know
that stars are formed in cold Giant Molecular Clouds of gas.
One first needs
to allow the gas to enter the haloes, but this process becomes inefficient in
haloes at the extremes of the mass function: 1) its entropy is too high to fall
into very
low-mass haloes \citep{Rees86}, 
2) its entropy is significantly raised when it is shock-heated near the
virial radius around high-mass \citep{Birnboim&Dekel03}.
Moreover, in dense environments, the outer gas can be stripped before it can fall
onto the disc and fuel the molecular clouds, from
a) the tides from the group/cluster
potential 
\citep*{Larson+80}, and
b) the ram pressure it feels from its motion relative to the hot intra-group/cluster gas
\citep{Gunn&Gott72}.
One then needs to retain the gas in the disc, against the feedback from
1) supernovae \citep{Dekel&Silk86} 
and 2) active galactic nuclei \citep{Silk&Rees98}.

The early realization that elliptical galaxies of increasing luminosity have
redder colours \citep{Sandage72} is now understood as partly due to the fact
that more massive ellipticals have older stellar populations
\citep{Thomas+05}, even if the colours of more massive ellipticals are also
redder because of their higher  metallicity  \citep{Faber73}.
This \emph{downsizing} trend of older stellar populations for massive
galaxies can be explained by a decrease in the efficiency of star formation
above some halo mass \citep{Cattaneo+06,Cattaneo+08}.

On the opposite end, the youngest galaxies should be those with the lowest
metallicities, as the neutral gas from which present stars are formed has not
been polluted by many previous generations of stars. Low-metallicity
star-forming objects possess strong emission-line spectra, characteristic of
HII regions and indicating the presence of an intense burst of star formation
(e.g. \citealp{Sargent&Searle70}).
Emission-line galaxies tend to be of low stellar mass (\nobreak{\citealp*{Mamon+01}}, who
used the near infrared $J$ band as a proxy for stellar mass). Similarly, the
strong positive correlations of metallicity with both the luminosity of ellipticals 
\citep{Faber73} and the stellar mass of irregular and blue compact dwarfs (BCDs,
\citealp{Lequeux+79} and with galaxies in general,
\citealp{Tremonti+04}) suggest that the youngest galaxies must be of low stellar
mass.

In particular, a prime candidate for a galaxy with a very young stellar
population is I~Zw~18, which has a very
low stellar mass around $10^7 {\rm M_\odot}$ \citep{Papaderos&Ostlin12,Izotov+18} and
an extremely low  metallicity (1/50th of solar).
Its spectrum shows strong emission-lines, indicative of active star
formation producing thousands of O stars emitting plenty of ionizing
radiation. Using the Hubble Space Telescope (HST) to resolve its stellar content
and construct its Hertzprung-Russell diagram, \cite{Izotov&Thuan04} found
that the bulk of the stellar population of I Zw 18 is younger than 500
Myr. Later, \cite{Aloisi+07} and 
\cite{ContrerasRamos+11} used  deeper HST 
imaging data of I~Zw~18 and detected an older
stellar population with age greater than  1~Gyr. 
However, the mass of the old stellar population is not known, as it depends on
the unknown star formation history of the galaxy. It is thus not clear
whether most of the stellar mass of I~Zw~18 was formed within the last
Gyr or earlier.
Another very young galaxy candidate  is J0811+4730, 
recently discovered by  \cite{Izotov+18}.
These authors found it to be even more metal-poor than I~Zw~18, and estimate
that
three-quarters of its
stellar mass is younger than (only) 5~Myr.

Motivated by the lack of galaxies whose bulk of stellar mass was
undoubtedly 
formed in
the last Gyr, and by the debate on the epoch when I~Zw~18 formed half of its
stellar mass, we are led to the following questions.
Can \emph{very young galaxies}
(hereafter VYGs) exist?
If yes, how frequent are VYGs in the
local Universe? 
In this article, we define VYGs as
\emph{galaxies in which more than half of the stellar mass  was formed in the last Gyr}.
This critical age of 1 Gyr is motivated by the debate on the age of I~Zw~18,
but is otherwise arbitrary, and we will also investigate how the frequency of
VYGs depends on this choice of critical age. 
Since I~Zw~18 \citep{Lelli+14} and J0811+4730 \citep{Izotov+18} are very
isolated galaxies, and since the physics of satellite galaxies is more
debated than that of centrals, we choose to focus on central galaxies, thus excluding
satellite galaxies.
However, it is not our aim, in this article, to model in
detail the properties of candidate VYGs I~Zw~18 and J0811+4730, but to
generally explore the fractions of VYGs among central galaxies
as a function of their $z$=0 stellar mass. 

We estimate the fraction of VYGs in bins of present stellar mass by using 
current models of galaxy
formation, both analytical and semi-analytical, and we also
compare the predictions between
a Warm Dark Matter cosmology and the standard $\Lambda$CDM.
In a companion article (Trevisan et al., in prep., hereafter Paper~II), we
estimate the fractions of VYGs as a function of stellar mass 
in the local Universe, using the Sloan Digital
Sky Survey (SDSS) spectral database, and compare them with the model
predictions presented here.

The models that we constructed to predict the frequencies of VYGs are  described
in
Sect.~\ref{sec:method} and tested in Sect.~\ref{sec:tests}.
In Sect.~\ref{sec:result}, we compute the fractions of VYGs, for the
different models.
These results are discussed in Sect.~\ref{sec:discuss} and summarized in Sect.~\ref{sec:concl}.

\section{Methods}
\label{sec:method}

\subsection{Basic considerations}
\label{sec:methods_basic}
Our choice of methods is guided by our requirement of producing large samples of
galaxies with sufficient mass resolution to form galaxies in a range of
stellar masses extending down to include the best two cases for VYGs, I~Zw~18
($\lm=7.38$, \citealp{Izotov+18})
and 
J0811+4730 ($\lm=6.3$, \citealp{Izotov+18}).
We must then resolve the much lower mass
progenitors of such $z$=0 galaxies.
One method is to use semi-analytical models of galaxy formation and evolution
(hereafter, SAMs), run on the halo merger trees derived from the dark matter
haloes extracted from high-resolution dissipationless cosmological $N$-body
simulations.
We analyse here the $z$=0 output of the recent state-of-the-art SAM of
\cite{Henriques+15}, run on 
two dissipationless cosmological $N$-body simulations: 
the Millennium Simulation (MS, \citealp{Springel+05}) and the Millennium-II
Simulation (MS-II,
\citealp{BoylanKolchin+09}).
Both simulations are re-scaled in time and space to the Planck 2014 cosmology, 
with $(\Omega_{\rm m},\Omega_\Lambda,h,\sigma_8)=(0.315,0.685,0.673,0.829)$
using the technique of \cite{Angulo&White10}, updated by
\cite{Angulo&Hilbert15}.

However, with  a particle mass of $9.5\times 10^6\, {\rm M_\odot}$,
the mass resolution of the  MS-II
is barely sufficient to resolve the haloes around our lower mass galaxies.
For example, galaxy I~Zw~18, whose halo log mass is 
$\lm = 8.9$ \citep{vanZee+98} or $\lm = 8.5$ \citep{Lelli+12}, both based upon
the distance of \citealp{Aloisi+07}),\footnote{We denote $M$ the halo mass and
  $m$ the galaxy stellar mass.} would only be resolved at $z=0$ with 90
particles
(the MS simulation, whose resolution is 125 coarser, clearly misses these
haloes).
Perhaps the halo mass of I~Zw~18 is underestimated as figure~3 of
\cite{Read+17} suggests a minimum value of
$10^{10}\,\msun$.
While no halo mass is available for galaxy J0811+4730, we infer from figure~3
of \citeauthor{Read+17} that its halo mass may be as low as $\lM=9.2$,
i.e. this galaxy's halo would be resolved with 170 particles.
However, the
progenitors of I~Zw~18 and J0811+4730 would be only marginally resolved in the MS-II.
This led us to also
consider Monte-Carlo halo merger trees rather than only rely on cosmological $N$-body
simulations to achieve adequate mass resolution for the haloes. So, in
addition to considering the SAM of \citeauthor{Henriques+15}, we also run
simple, single-equation, galaxy formation models on high mass-resolution Monte-Carlo halo
merger trees to derive the growth of the stellar masses of galaxies.

We discuss these Monte-Carlo halo merger trees in Sect.~\ref{sec:halotrees}
and present the
analytical modelling in Sect.~\ref{sec:anamodels} and the semi-analytical models in Sect.~\ref{sec:sam}.

\subsection{Halo merger trees}
\label{sec:halotrees}

\begin{figure*}
\centering
\includegraphics[width= 0.8\textwidth]{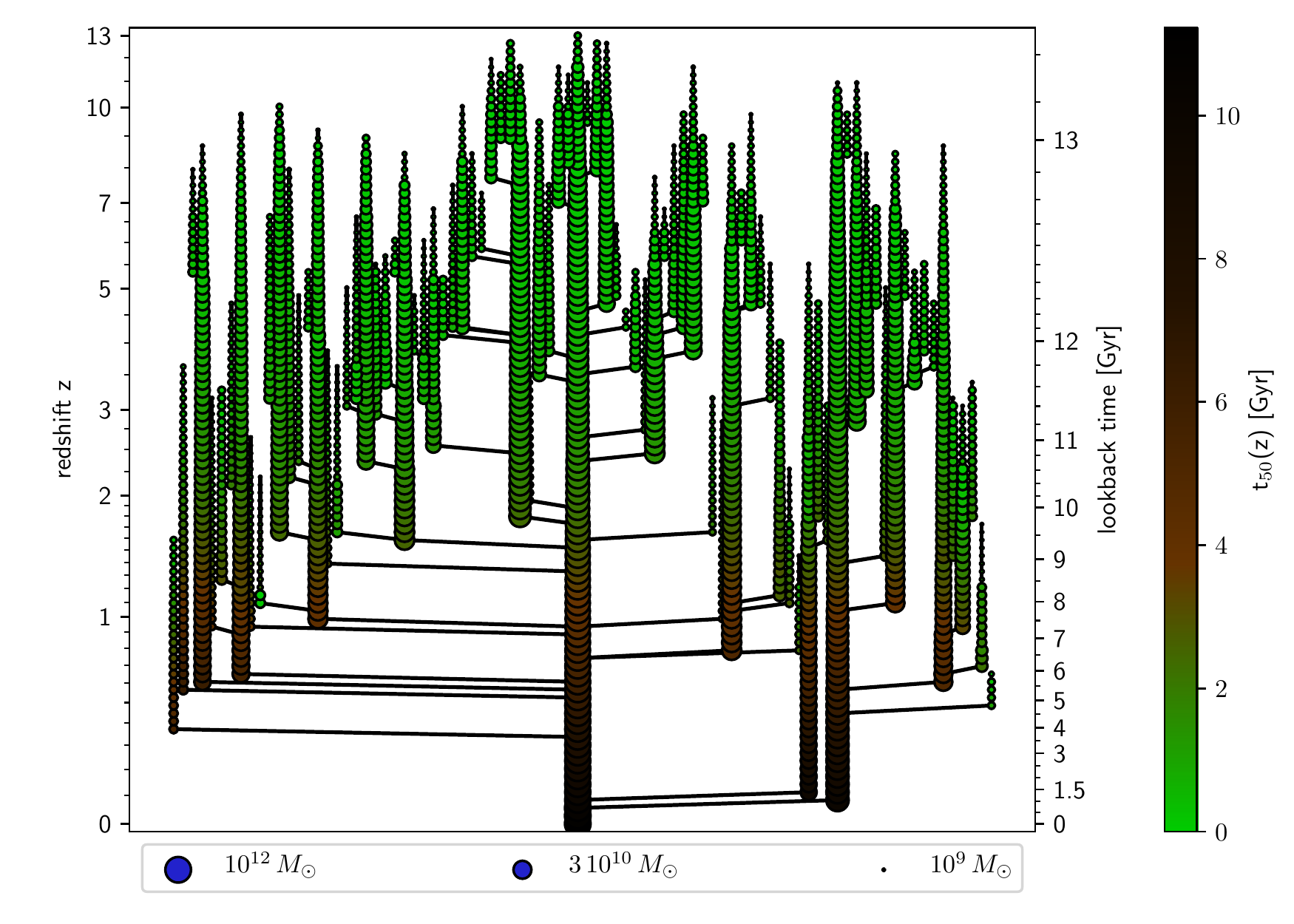}
\caption{Example of a halo merger tree, obtained with
the Monte-Carlo halo    merger tree code of Parkinson et al. (2008), 
where the $z$=0 halo mass is $10^{12}\, {\rm M_\odot}$. 
Each circle represents a halo, where the size of the symbol scales linearly
with halo log mass.
The colour code, from green to red to brown, 
scales as 
$t_{50} = t_f - t$, where $t_f$ is the lookback time when the most
massive progenitor of a halo has half its 
mass at lookback time $t$.
     Time flows from top to bottom. The \emph{central branch} shows the
     growth of the most massive progenitor.
For clarity, only branches above $2 \times  10^9\, {\rm M_\odot}$ and within
these, only haloes
above $10^9\,\msun$ are shown.
     \label{fig:mt}}
\end{figure*} 

Monte-Carlo halo merger trees are designed to generate realistic
merger histories of a given halo of mass $M$ at a redshift $z$ (usually 0).
These merger trees are built by
generating progenitor masses at a higher redshift, and iterating over those
progenitors.
For each
halo of mass $M_0$ at redshift $z_0$, 
the mass $M_1$ of the main progenitor at redshift $z_1$ is drawn according to a
probability distribution function that can be written as
$P_1(M_1|M_0,z_0,z_1)$. A secondary progenitor mass can then be drawn
following a probability distribution $P_2(M_2|M_1,M_0,z_0,z_1)$. For
some codes, multiple secondary progenitors can be drawn following
$P_2(M_i|\sum_{j=1}^{i-1} M_j, M_0, z_0, z_1)$. 
Monte-Carlo halo merger tree codes handle mass conservation in different ways,
either neglecting smooth accretion by imposing $\sum_{j}
M_j = M_0$, or incorporating diffuse mass growth, i.e. $\sum_{j} M_j \leq
M_0$. The difference $\Delta M = M_0 -\sum_{j} M_j$ can be interpreted as smooth accretion or unresolved
mergers. This process is iterated to increasingly higher redshifts, thus building the
branches of the halo merger tree down to a predefined mass resolution
or up to a maximal redshift. Since those probability distributions do
not depend on the previous (lower redshift) outcome, the entire process is
Markovian.

The first implementations of Monte-Carlo halo merger trees
\citep{Lacey&Cole93,Kauffmann&White93} used the extension of the
\cite{Press&Schechter74} model for the cosmic halo mass function, while modern
implementations use more accurate probability distribution functions.
\cite{Jiang&vandenBosch14} have recently compared 6 implementations of 4 halo merger
tree codes, for $z$=0 halo masses $M_0$ ranging from $10^{11}$ to $10^{15}\,h^{-1}\,\rm
M_\odot$, with branches of mass $>10^{-4}\,M_0$.
They concluded that the code of \cite*{Parkinson+08},
which is based on extended Press
Schechter theory \citep{Bond+91,Bower91,Lacey&Cole93}, 
with an additional term that is designed to achieve better mass
conservation,
reproduced best the mass assembly histories, merger rates and unevolved
subhalo mass functions
measured
in cosmological $N$-body simulations with 
the same cosmological parameters as previously used in the Millennium
simulations: $(\Omega_m,\Omega_\Lambda,h,\sigma_8) =
(0.25,0.75,0.73,0.9)$,
which is close to the WMAP $1^{\rm st}$ year
cosmology \citep{Spergel+03}.
\citeauthor{Parkinson+08} calibrated the two free parameters of
their algorithm to match the conditional mass functions, as well as the distribution
of the epochs of most recent major mergers, both measured in the Millennium
simulation. 
We therefore adopted the code of \cite{Parkinson+08}.

The \citeauthor{Parkinson+08} code creates a binary tree
with very fine time resolution,
so 
that only the main and one secondary halo are drawn for a given halo
mass.
This code enables the user to 
adopt a custom, coarser, output time resolution of the merger tree.
Thus, our tree outputs may contain non-binary mergers, but these are built
from binary mergers at the fine internal resolution of the code. We chose 101
output timesteps in equal increments of 
$\log(1\!+\!z)$ from redshift $z=0$ to redshift $z=13$.
Our first non-zero redshift is $z=0.0267$, corresponding to a lookback time
of 350 Myr.

Figure~\ref{fig:mt} illustrates the Monte-Carlo halo merger tree obtained
with the \citeauthor{Parkinson+08}
halo merger tree code. 
While mergers of branches are clearly seen along the main branch, they also 
 occur within the
secondary branches. The figure shows the diversity of halo formation
times (summed over the progenitors): before being merged into more
massive haloes, massive haloes always form early (brown) while low-mass haloes form
early (brown) or late (green).  
A 
 halo
can have only one
\emph{descendant} per timestep (halo fragmentation is not allowed). 
However, the number of \emph{progenitors}
a
given halo can have is zero (a newborn
halo), one (a quiescent halo), or several
(mergers occurring between time outputs).

We ran the \citeauthor{Parkinson+08} halo merger tree code 
with the cosmological parameters of the Millennium simulations, i.e.,
$(\Omega_m,\Omega_\Lambda,h,\sigma_8) =
(0.25,0.75,0.73,0.9)$. We adopted final halo
log masses of
$\log (M_0/{\rm M}_\odot) = 7$ to 14 in steps of 0.025 dex, and a mass
resolution $M_{\rm min} = 10^{-4} M_0$.
The minimum halo mass of $10^7\,\msun$ is chosen to ensure that we include
the halo mass of I~Zw~18 (possibly as low as $10^{8.5}\,\msun$
\citealp{Lelli+12}).
By going to such low halo masses, we are assuming that
the \citeauthor{Parkinson+08} tree code remains valid 4 dex below where it was tested. 
For each value of $\log M_0$, we have run the halo merger tree code 1000 times with different
random seeds. In total, we have generated $1000\times [(14-7)/0.025+1] = 281$\,$000$
halo merger trees.

These merger trees are qualitatively very similar
to analogous merger trees extracted from (cosmological) $N$-body simulations. 
Our Monte-Carlo trees have superior mass resolution in comparison with the
halo trees extracted from the MS or even from the MS-II: the final halo
masses extend 2 dex lower than the haloes resolved by MS-II with 100
particles, and the progenitors of our haloes have 4 extra dex of resolution,
so that our Monte Carlo halo merger trees reach a progenitor mass resolution
that is 6 orders of magnitude better than that of the MS-II haloes. 

Note that, while the haloes in $N$-body simulations can decrease in mass from one step to the
next (because of tidal forces during close interactions),
our haloes cannot lose mass, by construction.
 
\subsection{Analytical galaxy formation models}
\label{sec:anamodels}

\subsubsection{Basic formalism}
\label{sec:anabasic}
We first 
consider very simple galaxy formation models 
where stellar masses $m_{\rm stars}$ are assigned to
haloes with a \emph{star formation efficiency} (SFE) that depends on halo
mass $M$ and redshift $z$, according to
\begin{equation}
m_{\rm stars}\equiv \widetilde m(M,z) = f_{\rm b}\,M\,F_{\rm SFE}(M,z) \ ,
\label{mofMz_generic}
\end{equation}
where
the tilde sign
is to denote that this is a model.
In equation~(\ref{mofMz_generic}), $f_{\rm b} = \Omega_{\rm b}/\Omega_{\rm
  m}$
is
the universal baryonic
fraction ($f_{\rm b} = 0.18$ for our adopted $\Omega_{\rm b} = 0.045$ as in the Millennium
simulations), while
$F_{\rm SFE}$ represents the ratio between the stellar mass of a halo and the
total mass in baryons expected within the halo.
Our analytical models are entirely based on our choice for $F_{\rm
  SFE}(M,z)$.
They predict the stellar mass, but not the gas mass.

We consider a physically-motivated model, as well as two empirical
ones based on abundance matching (see Sect.~\ref{sec:Moster}).
We  also  use another empirical model, based on an equation
similar to equation~(\ref{mofMz_generic}), but where masses are replaced by
mass variations.
This use
of several different galaxy formation 
models allows us to gauge the dependence of our results on the uncertainties
of galaxy formation.

If one specifies a form for $F_{\rm SFE}(M,z)$, one can derive the stellar
mass history of every halo. Following \cite{Cattaneo+11}, who pioneered the
use of equation~(\ref{mofMz_generic}),  and \cite{Habouzit+14}, 
we \emph{forbid stellar masses to decrease}, i.e. the stellar mass of the 
(central) galaxy has to be greater or equal to the sum of its galaxy
progenitors. 

\subsubsection{Physical analytical model: \citet{Cattaneo+11} with \citet{Gnedin00}  at the low-mass end}
\label{sec:Cattaneo}
\citet{Cattaneo+11} (hereafter C11) have presented a quasi-physical model for $m(M,z)$
that combines 1) supernova feedback, 2) a gentle cutoff at the 
high-mass end caused by the virial shock around high-mass haloes \citep{Birnboim&Dekel03,Dekel&Birnboim06},
and 3) a sharp cutoff at the low-mass end  (motivated by
early hydrodynamical simulations of \citealp{Thoul&Weinberg96}) due to the
\emph{entropy barrier} that prevents high entropy gas from collapsing onto
its halo (given that its entropy cannot decrease).
This is written as $F_{\rm SFE} \propto 1-v_{\rm reion}^2/v_{\rm c}^2$, where
$v_{\rm reion}$ is a constant, while
$v_{\rm c}^2$ is the squared circular velocity of the halo
\begin{equation}
v_{\rm c}^2 = \left [{\Delta(z)\over 2}\right]^{1/3}\,
\left[ H(z)\,G\,M \right]^{2/3} \ ,
\label{vcofM}
\end{equation}
where $\Delta(z)$ is the ratio of the mean density within the virial radius to
the critical density of the Universe, while $H(z)$ is the Hubble constant.

The supernova feedback prescription of \citeauthor{Cattaneo+11} is physically
motivated: it is 
based on the idea that a fraction of the accreted gas that is processed into
stars is rapidly ejected as supernova winds, whose velocity matches the
virial velocity of the halo and whose energy is assumed to be entirely
mechanical and proportional to the remaining
stellar mass. This yields $F_{\rm SFE} \propto 1/(1+v_{\rm SN}^2/v_{\rm
  c}^2)$,
where $v_{\rm SN}$ is another constant.
At large halo masses,
the virial shock quenches the infall of cold gas filaments by heating them up
near the virial radius. 
Star formation in the disc is then limited by the
much longer cooling time of the shock-heated gas.
This is assumed to yield 
 $F_{\rm SFE} \propto 1/(1+M/M_{\rm shock})$,
where $M_{\rm shock}$ is a third constant.

This leads to equation~(8) of \cite{Cattaneo+11}:
\begin{equation}
F_{\rm SFE}^{\rm C11}(M,z) = {1-v_{\rm reion}^2/v_{\rm c}^2 \over 1 +
  M/M_{\rm shock}}
\,\left (1+v_{\rm
    SN}^2/v_{\rm c}^2\right)^{-1} \ ,
\label{C11}
\end{equation}
where the constants
 $v_{\rm reion}$,
$v_{\rm SN}$, 
and 
$M_{\rm shock}$  respectively 
represent the minimum circular velocity for  gas to overcome the entropy
barrier and collapse with the dark matter and subsequently form stars,
the impact of supernova feedback, 
and
the characteristic minimum mass
for the occurrence of virial shocks.
The first term of equation~(\ref{C11})  describes the ability to accrete
gas, while the second term describes the ability to retain this accreted
gas. 
At
high halo masses, galaxy mergers produce galaxies with higher
masses than predicted with  $F_{\rm SFE}(M,z)$ (see \citealp{Cattaneo+11}),
but the present work focuses on intermediate- and low-mass galaxies.

Given the expected sensitivity of the fraction of VYGs to the form
of the threshold of star formation efficiency, $F_{\rm SFE}$,
we improve on the model of \cite{Cattaneo+11} by introducing, at the low-mass end,  a 
 smoother cutoff, derived from hydrodynamical 
simulations \citep*{Gnedin00,Okamoto+08}. The accreted mass (disregarding for now
the virial shock affecting higher masses) is no longer $m_{\rm accr} = f_{\rm b} M (1- v_{\rm
  reion}^2/v_{\rm c}^2)$ but instead
\begin{equation}
  \label{reion_gnedin}
  m_{\rm accr} = \left[1+\left(2^{2/3}-1\right)\left(\frac{v_{\rm
        reion}}{v_{\rm c}}\right)^{6}\right]^{-3/2}
  f_{\rm b}\,M \ .
\end{equation}
Thus, $v_{\rm reion}$ is no longer the halo circular velocity below which no
star formation can occur, but instead the halo circular velocity where the
accreted mass is reduced by a factor 2 by the entropy barrier. There
is no longer a sharp cutoff of star formation at $v_{\rm c} = v_{\rm reion}$
as in (eq.~[\ref{C11}]), but instead the stellar mass
increases as $v_{\rm c}^9$ (much steeper than the Tully-Fisher relation). 

Using equation~(\ref{reion_gnedin}), equation~(\ref{C11}) becomes
\begin{equation}
  \label{SFE_C11G00}
  F_{\rm SFE}^{\rm  C+G}(M,z) = {\left[
1+\left(2^{2/3}\!-\!1\right) \,
\left( v_{\rm reion}/v_{\rm c}\right)^6
\right]^{-3/2}
\over \left (1+M/M_{\rm shock}\right)\,
\left(1+v_{\rm SN}^2/v_{\rm c}^2\right)} \ .
\end{equation}

While the terms for the effects of supernovae and the entropy barrier are
respectively motivated by physical principles and hydrodynamical simulations, 
the term involving the virial shocks is, admittedly,  empirical
(which is why we dub this model `quasi-physical', but later call it
`physical' to distinguish it from the fully empirical models that we will discuss below).

In this `Cattaneo+Gnedin' model (hereafter, C+G), we refer to the values of
$v_{\rm reion}$ before and after 
reionization as $v_{\rm pre-reion}$ and $v_{\rm post-reion}$, respectively.
We adopt the following parameters, which produce a good fit to the
present-day mass function of galaxies (see Figure~\ref{fig:fofm} below):
$v_{\rm SN} = 300 \, \rm km \, s^{-1}$,
  $v_{\rm pre-reion} = 18 \, \rm km \, s^{-1}$,
  $v_{\rm post-reion} = 50 \, \rm km \, s^{-1}$,
  $M_{\rm shock} = 10^{12} {\rm M_\odot}$.\footnote{There was no
  pre-reionization velocity in \cite{Cattaneo+11}, who had assumed that
  Universe was fully reionized from the start.}
According to equation~(\ref{reion_gnedin}), gas accretion is suppressed by a factor
10 if $v_{\rm c} = 0.74\, v_{\rm reion}$. Since $v_{\rm c} = \sigma_v/0.7$
for NFW halos (derived from eqs. [22] and [24] of \citealp{Lokas&Mamon01}),
$v_{\rm pre-reion} = 18 \, \rm km \, s^{-1}$ corresponds to a halo
temperature of 
$10\,500\,\rm K$, i.e. the temperature of atomic Hydrogen cooling. Similarly, 
$v_{\rm post-reion} = 50 \, \rm km \, s^{-1}$ corresponds to $T=82\,000\,\rm
K$. 
Our adopted value of $M_{\rm shock}$ is double that of
\cite{Dekel&Birnboim06} and 10~per cent lower than the value employed by
\cite{Cattaneo+11}.
Finally, our adopted value of $v_{\rm SN}$ is chosen to roughly fit the $z$=0
stellar mass function.

The effects of these parameters on the stellar mass function is shown in
preliminary versions of this work \citep{Mamon+11tmp,Mamon+12}. 
The baryonic  Tully-Fisher relation is well reproduced by the \CG\ model  \citep{Silk&Mamon12}.

We could have assumed that the Universe has reionized instantaneously at a fixed
redshift $z_{\rm reion}$, somewhere between 6 and 12.
Although reionization fronts are thought to have spread fast throughout
the Universe \citep{Gnedin&Ostriker97}, there is observational evidence that reionization
took a time comparable to the age of the Universe at that epoch. This is
suggested  
by the redshift difference between the epoch $z\approx 9$ where the optical depth to
neutral Hydrogen was unity
\citep{Planck_collaboration+15_cosmopars}
 and the latest
epoch, $z\approx 6$,
when evidence of a substantially neutral intergalactic medium is seen
\citep{Becker+01}.
For this reason, we  have assumed instead that the Universe reionizes
\emph{stochastically}.
Thus,
for each halo merger tree, we have drawn $\log (1\!+\!z_{\rm reion})$ from a Gaussian
of mean corresponding to a median reionization redshift of
$\langle z_{\rm reion}\rangle = 7.5$ 
and a standard deviation $\sigma[\log(1\!+\!z_{\rm reion})] = 0.05$.
This stochastic reionization redshift is applied to all progenitors
within a merger tree. We suppose that merging haloes lie in the same region of
the Universe and thus are
reionized at the same redshift.
The mean SFE (only used for illustrative figures, see Fig.~\ref{fig:sfevsz}
below) is then
\begin{eqnarray}
\left\langle F_{\rm SFE}^{\rm C+G}(M,z)\right\rangle &=&
F_{\rm reion}(z) \,F_{\rm SFE-pre}^{\rm C+G}(M,z) \nonumber \\
&\mbox{}&  + 
\left [1-F_{\rm reion}(z)\right]\,F_{\rm SFE-post}^{\rm C+G} (M,z) 
\ ,
\label{meansfeC11G00}
\end{eqnarray}
where
$F_{\rm SFE-pre}^{\rm C+G}$ and $F_{\rm SFE-post}^{\rm C+G}$ are the SFEs for
$v_{\rm reion} = v_{\rm pre-reion}$ and $v_{\rm reion} = v_{\rm
  post-reion}$, respectively,
while
\begin{equation}
F_{\rm reion}(z) = {1\over2}\, \left\{1+{\rm erf} \left[{\log(1\!+\!z)-\log(1\!+\!\left\langle z_{\rm
      reion} \right\rangle)\over \sqrt{2}\,\sigma\left[\log\left(1\!+\!z_{\rm
      reion}\right)\right]}\right]\right\}
\label{CDFzreion}
\end{equation}
is the probability that reionization occurs before redshift $z$.
Equation~(\ref{CDFzreion}) implies that 
 90~per cent of the
reionization occurs between $z=9.3$ and 6.0.

\subsubsection{Empirical model: \citet{Moster+13}}
\label{sec:Moster}
The relation between stellar and halo mass can also be determined empirically, by
comparing the cumulative distribution function (CDF) of the observed stellar mass
function with the CDF of the halo mass function predicted from cosmological
$N$-body simulations, a method known as
\emph{abundance matching} \citep{Marinoni&Hudson02}.

\citeauthor*{Moster+13} (\citeyear{Moster+13}, hereafter MNW) have proposed the following form for the star formation efficiency:
 \begin{equation}
F_{\rm SFE}^{\rm MNW}(M,z) = 2\,{N\over f_{\rm b}} \,\left[
\left ({M\over M_1}\right)^{-\beta}+
\left ({M\over M_1}\right)^\gamma
\right]^{-1} \ ,
\label{SFE_Moster}
\end{equation}
where $F_0$, $M_1$, $\beta$ and $\gamma$ are positive constants.
\citeauthor{Moster+13} 
computed the halo and subhalo mass functions in  the
MS and MS-II
rescaled to WMAP-7
cosmological parameters ($\Omega_{\rm m}=0.272$, $h=0.704$, $\sigma_8=0.81$).
Since subhaloes are tidally stripped, \citeauthor{Moster+13} computed the subhalo masses at the
time they entered their halo, and accounted for \emph{orphan subhaloes} that
were no longer resolved in the simulations but should have survived given
their long expected orbital decay time from dynamical friction.
Performing abundance matching at several 
redshifts from 0 to 4, \citeauthor{Moster+13} determined
\begin{eqnarray}
N &=& 0.0351 - 0.0247 \,z/(1\!+\!z) \ , \nonumber\\
M_1 &=& {\rm dex} \left [11.59 + 1.195\,z/(1\!+\!z) \right] \ ,\nonumber\\
\beta &=& 1.376 - 0.826\, z/(1\!+\!z) \ ,\nonumber\\
\gamma &=& 0.608 + 0.329\,z/(1\!+\!z)\ .
\label{pars_Moster}
\end{eqnarray}

\subsubsection{Empirical model: \citet{Behroozi+13}}
\label{sec:Behroozi}
\citeauthor*{Behroozi+13} (\citeyear{Behroozi+13}, hereafter BWC)  
have also (independently) performed abundance matching, with however several
differences as compared to
\cite{Moster+13}.
First, their halo mass functions are derived from a different
dissipationless cosmological $N$-body simulation, Bolshoi \citep*{Klypin+11}, with
cosmological parameters ($\Omega_{\rm m}=0.27$, $h=0.70$, $\sigma_8=0.82$)
consistent with WMAP5 \citep{Komatsu+09}.
The second difference is the more complex form for $F_{\rm SFE}(M,z)$:
\begin{equation}
F_{\rm SFE}^{\rm BWC}(M,z) = 
 \left ({\epsilon M_1\over
f_{\rm b} M}\right) \,
{\rm dex} \left \{ 
f\left [\log \left ({M\over M_1}\right)\right]\!-\!f(0)
\right\} ,
\label{SFE_Behroozi}
\end{equation}
where
\begin{equation}
f(x) = -\log \left (1+10^{\alpha x}\right) + \delta\,
{\left\{\log\left[1+{\rm e}^x\right]\right\}^\gamma\over
  1+\exp\left(10^{-x}\right)} \ .
\label{f_Behroozi}
\end{equation}
\citeauthor{Behroozi+13} argue that their more refined shape for $F_{\rm SFE}(M,z)$
  improves its accuracy by a factor 4. 

The third difference with \MNW$\,$ is that
\citeauthor{Behroozi+13}
fitted their model and the halo+subhalo mass function to, not only the
  stellar mass function at different redshifts, but also to the 
  dependence with mass of the specific star formation rate at different redshifts, as
  well as to the evolution of the cosmic star formation rate  with
  redshift. 
The fourth difference is that the \citeauthor{Behroozi+13} model is fit to 
observational data at sufficiently high redshifts to start probing
the epoch of reionization:  from 0 to 8 (while \MNW$\,$ stops at $z=4$).
\citeauthor{Behroozi+13}  derive the halo+subhalo mass function by assuming a model for the
dependence with halo/subhalo mass of the 
fraction of haloes or subhaloes that are subhaloes (which they calibrate at high
masses to measurements from simulations).

Their analysis yields the best-fit set of parameters:
\begin{eqnarray}
\epsilon &=& {\rm dex} \left [-1.777 - 0.006\,(a\!-\!1)\,\nu -0.119\,(a\!-\!1)\right]  \ , \nonumber\\
M_1 &=& {\rm dex} \left \{11.514 -\left[1.793\,(a\!-\!1)+0.251\,z \right]\,\nu\right\} \ , \nonumber\\
\alpha &=& -1.412 + \left[0.731\,(a\!-\!1)\right]\,\nu  \ , \nonumber\\
\delta &=& 3.508 + \left[2.608\,(a\!-\!1)-0.043\,z\right]\,\nu \ , \nonumber\\
\gamma &=& 0.316 + \left[1.319  \,(a\!-\!1)+0.279\,z\right]\,\nu \ ,
\label{pars_Behroozi}
\end{eqnarray}
where
\begin{eqnarray}
a &=& {1\over 1\!+\!z} \ ,\nonumber\\
\nu &=& \exp \left [-{4\over (1\!+\!z)^2}\right] \ .
\label{pars2_Behroozi}
\end{eqnarray}

\subsubsection{Empirical model: \citet{Mutch+13}}
\label{sec:Mutch}
The model of \citeauthor*{Mutch+13} (\citeyear{Mutch+13}, hereafter MCP)
replaces the stellar and halo masses of 
equation~(\ref{mofMz_generic}) by their time derivatives (see also \citealp*{Moster+17}):
\begin{equation}
{{\rm d}m\over {\rm d}t} = f_b {{\rm d}M\over {\rm d}t}\,F_{\rm SFE}^{\rm MCP}(M,z) \ .
\label{mutch}
\end{equation}
\citeauthor{Mutch+13} assume lognormal functions for $F_{\rm SFE}(M)$ at
given $z$, and we adopt their evolving, ``$M_{\rm vir}$''-based model, which in our
notation is
\begin{equation}
F_{\rm SFE}^{\rm MCP}(M,z) = {\cal E}(z) 
\exp\left [
  -\left ({ \log M\!-\!\log M_{\rm peak}(z)\over \sigma(z)}\right)^2
  \right] ,
\label{SFE_Mutch}
\end{equation}
where
\begin{eqnarray}
{\cal E}(z) &=& 0.90\,(1\!+\!z)^{-0.74} \ ,\nonumber \\
\log M_{\rm peak}(z) &=& 11.6\,(1\!+\!z)^{0.03} \ ,  \nonumber \\
\sigma(z) &=& 0.56\,(1\!+\!z)^{0.25} \ .
\label{pars_Mutch}
\end{eqnarray}
The parameters of the \MCP\  model were fit to constrain both the $z$=0 SMHM and
the stellar mass functions at redshifts up to $z=3$, split between red and blue galaxies.
\citeauthor{Mutch+13} obtained these parameters by running their model on
halo merger trees from the MS, with the same cosmological parameters (but
they express their halo masses using
$h=0.7$ instead of $h=0.73$, a 4~per cent relative difference that we, hereafter, neglect).

The \MCP\ model of equation~(\ref{mutch}) has the advantage 
of producing, through  stellar mass growth, a natural scatter
in the SMHM from the stochasticity of the halo mass growth. 
Moreover, by construction, stellar mass cannot decrease in the \MCP\ model.
Since the \MCP\ model is based on
time derivatives, it does not predict a minimal stellar mass, but
a minimal mass growth. Therefore, higher stellar mass growth can
only be caused by mergers. 
We
decreased their normalization ${\cal E}$ by 30~per cent  (as they suggest)
to account for the
loss of stellar mass by supernova explosions.

\subsubsection{Comparison of the analytical galaxy formation models}
\label{sec:compare}

\begin{table*}
\caption{Galaxy formation models}
\centering
\tabcolsep 1.75pt
\begin{tabular}{llllcccccccrlll}
\hline
Abbreviation & Reference & Nature & Relation & Calibration &
Power & 
\multicolumn{3}{c}{Cosmology} & &
\multicolumn{2}{c}{$z$=0 haloes} & &
\multicolumn{2}{c}{Halo merger trees} \\
\cline{7-9}
\cline{11-12}
\cline{14-15}
        &           &        &          & max $z$
& spectrum & $\Omega_{\rm m}$ & $h$ & $\sigma_8$ & &
$\log M_0^{\rm min}$& \multicolumn{1}{c}{number} & &
method & resolution \\
\multicolumn{1}{c}{(1)}&
\multicolumn{1}{c}{(2)}&
\multicolumn{1}{c}{(3)}&
\multicolumn{1}{c}{(4)}&
\multicolumn{1}{c}{(5)}&
\multicolumn{1}{c}{(6)}&
\multicolumn{1}{c}{(7)}&
\multicolumn{1}{c}{(8)}&
\multicolumn{1}{c}{(9)}&
&
\multicolumn{1}{c}{(10)}&
\multicolumn{1}{c}{(11)}&
&
\multicolumn{1}{c}{(12)}&
\multicolumn{1}{c}{(13)}
\\
\hline
\CG\ & Cattaneo+11 & physical & $m=f(M,z)$ & 0.1 & CDM & 0.28 & 0.70 & 0.82 &
& 7$\ \ \,$& 281\,000 &&
Monte-Carlo & $10^{-4}$\\
\MNW\ & Moster+13 & empirical &  $m=f(M,z)$ & 4$\ \ \,$&CDM & 0.27 & 0.70 &
0.81 & & 7$\ \ \,$& 281\,000 &&
Monte-Carlo & $10^{-4}$\\
\BWC\ & Behroozi+13 & empirical &  $m=f(M,z)$ & 8$\ \ \,$&CDM & 0.27 & 0.70 &
0.82 & & 7$\ \ \,$& 281\,000 &&
Monte-Carlo & $10^{-4}$\\
\MCP\ & Mutch+13 & empirical &  $\dot m=f(\dot M,z)$ & 3$\ \ \,$&CDM & 0.25 &
0.70 & 0.90 & & 7$\ \ \,$&
281\,000 &&
Monte-Carlo & $10^{-4}$\\
\Hen\ & Henriques+15 & SAM & complex & 3$\ \ \,$& CDM & 0.31 & 0.67 & 0.83 &
& 7.8& 1\,169\,786 &&
MS-II & $10^9 \,\rm M_\odot$  \\
\hline
\Hen\ & Henriques+15 & SAM & complex & 3$\ \ \,$& CDM & 0.31 & 0.67 & 0.83 &
& 9.9& 1\,514\,920 &&
MS & $10^{11.1} \,\rm M_\odot$  \\
Menci & Menci+08,14 & SAM & complex & none & CDM & 0.30 & 0.70 & 0.90 & & 9.5 & 2\,100 &&
Monte-Carlo & $10^{7.7}\,\rm M_\odot$\\
Menci & Menci+08,14 & SAM & complex & none & WDM & 0.30 & 0.70 & 0.90 & & 9.5 & 2\,100 &&
Monte-Carlo & $10^{7.7}\,\rm M_\odot$\\
\hline
\end{tabular}
\parbox{\hsize}{Notes: the columns are as follows. 
 (1): model abbreviation used in text; 
 (2): reference; 
 (3): nature (quasi-physical analytical, empirical
analytical, or semi-analytical [SAM]); 
 (4): relation between stellar and halo mass;
 (5) maximum redshift for calibration
to observational data; 
 (6): primordial power spectrum;
 (7): cosmological density parameter;
 (8): dimensionless Hubble constant;
 (9): standard deviation of primordial density fluctuations at scales of $8
\, h^{-1} \, \rm Mpc$  (both linearly
extrapolated to $z=0$); 
 (10): minimum final halo log mass (solar units);
 (11): number of Monte Carlo merger trees or of $z$=0 haloes more massive than $10^9 M_\odot$ (for the Henriques SAMs);
 (12): tree method;
 (13):  minimal halo mass in tree (the first four rows are relative to the $z$=0 mass;
the values for the Henriques SAMs are for haloes resolved with 100 particles).
The first five models are the major ones of this study, while the latter
three are for specific analyses. 
}
\label{tab:models}
\end{table*}
Table~\ref{tab:models} summarizes the analytical models used here, as well
as the semi-analytical models presented in Sect.~\ref{sec:sam}.
The cosmological parameters are those of the
simulations that were used to calibrate to the observations.

\begin{figure}
% SM: set zvec = {0 1 2 4 6 8}                            DO NOT ERASE THIS COMMENT!
% SM: SFEmodels zvec     50 18 300 1e12 7.5 0.05 file     DO NOT ERASE THIS COMMENT!
% SM: SFEmodels_vsM zvec 50 18 300 1e12 7.5 0.05 file     DO NOT ERASE THIS COMMENT!
\centering
\includegraphics[width=\hsize,viewport=1 160 600 690]{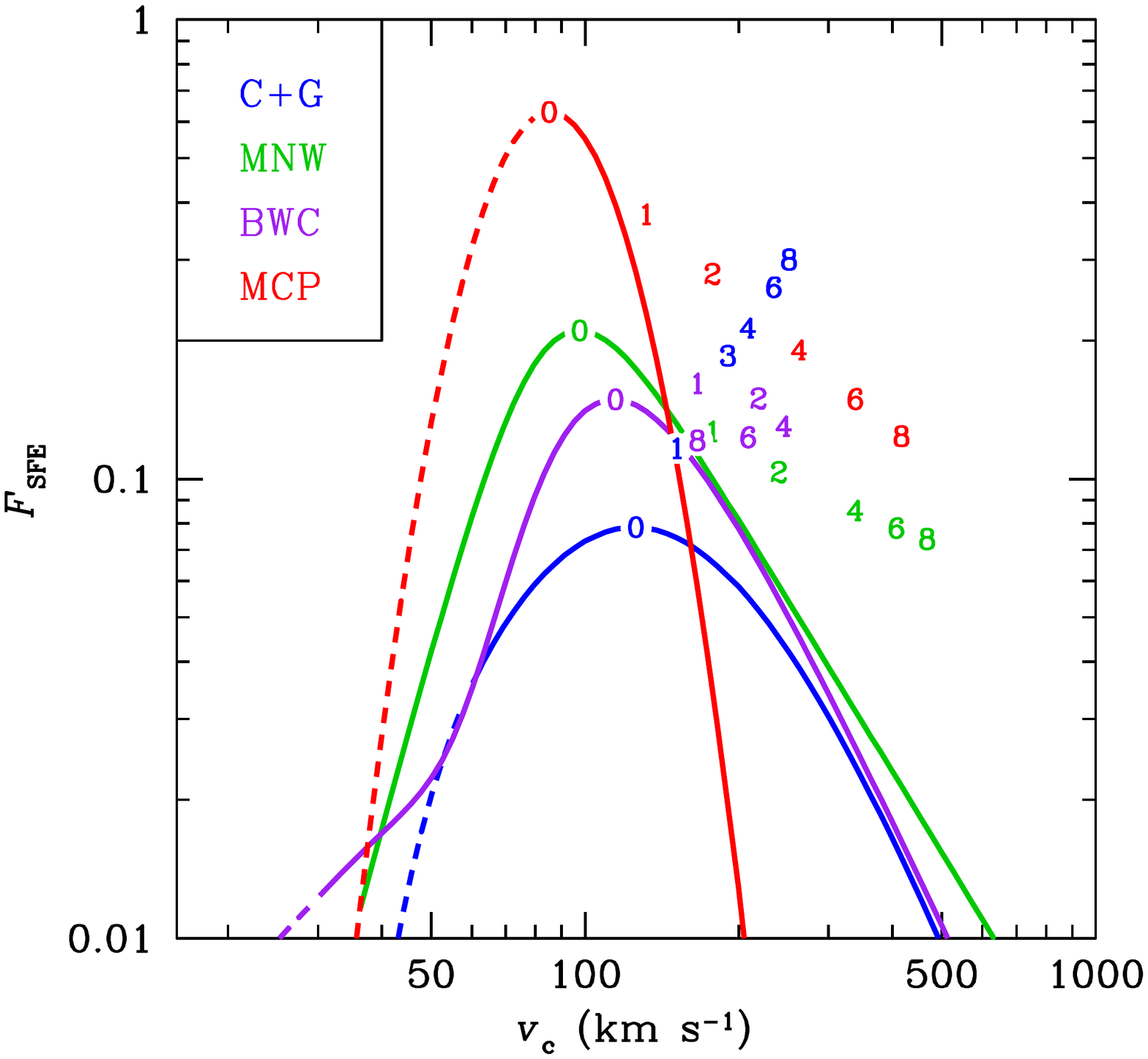}
\includegraphics[width=\hsize,viewport=1 180 600 690]{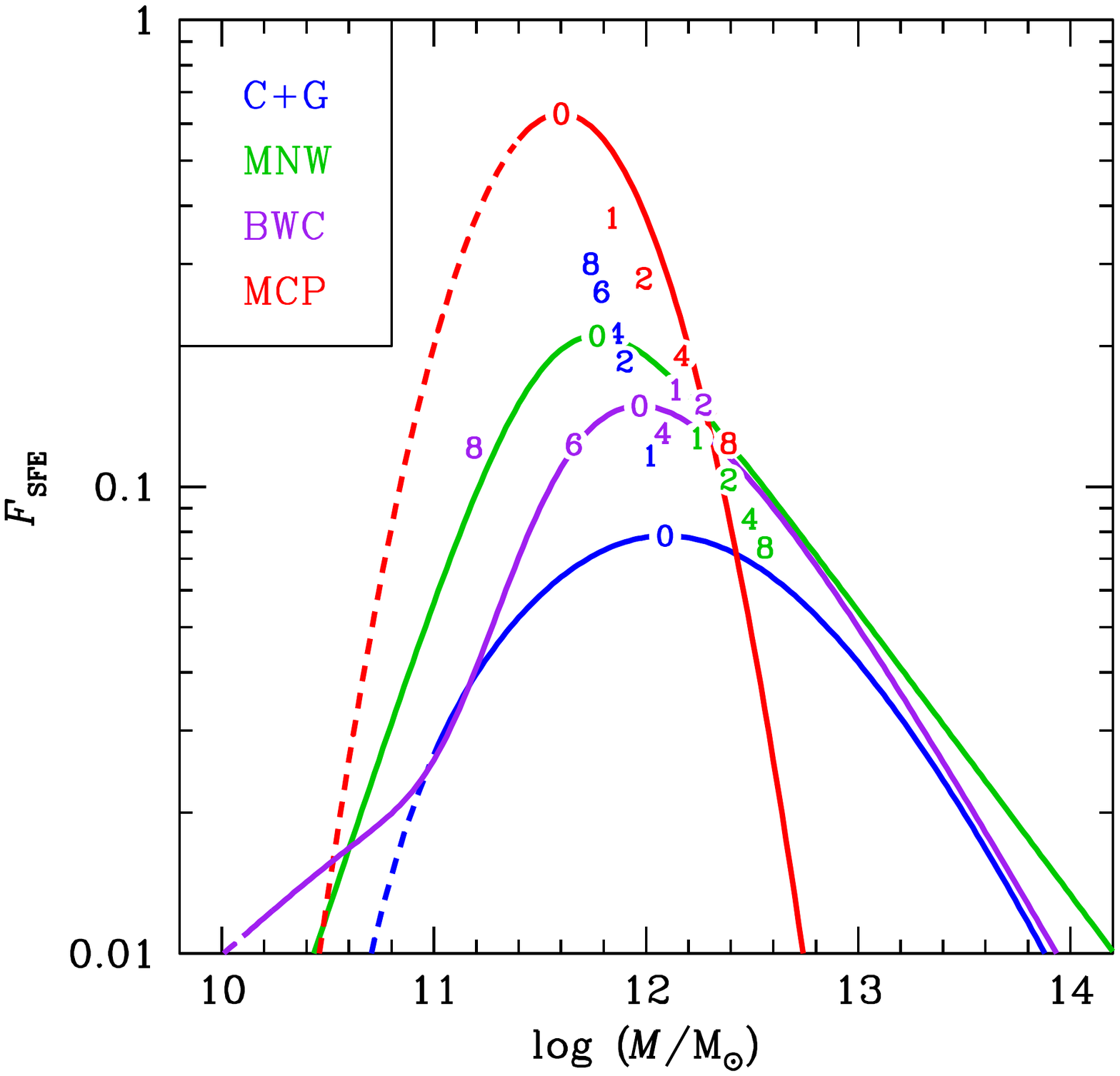}
\caption{Star formation efficiency, $F_{\rm SFE}$, 
of our analytical galaxy formation
    models versus halo circular velocity (\emph{top}) and mass
    (\emph{bottom}) at $z=0$.
The \emph{blue}, \emph{green} and \emph{purple curves} respectively show
$F_{\rm SFE}=m/(f_{\rm b} M)$     for models 
    \CG\ (\citealp{Cattaneo+11} with \citealp{Gnedin00}, 
    eqs.~[\ref{SFE_C11G00}], [\ref{meansfeC11G00}] and [\ref{CDFzreion}]),
    \MNW\ (\citealp{Moster+13}, eqs.~[\ref{SFE_Moster}] and [\ref{pars_Moster}]),
     and \BWC\ (\citealp{Behroozi+13}, eqs.~[\ref{SFE_Behroozi}],  [\ref{f_Behroozi}],
    [\ref{pars_Behroozi}], and  [\ref{pars2_Behroozi}]).
For model \MCP\
(\citealp{Mutch+13},
    eqs.~[\ref{SFE_Mutch}] and [\ref{pars_Mutch}],  \emph{red curve}), 
    $F_{\rm SFE} = 0.7\,\dot m / (f_{\rm  b}\dot M)$.
The \emph{dashed} portions of the curves indicate where the models are extrapolations.
The \emph{numbers} indicate the position of the peak star formation efficiency at
    the corresponding redshift.
    \label{fig:compare_sfe}} 
\end{figure} 

Figure~\ref{fig:compare_sfe} shows the comparison of the 4 models at redshifts
$z=0$ and the peak SFE at redshifts 1, 2, 4, 6, and 8. Note that only the BWC
model was fit by the authors up to $z=8$. 
The four 
models agree to first order, although 
there are some notable differences.
While the \MCP\ model, by construction, has a symmetric $F_{\rm SFE}$ in $\log v_{\rm circ}$ and in $\log M$,
all other models are asymmetric, with a relatively faster decrease of
SFE at low circular velocity or halo masses, except for the
shallow low-end tail in the
\BWC\ model at low redshift.
But recall that the $F_{\rm SFE}$ of the \MCP\ model is defined in a
different fashion (eq.~[\ref{mutch}] vs. eq.~[\ref{mofMz_generic}]). 

The peak SFE 
decreases with time in the \CG\  model,
increases with time in the \MNW\  and \MCP\ models, and is roughly independent of redshift
in the \BWC\  model.
The circular velocity (respectively, halo mass) where this peak SFE is reached 
decreases (increases) slightly with time in the \CG\  model,
increases sharply with time in the \MNW\ and \MCP\ models,
and increases moderately with time in the \BWC\ model up to $z=4$ ($z=2$) then
reverses and decreases with time increasingly faster.

At the high-mass end,
the decrease of $F_{\rm SFE}$ with halo circular velocity at $z=0$
follows a  similar
slope (--2.5) in the physical \CG\ and empirical \BWC\ 
models,\footnote{Admittedly, the high-end slope of SFE versus $M$ in the
  C11 and C+G models (e.g., the denominator in eq.~[\ref{C11}])  was a guess, but it is
 remarkably close to the 
  empirical relation of \BWC.} while the \MNW\ model shows a somewhat shallower
decrease (slope --2). In contrast, 
the \MCP\ model shows an increasingly rapid decrease, with a much steeper slope,
close to $-9$ at 1 percent of peak SFE.

At the low-mass end,
the rise of $F_{\rm SFE}$ with halo circular velocity is fastest for the
\CG\  model ($F_{\rm SFE}$ rises as $v_{\rm c}^9$ according to
  eq.~[\ref{SFE_C11G00}]) and the \MCP\ model, for which the slope is even
  steeper than 9 at levels where SFE is at 1 percent of its peak value.
At $z=0$, the \MNW\  model gives a low-end slope of 4.1 for $F_{\rm SFE}$ vs. $v_{\rm
  c}$, the \BWC\  model has a much shallower slope
of 1.2, while that of the \MCP\ model
slowly tends to
infinity (because of the lognormal SFE vs. halo mass), and at 1~per cent of the peak
SFE, the slope is 3.3 at $z=0$ and 2.1 at $z=5$.
At higher redshifts, while the \CG\  model retains the same steep slopes at
the low-end of $F_{\rm SFE}$ vs. $v_{\rm c}$, the \MNW\  model becomes shallower
with increasing redshift, while the \BWC\  model becomes steeper with redshift.
Thus, at $z=5$, the \BWC's  $F_{\rm SFE}$ vs. $v_{\rm c}$ relation is steeper
(slope 2.9)
than that of \MNW\ (slope 2.1).

\begin{figure}
\centering
\includegraphics[width=\hsize,viewport=0 160 600 690]{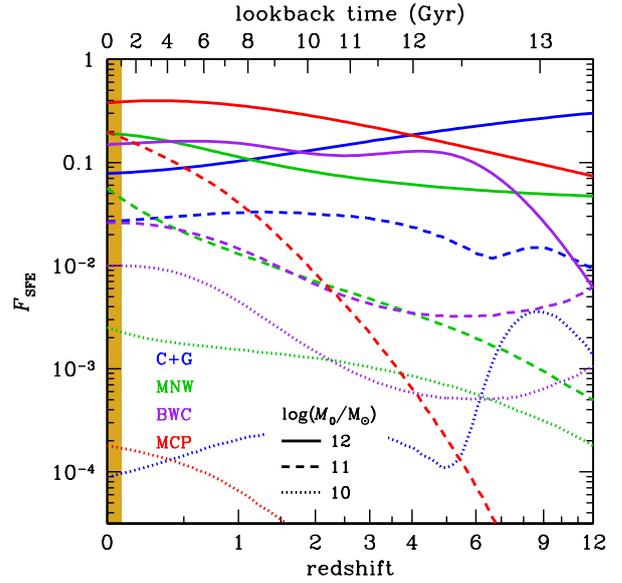} 
\caption{Evolution of the star formation efficiency, at fixed halo mass, 
for the four analytical models (see Fig.~\ref{fig:compare_sfe}),
for final
log halo masses of 10, 11 and 12 (solar units).
The \emph{shaded gold region} highlights the final Gyr.
\label{fig:evolsfe}
}
\end{figure}

Figure~\ref{fig:evolsfe} compares directly the evolution of $F_{\rm SFE}$
between the 4 analytical models.
In particular, the \CG\ model has decreasing SFE in the last 8 or more Gyr
(depending on final halo mass), whereas the \BWC\ model shows a flat SFE in the last
Gyr, while the \MNW\ and \MCP\ models appear to have increasing SFE at all times.
In models with $F_{\rm SFE}$ that decrease with time such as \CG\ and
\BWC\ at late times, a galaxy whose
halo mass remains constant in time (from an unlucky lack of mergers that would otherwise
make it grow) will see a decreasing $F_{\rm SFE}$ at constant $M$, hence a
decrease in stellar mass.
In these cases, our forcing the stellar mass not to decrease comes into effect.
In contrast, the \MCP\ model, based on time derivatives, is the only one
where stellar mass is guaranteed to
never decrease in time.

\begin{figure*}
% SM:
%   SFEvsMandz_pretty CG SFEvsMandz_CGupd_nus.pdf
%   SFEvsMandz_pretty MNW SFEvsMandz_CGupd_nus.pdf
%   SFEvsMandz_pretty BWC SFEvsMandz_CGupd_nus.pdf
%   SFEvsMandz_pretty MCP SFEvsMandz_CGupd_nus.pdf
\centering
\includegraphics[width=0.40\hsize,angle=-90,viewport=0 0 607 730]{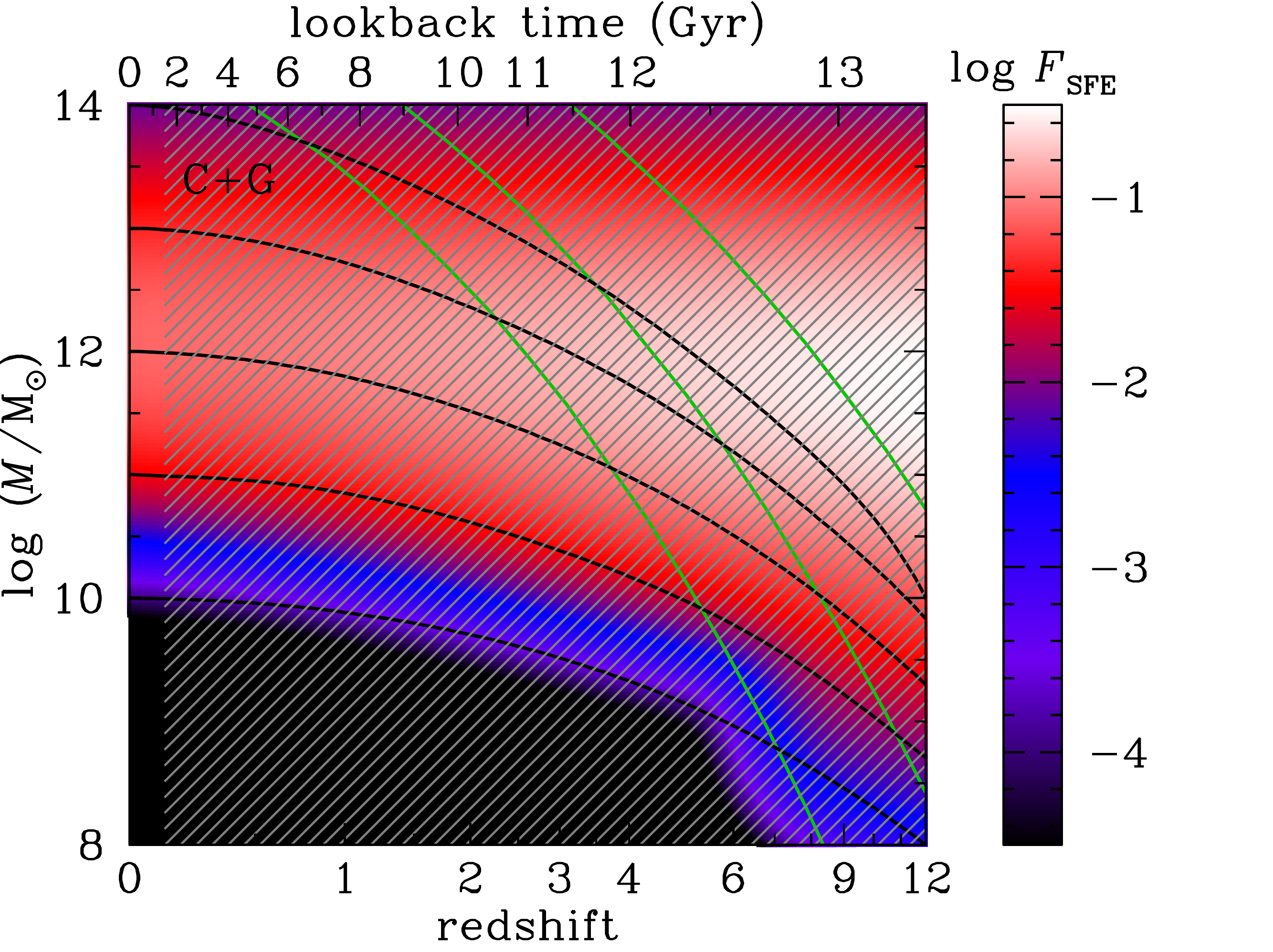}
\includegraphics[width=0.40\hsize,angle=-90,viewport=0 0 607 730]{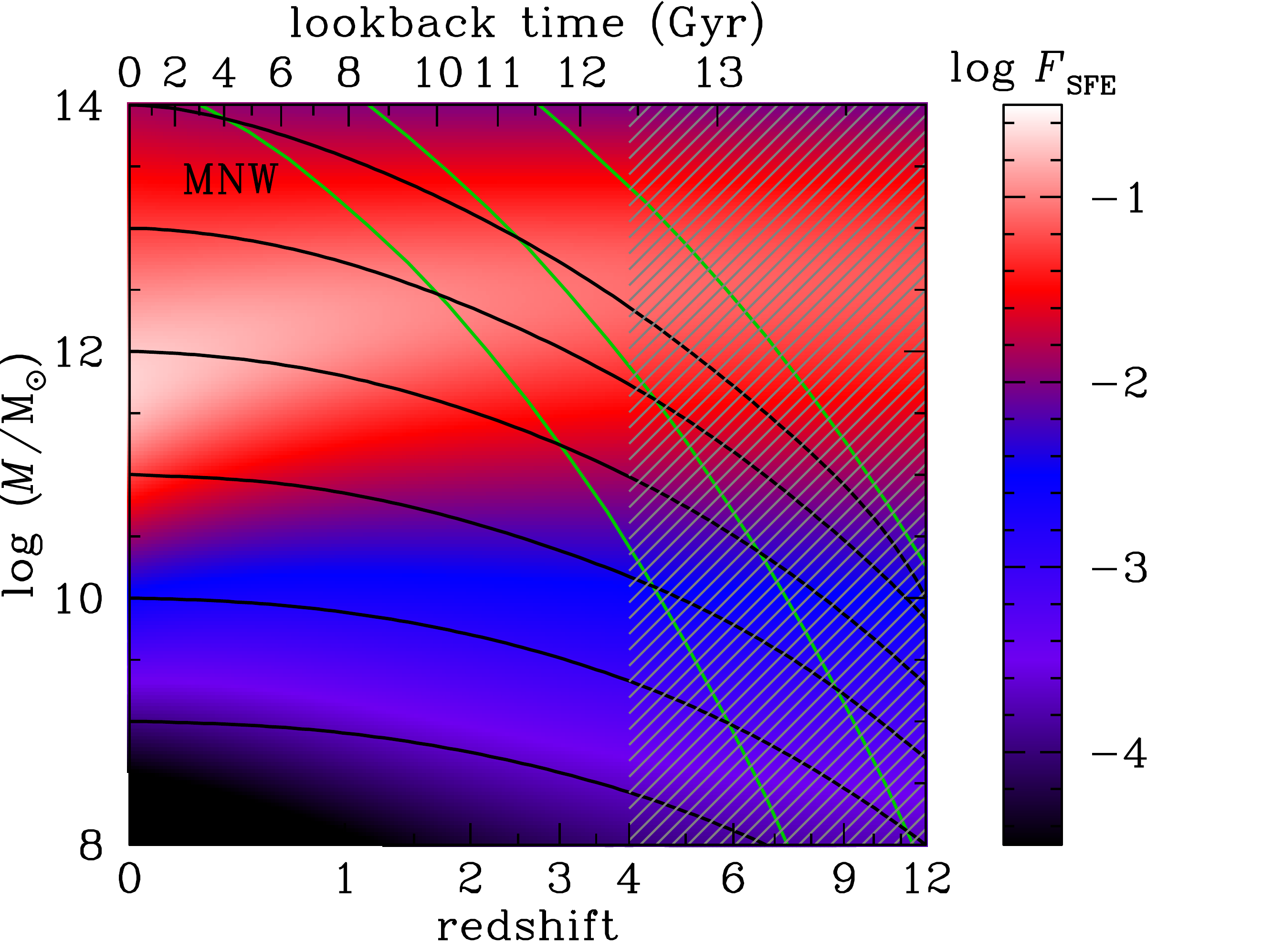}
\includegraphics[width=0.40\hsize,angle=-90,viewport=0 0 607 730]{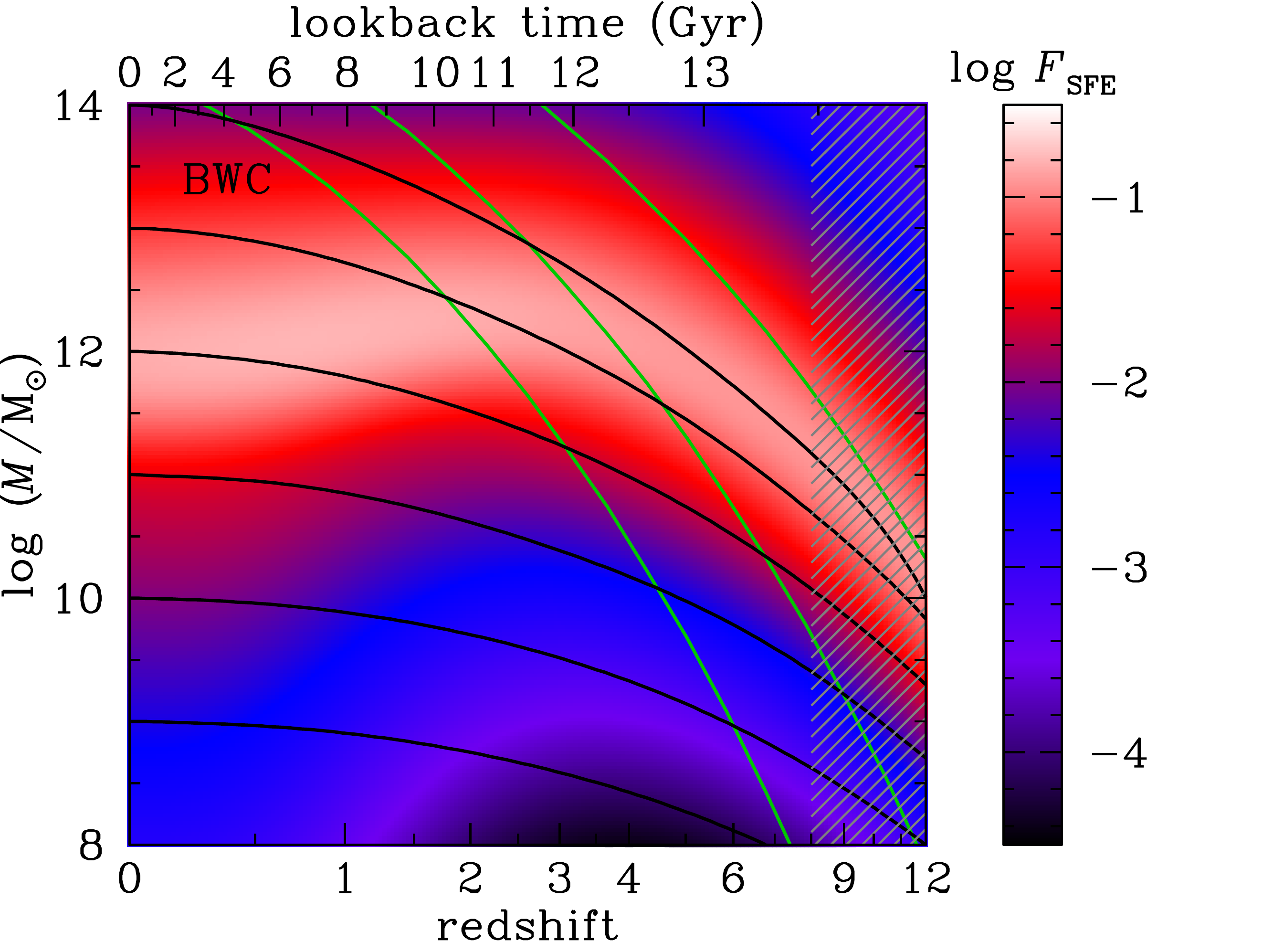}
\includegraphics[width=0.40\hsize,angle=-90,viewport=0 0 607 730]{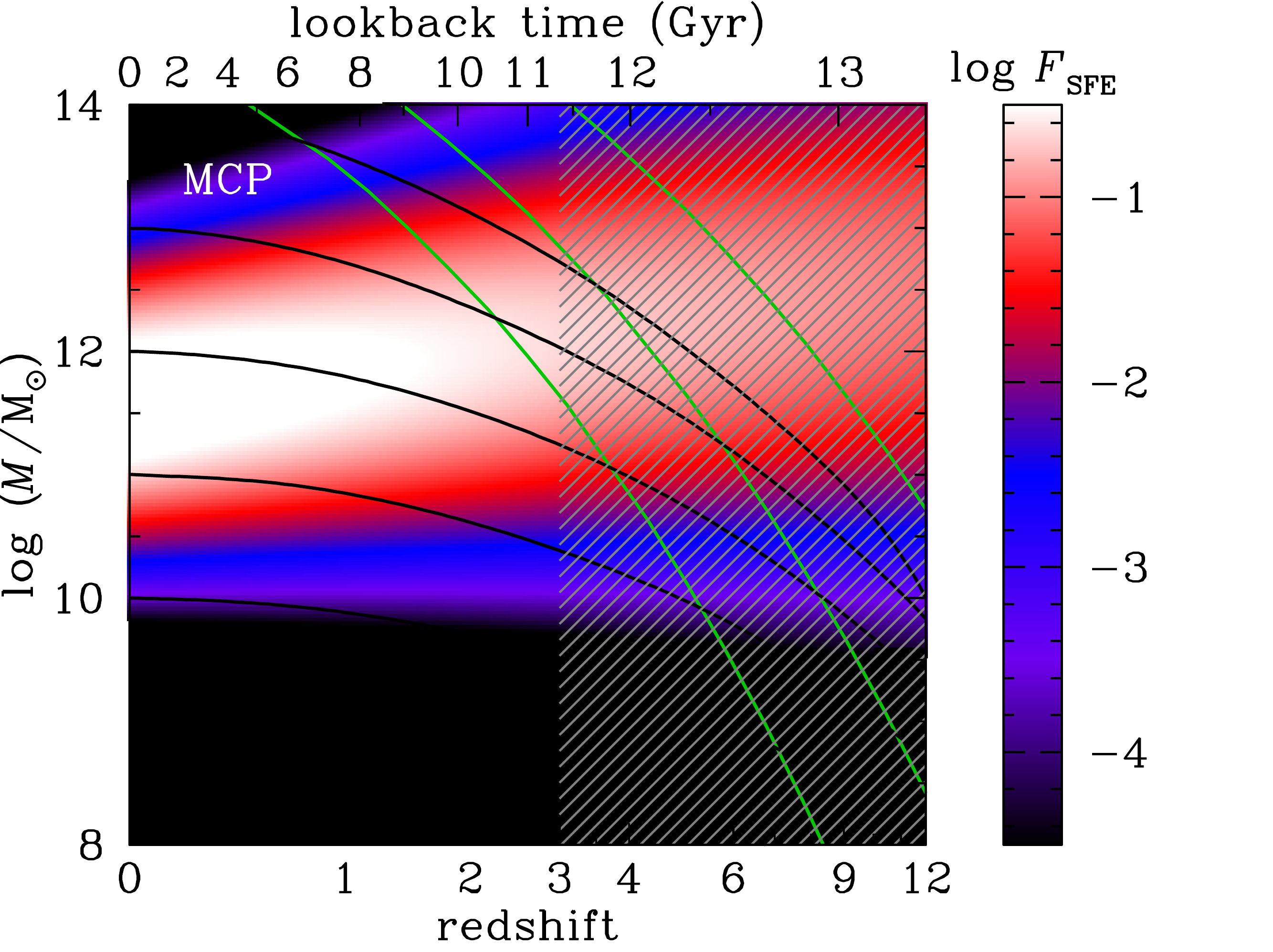}
\caption{Variation of the star formation efficiency with halo mass and epoch
  for the four analytical models (see Fig.~\ref{fig:compare_sfe}).
The \emph{slowly decreasing black curves} indicate the median mass assembly
of the main progenitors of the haloes. 
The 3 \emph{rapidly decreasing green curves} indicate 2, 3 and 5$\,\sigma$ fluctuations (going upwards), so
that the upper right portions of each panel are irrelevant, since such massive
haloes are extremely rare at high redshift.
The \emph{grey hatched} regions indicate the redshifts where the models are extrapolated.
The colour scales of the 4 panels are identical.
Given the different definition of $F_{\rm SFE}$ of the \MCP\ model, it
is not directly comparable to the other three models.
Note the different cosmologies used for each model to link lookback time and redshift.
\label{fig:sfevsz}
}
\end{figure*}

Figure~\ref{fig:sfevsz} illustrates how $F_{\rm SFE}$ varies with both halo mass and
redshift for each of the 4 analytical models. Naturally, the essential
features of Figures~\ref{fig:compare_sfe} and \ref{fig:evolsfe} are recovered:
The higher peak $F_{\rm SFE}$ at low redshift for the \MCP\ model shows as a
whiter colour in Figure~\ref{fig:sfevsz}.
The sharp drops of $F_{\rm SFE}$ at low halo mass of the \MCP\ model at $z=0$ seen in
Figure~\ref{fig:compare_sfe} are visible here as rapid transitions to black
colour at all redshifts (again, recall that $F_{\rm SFE}$ is defined in a different fashion
for the \MCP\ model). This is also the case for the \CG\ model. However, at high mass, the
drop of $F_{\rm SFE}$ with mass at $z=0$ is only pronounced for the
\MCP\ model (again seen at all redshifts).

Given that the intergalactic medium should be fairly cold before reionization
and then is gradually heated by photoionization from galaxies and quasars,
one expects that star formation  at low circular velocities (masses) should
only occur at 
higher redshifts.
The
\CG\ model has this low-mass effect built-in, with $v_{\rm post-reion} > v_{\rm
  pre-reion}$. But, a striking feature of Figure~\ref{fig:sfevsz} 
is that the \BWC\ model is the only one for which the halo mass at peak
$F_{\rm SFE}$ has a clear maximum at $z=3$ (see also
Fig.~\ref{fig:compare_sfe}).
In other words, the
\BWC\ empirical model suggests that the effects of photoionization apply
to all masses, as the entire $F_{\rm SFE}(M)$ is shifted to lower masses at
higher redshifts. 
This ``reionization of the Universe'' feature strongly seen in \BWC\ and also
visible at the low end of the \CG\ model, is not seen in the \MNW\ or
\MCP\ models, because they were tuned to observations that did not extend far
enough in redshift, so the extrapolation to higher redshifts  (grey shaded
regions of Fig.~\ref{fig:sfevsz}) is incorrect.

Figure~\ref{fig:sfevsz} also expands on Figure~\ref{fig:evolsfe} by showing
that not only do models \CG\ and \BWC\ show halo masses with decreasing
$F_{\rm SFE}$, but
models \MNW\ and \MCP\ also show decreasing $F_{\rm SFE}$ for high halo
masses ($\lM \ga 12$).

Moreover, Figure~\ref{fig:sfevsz}  shows that the \BWC\ model is the only one that
displays a decrease at late times of the lower envelope of the halo masses
with efficient star formation. By definition, VYGs require a very late
jump in stellar mass, which should usually come together with a sudden jump
in halo mass (we will discuss this in more detail in
Sect.~\ref{sec:merggrowth}). 
One would therefore expect that, contrary to the other 3 analytical
models, the \BWC\ model will lead to fewer VYGs, because, at relatively late
times, the
SFE will increase as the halo mass rises, causing star formation in the last
few Gyr before
$z=0.08$, which should make it difficult to have a very young stellar
population at $z=0$. 

\subsubsection{Galaxy merging in analytical models}
\label{sec:delayedMergers}
The galaxies in our analytical models can grow in a quiescent mode, via
equation~(\ref{mofMz_generic}) (or (eq.~[\ref{mutch}] for the \MCP\ model),
or by galaxy mergers.
The new stellar mass of the galaxy in halo $i$ of mass $M_i$
is the maximum value of the masses given by the
quiescent growth and by the
growth by galaxy mergers:
\begin{eqnarray}
  m(i,s) &\equiv& {\rm Max} \left[\widetilde m(M_i,s), \sum_{\rm prog} m_{\rm
    old} \right ] \nonumber \\
&=& {\rm Max} \left[\widetilde m(M_i,s),
\sum_{j}m(j,s\!-\!1)\right] \ ,
\label{mnew}
\end{eqnarray} 
where the $j$s are the indices of the progenitors of halo $i$ from the
previous timestep, while
$\widetilde m(M,s)$  
is the model stellar mass at timestep $s$ for a galaxy in
a halo of mass $M$ (eq.~[\ref{mofMz_generic}]).
For the \MCP\ model, the model mass in equation~(\ref{mnew}) is understood to be
$\widetilde m = m_{\rm old} + f_{\rm b} \,F_{\rm SFE}(M,z)\, \Delta M$.

While our different analytical galaxy formation models generally predict
different stellar masses associated to both progenitors, they are run on the
same halo merger trees, hence  galaxy
mergers occur at the same time in each model (but involve different stellar masses).

Our analytical models cannot directly handle possible starbursts in
merging gas-rich galaxies. Instead, we
have implemented 2 different schemes to handle halo mergers that roughly
reproduce the situations of gas-rich and gas-poor mergers.

In our \emph{bursty galaxy merging} scheme,
we delay subhalo (galaxy) merging by a suitable dynamical friction time,
$t_{\rm df}$
(see eq.~[\ref{tdf}] below), measured from the last time when the two
haloes were distinct in the halo merger tree.
We do not model the subhalo
mass evolution (its decrease by tidal stripping), and thus the
\emph{satellite galaxy} that originated from  this branch has a constant stellar mass
until it
merges with the \emph{central galaxy}.
We do not consider satellite galaxies that merge later than $z=0$
because of disk storage issues and because our main focus is on central
galaxies.

With bursty galaxy merging,  
the central galaxy sees a boost in stellar mass at the
time of the halo merger, when the first term within the brackets of
equation~(\ref{mnew}) dominates the second one.
It recovers the masses of the satellites only  after the dynamical
friction time has elapsed.
This
scheme thus creates a burst of star formation at
the first pericentre.
Note that hydrodynamical simulations of
merging galaxies indicate that while a starburst occurs at first pericentre,
there is usually another stronger one at the second pericentre, when the 2 galaxies
complete their merger (e.g. \citealp{Cox+08,DiMatteo+08}), but this cannot be
modelled in our code.

In our \emph{quiet galaxy merging} scheme,
we not only delay the galaxy merger by the dynamical friction time, but we also
modify the stellar mass at the time of the halo merger
(eq.~[\ref{mnew}]).
For this, we subtract from the new halo mass the total mass $\sum M_k$ of all the branches
that 1) merge at the current timestep or before, and 2) contain galaxies
turning into satellites that do not have time to merge into the central
galaxy (after dynamical friction) by the current timestep.
The new central galaxy mass is then
\begin{equation} 
m(i,s) = {\rm Max} \left[\widetilde m\left(M_i\!-\!\sum M_k,s\right),
\sum_{j}m(j,s\!-\!1)\right] \ ,
\label{mnew_quiet}
\end{equation}
where the  $j$s are again (as in the bursty merger scheme) the indices of the progenitors of halo $i$
from the
previous timestep.
With equation~(\ref{mnew_quiet}), there is no boost of stellar mass at the time
of the halo merger, hence our quiet scheme represents well the situation of
dry galaxy mergers.

Following \cite{Jiang+08}, we adopt a dynamical friction time of
\begin{equation}
  t_{\rm df}(M_1,M_2,z) = 
  A\,
  \frac{M_1/M_2}{\ln(1+M_1/M_2)} \,{r_{\rm v}\over v_{\rm v}}\ ,
  \label{tdf}
\end{equation}
where $r_{\rm v}/v_{\rm v}=\sqrt{2/\Delta(z)}/H(z)$ is the orbital circular time divided by $2{\rm \pi}$ at
radius $r_{\rm v}$, with $\Delta(z)$ being the overdensity relative to critical for
collapse, for which we adopt the approximation of \cite{Bryan&Norman98}.
In equation~(\ref{tdf}), $A$ is a dimensionless constant that depends on
orbit eccentricity.
There is some debate on the value of $A$. 
\citet{Jiang+08} calibrated $A=1.43$ using hydrodynamical
cosmological simulations, while 
other values of $A$ range from 0.58
\citep{Springel+01} to 2.34 \citep{DeLucia&Blaizot07,Guo+11,Henriques+15}, with
the latter value adjusted to better fit the observed galaxy optical
luminosity functions with SAMs.
We adopted $A=2.34$ to be consistent with the \Hen\ SAM and the \MNW\ model.

The dynamical friction time of equation~(\ref{tdf}) is always longer than a
few Gyr. It is shortest for equal mass mergers, for which it
is $A/(\ln 2)\,\sqrt{2/\Delta}/H > 0.36/H$ at all times, and is $>0.47/H$  and 
$>6.2\,\rm Gyr$ for lookback times less than
1 Gyr (where we assumed the Millennium cosmology). Therefore, galaxies merge
several Gyr (hence 
many timesteps) after their halos merge. 

Since our focus is on VYGs, which should be gas-rich, we adopt the bursty
merging as
our primary galaxy
merging scheme, but will later compare to the quiet scheme.

\subsubsection{Practical issues for the analytical models run on the Monte-Carlo
  merger trees}

We first run the halo merger tree code, which writes the trees to
files. The analytical galaxy formation models are applied
using a second code that is run on
each halo within each tree.  
This code starts at the highest
redshift of $z=13$ and works forward in time, producing the stellar mass assembly
history within the halo mass assembly history. 
We assign initial stellar masses to each halo according to the model mass of
equation~(\ref{mofMz_generic}) for the first three models, and to  zero for the \MCP\ model.

We analyse the stellar mass history summed up over all
progenitors of a particular halo at redshift zero. Had we only considered the growth of stellar mass of
the main progenitor, we would have measured the stellar mass assembly instead
of the star formation history. We discard all $z$=0 stellar masses below $10^3\,\msun$.

Ideally, since the 
 analytical models have been calibrated on halo
mass functions from different cosmologies, 
it would have been best for us to run the halo merger tree code separately  
for each of our 
analytical models, adapting the cosmological parameters on
which each model was built. This proved difficult to do in practice because
of the large disc space required, hence we ran the halo merger tree code only
once, i.e. with the same cosmological parameters (see
end of Sect.~\ref{sec:halotrees}) for all 4 analytical models.

The different models employ slightly different definitions of halo mass.
Our Monte-Carlo halo merger tree is built with the 
\cite{Parkinson+08} algorithm, which is based on masses of the Friends of Friends
halo membership in the cosmological simulation (S. Cole, private communication).
However, in the halo to stellar mass relations obtained from the empirical
abundance matching technique, the halo mass is defined at the spherical
overdensity of 200 times critical (\MNW) or at \citeauthor{Bryan&Norman98}'s
(\citeyear{Bryan&Norman98}) virial value
(\BWC, \MCP).
The mass used with the \CG\ model is simply that of the Monte-Carlo
halo merger tree, on which the model parameters were roughly calibrated, with
no pre-calibration on halo mass functions. 
These inconsistencies in the mass definition should have no
significant 
effect on the resulting analysis (e.g., $\log (M_{\rm vir}/M_{200})$ is only  0.1).

\subsection{Semi-analytical galaxy formation models}
\label{sec:sam}
We also considered two semi-analytical models (SAMs) of galaxy formation.
In SAMs, galaxies are modelled as discs plus bulges, each with their
characteristic sizes.
These codes include complex physical recipes to incorporate
gas cooling, star formation,
stellar evolution, 
feedback from supernovae and active galactic nuclei (AGN).
These models should be more realistic than the physical analytical models,
in particular because they usually
model the galaxy positions from those of the subhaloes in
the Millennium simulations.
However, nearly all SAMs fail to reproduce many of the
relations observed for galaxies, while the empirical analytical models are
calibrated to these observations. 

\subsubsection{\citet{Henriques+15}}
\label{sec:Henriques}

We first analysed the recent, state-of-the-art SAM of
\cite{Henriques+15} that they ran on the halo merger trees
of the MS-II, rescaled to the Planck cosmology.
The \Hen\ SAM continues the series of SAMs developed by the Munich team
\citep{Croton+06,DeLucia&Blaizot07,Guo+11}.
It also includes ram pressure stripping of satellites in cluster-mass haloes
and
 the formulation of \cite{Gnedin00} (eq.~[\ref{reion_gnedin}]) to prevent gas
accretion on low-mass haloes. 

\citeauthor{Henriques+15} tuned the free parameters  of their model to match
as best as possible the $z=0$ to 3 observations of the stellar mass functions
and the fractions of passive
galaxies. Such a tuning solves several problems in previous implementations of
the Munich model.
In particular, low-mass galaxies no longer form too early (in contrast with
\citeauthor{Guo+11}), thanks to longer timescales for the galactic winds to
fall back into the discs they originated from.

We
extracted  from the {\tt Henriques2015a} table of the Virgo -- Millennium database of the 
German Astrophysical Virtual 
Observatory (GAVO)\footnote{http://gavo.mpa-garching.mpg.de/Millennium/Help}  
the following parameters:
the halo mass (corresponding to $M_{200}$) {\tt centralMvir}, 
stellar mass {\tt stellarMass},
and the \emph{mean} (mass-weighted) age {\tt massweightedAge}, all at $z=0$ with
$m>10^6 \,{\rm M_\odot}$.\footnote{The \cite{Henriques+15} model galaxy ages are
  not directly comparable to those of the other models, because
  \citeauthor{Henriques+15} consider \emph{mean} stellar age, while for the
  other models, we consider the \emph{median} age.}
This extraction yielded $1\,175\,987$ galaxies.
We, hereafter, refer to this virtual catalogue and the corresponding SAM model as \Hen.
Since the analytical models described above fail to consider satellite
galaxies, we limit the \Hen\ catalogue to the central galaxies.

\subsubsection{\citet{Menci+08}}
\label{sec:menci}
Finally, we considered a second SAM, by \cite{Menci+08}, where
the baryonic physics is built on top of Monte-Carlo halo merger trees.
We chose this  SAM, because it 
has been run on two sets of merger trees: one \citep{Menci+08} for the
$\Lambda$CDM cosmology (with $\Omega_{\rm m}=0.3$, $h=0.7$, $\Omega_{\rm
  b}=0.04$, and $\sigma_8=0.9$) and another (\citealp*{Menci+12,Calura+14};
\citealp{Menci+14}) for  
a $\Lambda$ Warm Dark Matter ($\Lambda$WDM or simply WDM) cosmology.
We will only discuss this SAM in the context of estimating the effects of WDM
on the fraction of VYGs.

WDM
is analogous to $\Lambda$CDM, but with the power spectrum truncated at high
wavenumbers, corresponding to 
scales below $\sim$ 1 Mpc, 
as expected for a $m_\chi\sim 0.75$ keV thermal relic particle. 
Note that this choice for $m_\chi$ is fairly extreme, as the observed
abundance of dwarf galaxies around the Milky Way
leads to
$m_\chi >  2.3\,\rm keV$ \citep{Polisensky&Ricotti11,Menci+16}.
This low value of $m_\chi$  sets an upper limit to the effects of
the suppression of primordial density fluctuations at high wavenumber.

For both
$\Lambda$CDM and $\Lambda$WDM cosmologies,  
the merger trees were run
using 21 final, equal-spaced  halo
log masses ($9.5 < \log (M_0/{\rm M}_\odot) \leq 14.5$), with a mass resolution
of $5\times 10^7\,\msun$ for all final halo masses $M_0$. Each $z$=0 halo mass was run
100 times, leading to a total of 2100 trees (see Table~\ref{tab:models}).

In this SAM, gas is converted into stars through two main
channels: a quiescent accretion mode, occurring on long timescales ($\sim$ 1 Gyr), 
and an interaction-driven mode, where gas -- destabilized during
major and minor mergers and fly-by events -- turns into stars on
shorter timescales (typically $\sim 10^7$ yr). 
AGN activity triggered by galaxy
interactions and the related feedback processes are also included.

The \citeauthor{Menci+08} SAM was run on the CDM and WDM trees with the exact
same recipes and parameters for the baryonic physics.  We do not include it
in our standard analyses and figures, so as not to overcrowd our figures,
given that
it uses 100 times
fewer trees than our standard ones and the lower mass ones are poorly
resolved,
and is a much less used SAM 
than that of \cite{Henriques+15}, whose
output is publicly available, and which has the additional advantage of
being calibrated to the stellar mass functions at redshifts up to 3.

\section{Tests of the models}
\label{sec:tests}

\subsection{Stellar versus halo mass}

\begin{figure}
  \centering
% SM: loop_mvsM_Apr17 mvsM_Nov17_dmnew 0 1 1 dmnew 1       DO NOT ERASE THIS COMMENT!!!
\includegraphics[width=\hsize,viewport=0 30 530 755]{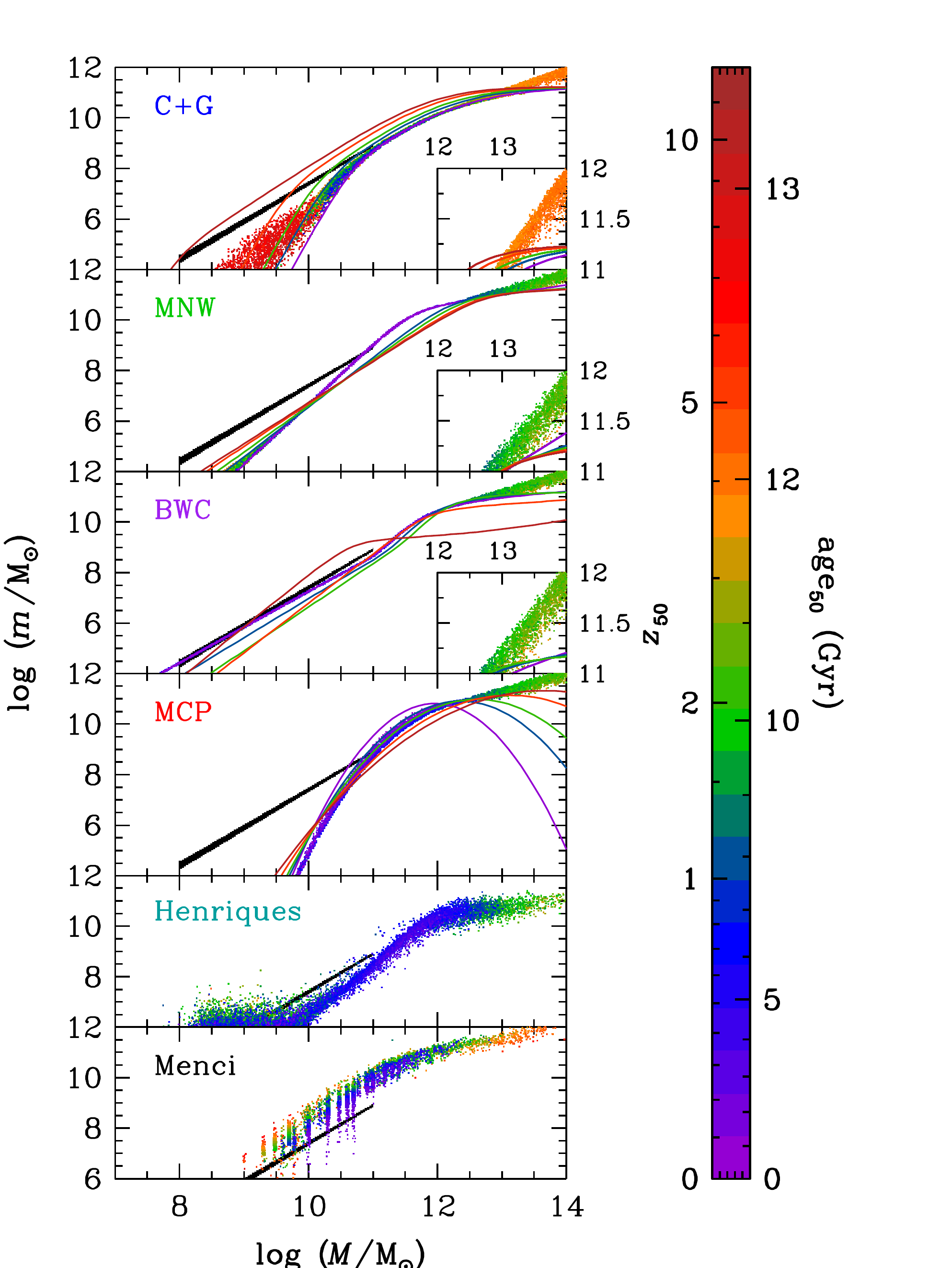}
  \caption{Stellar mass versus halo mass at $z=0$ for the four analytical
    models (with bursty galaxy merging, top panels,  see Fig.~\ref{fig:compare_sfe})
    and for the  SAMs of
     \Hen\ and Menci (\emph{bottom 2 panels}),
    restricted to centrals with galaxy stellar masses $m > 10^6\, {\rm
      M_\odot}$ for lack of mass resolution). 
    Each \emph{point} corresponds to a galaxy extracted from the
    analytical or semi-analytical models run on the halo merger trees.
    The points are 
colour-coded by the median (analytical models) or mean (\Hen\ SAM)
    redshift, $z_{50}$,
    and are plotted in order of decreasing age, with violet colours
    for the galaxies younger than 1.3 Gyr.
    For clarity, only 500 galaxies are shown per 0.25~dex halo mass interval.
    The \emph{curves} show the stellar mass -- halo mass relations 
 directly predicted by the analytical models, with 
    equation~(\ref{mofMz_generic}) for the \CG, \MNW, and \BWC\ models
and equation~(\ref{mutch}) for the \MCP\ model, using the star formation
efficiencies illustrated in Fig.~\ref{fig:compare_sfe}.
These curves are given at redshifts 0,
    1, 2, 5, and 10, following the same colour code as for $z_{50}$.
    The insets show 4$\times$ vertical zooms of the high-mass end.
The \emph{black lines} show the observations of isolated dwarf
irregular galaxies compiled by Read et al. (2017), with uncertainties
illustrated by the line thickness.
    \label{fig:mvsM}
  }
\end{figure}

In this section, we test our analytical models, by
analysing the $z$=0 stellar masses ($m$), halo masses ($M$),   and the
redshift $z_{50}$ (the corresponding stellar age or lookback time,
is called $\rm age_{50}$) when half the mass in stars is formed.
For the \Hen\ SAM, which does not easily provide the epochs when half the
stellar mass is formed, we consider instead
the mass-weighted mean age of the stellar
population, which we also denote as $z_{50}$.

The  stellar mass -- halo
mass (SMHM) relation is a fundamental test of galaxy formation models.
Figure~\ref{fig:mvsM} illustrates  the SMHM, according to the 4 analytical models
as well as the
Henriques and Menci SAMs,
and how galaxy ages depend on their location in the SMHM relation.
Note that the Menci SAM uses 0.25~dex steps of $\lM$, but their output is
shown at $z=0.12$ instead of $z=0$, which leads to somewhat less discretization.

The first aspect to note is that only the \BWC\ model leads to an SMHM
that matches well the observations of 11 nearby, highly-inclined,
isolated dwarf irregulars 
(black line in Fig.~\ref{fig:mvsM})
as compiled by \cite{Read+17}.
The \CG, \MNW, and especially \MCP\ models
predict lower stellar masses at given halo mass than observed by
\citeauthor{Read+17}, as also does the \Hen\ SAM.
On the other hand, the
Menci SAM predicts higher stellar masses at given halo masses than found by
\citeauthor{Read+17}.

The better match to observations with the \BWC\ model appears to be
a consequence of its inclusion of
a change in slope of the SMHM at the low end, which was
not incorporated in the other analytical models. Interestingly, 
\cite{Read+17} noticed that the SMHM they obtained for their sample of 11
dwarf  galaxies follows the
SMHM related predicted from the extrapolation to low masses of the abundance matching of
the SDSS \emph{stellar mass function} (SMF) of \cite{Baldry+08}
with the cosmic halo mass function. This explains the agreement between
their SMHM and that found for \BWC, who had also used the SMF
of \cite{Baldry+08}. In comparison, \MNW\ adopted the same SMF, but
adopted a too restrictive single-slope low-end SMHM parameterization.
Finally, \MCP\ adopted a different set of SMFs.

The second interesting feature is that the most massive galaxies have formed
their stars earlier than the other galaxies (except for the very low-mass
galaxies in the \CG\ model),
i.e. one is witnessing \emph{downsizing}. In contrast the \MNW, Menci,
and especially the \CG\ model also predict that the
smallest galaxies are old, i.e. \emph{upsizing}.
We will analyse in more detail in Sect.~\ref{sec:zvsm} how the typical stellar ages of  galaxies
vary with their $z$=0 stellar mass.

The third interesting aspect of Fig.~\ref{fig:mvsM} is that, at intermediate masses,
the galaxy stellar masses  derived by running the analytical models on the Monte-Carlo
halo merger trees (points) are often close to those directly obtained from the
models for the redshift corresponding to $z_{50}$ (curves of similar colour).
However, 
at the high-mass end, the stellar masses predicted from running the models on
our halo merger trees
are greater than 
those predicted by these two models at $z=z_{50}$ (or at any $z$). For the
\CG, \MNW, and \BWC\ models, this extra stellar mass at high halo mass, which
is clearly seen in the 3 corresponding insets of Fig.~\ref{fig:mvsM},
suggests that we are in a regime where the analytical stellar mass decreases in time,
which our implementation forbids (see end of Sect.~\ref{sec:compare}). 
In Figure~\ref{fig:sfevsz}, the ridge of highest SFE occurs at decreasing
halo mass at later times before $z=0$, except in the \CG\ model (where it is
roughly constant). However, in this physical \CG\ model,
galaxy mergers can boost the stellar mass beyond the analytical prediction
\citep{Cattaneo+11}. 
Thus, the \CG\ model highlights the role of mergers at high masses, while
showing negligible effects of mergers at lower masses, where the analytical
prediction matches well the outcome of the model run on the halo merger
trees.
The importance of mergers at high masses was previously noted by
\citet{Guo&White08} and \citet{Hopkins+10}, who studied the merger rates
of galaxies, by \cite{Cattaneo+11} who studied  (at better mass
resolution)
the fraction of
stellar mass
acquired by mergers, as well as 
by \cite{Bernardi+11} who analyzed the observations of galaxy properties.

The fourth interesting feature of Figure~\ref{fig:mvsM} is that the scatter
in the SMHM relation strongly depends on the model.
The \BWC\  and \MNW\ models produce virtually no scatter for halo
mass below $10^{12} \,\msun$.
In contrast, the \CG\ model and Menci SAM produce a large scatter in the SMHM relation at
low mass. The \Hen\ SAM produces scatter at both low mass (where mass
resolution effects become significant) and high mass (perhaps from the
stochasticity of galaxy mergers).
While the scatter of the stellar mass of central galaxies in the \Hen, and Menci SAMs are
low at intermediate halo masses, it is 
even lower in the EAGLE hydrodynamical simulations \citep{Guo+16}. 
The large scatter in the \CG\ SMHM  is clearly due to its higher SFE of low-mass
haloes before reionization  (see Fig.~\ref{fig:sfevsz}) combined with the
stochasticity in recombination epochs in the \CG\ model. 
The  \MCP\  model  produces some scatter at all masses, as expected from its
use of mass evolution rates instead of masses themselves.

\subsection{Stellar mass function}
\label{sec:SMF}

In theory, the SMF, ${\rm d}n/{\rm d}m$,
can be estimated by integrating over
the cosmic $z$=0 halo
mass function (known from cosmological simulations), ${\rm d}n/{\rm d}M$:
\begin{eqnarray}
{{\rm d}n\over {\rm d}\log m} &=& 
\int {\rm d}\log M\,
{{\rm d}^2n\over {\rm d}\log M\,{\rm d}\log m} \nonumber \\
&=& 
\int {\rm d}\log M\,\left({{\rm d}n\over {\rm d}\log M}\right)\,
\left({{\rm d}n\over {\rm d}\log m}\right)_M \ .
\label{dndlogm}
\end{eqnarray}
Our choice of equal numbers of haloes (trees) in logarithmic bins of mass
produces an unrealistic halo mass function and would thus lead to an
unrealistic galaxy SMF.
We can nevertheless compute
the SMF (in bins $j$ of log stellar mass)
as a sum over bins ($k$) of halo mass, by
normalizing our flat halo log mass function by the cosmic halo mass function.
We thus weight the galaxies $i$ as the ratio of the $z$=0 cosmic halo mass function
to the available $z$=0 halo mass function returned from the Monte-Carlo
trees, i.e.
\begin{equation}
w_i(M_k) = {\left({\rm d}n/{\rm d}\log M\right)_{M=M_k}
\over \Delta n / \Delta \log M_k} \ .
\label{weight}
\end{equation}
We can then write the measured SMF in bins of constant $\Delta
\log m_j$ as
\begin{eqnarray}
  g(\log m_j) &\!\!\!\!\equiv\!\!\!\! &
\left ({{\rm d}n\over {\rm d}\log m}\right)_{m=m_j}\nonumber \\
&\!\!\!\!=\!\!\!\!& 
\Delta \log M_k \sum_k \left({\Delta n\over \Delta \log M_k}\right)w_i(M_k)
\left({\Delta n\over \Delta \log m_j}\right)_k \nonumber \\
&\!\!\!\!=\!\!\!\!& 
{
1
\over \Delta \log m_j}\,\sum_{i \in j} w_i(M_k) \ .
\label{mfunfromdata}
\end{eqnarray}
The second equality of equation~(\ref{mfunfromdata}) makes use of
equations~(\ref{dndlogm}) and (\ref{weight}) and of the constant
value of $\Delta \log M_k$.

We adopted the $z$=0 cosmic halo mass function of \citet{Warren+06}, 
computed using {\sc HMFCalc}\footnote{http://hmf.icrar.org} \citep*{Murray+13}
with
the cosmological parameters used in the Millennium Simulations, and for an
overdensity of 200 relative to critical.
\footnote{The choice of the model for the halo mass function has little
  impact on the \emph{relative} weights used to compute the fraction of young
    galaxies in Sect.~\ref{sec:method2} below; it only affects the SMF in
    Fig.~\ref{fig:fofm},
    and only slightly so.}
\footnote{\cite{Benson17} has shown that the weights of
  equation~(\ref{weight}) are slightly incorrect, because backsplash
  (sub)haloes are double counted. The effects are relatively small (at most 10 per
  cent of haloes of a given mass above $10^{11}\,\msun$ are backsplash
  according to fig.~4 of \citeauthor{Benson17}).} 

\begin{figure}
\centering
% SM: fofm_Sep17 file dmnew
\includegraphics[width=\hsize,viewport=1 160 580 680]{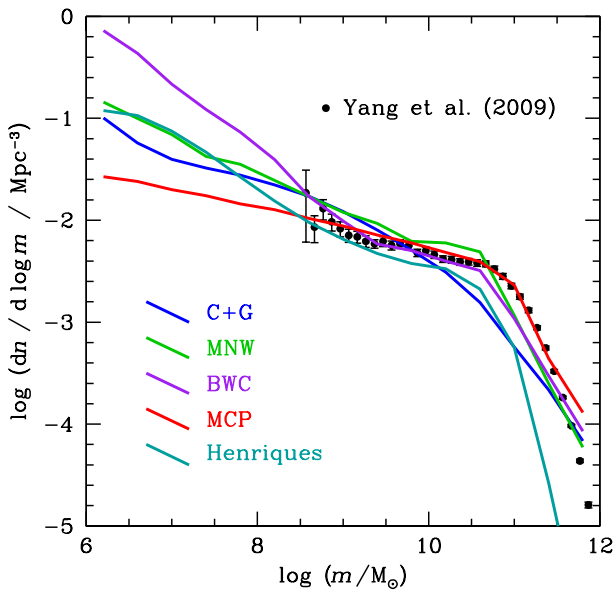} 
 \caption{Galaxy stellar mass functions at $z=0$ for the different galaxy
   formation 
   models: the 4 analytical models (with bursty galaxy merging,
   see Fig.~\ref{fig:compare_sfe}),
  as well as for the
  semi-analytical model of
Henriques et al. (2015,
\emph{cyan}), 
  for centrals only.
  The symbols represent the observed central galaxy stellar mass function
of Yang et al. (2009),
  converted to $h=0.7$ and to the 
Chabrier (2003) initial mass
  function.
\label{fig:fofm}
}
\end{figure} 

\nocite{Chabrier03}
\nocite{Yang+09}

The resultant $z$=0 galaxy SMFs produced by our Monte-Carlo
runs for the analytical models
are shown in Figure~\ref{fig:fofm}, as well as that of the \Hen\ SAM
(restricted to central galaxies) and the observed  central galaxy
stellar mass function of \cite*{Yang+09}.
The 
figure indicates that the \CG\ model fails to reproduce
the knee of the SMF at $\log(m/{\rm M_\odot}) \simeq 10.8$. 
In the range $7 \leq \lm \leq 8$, the logarithmic slopes of the model SMFs
are
$\approx -1.28$, --1.36, --1.66,  $-1.18$, and $-1.51$ for the \CG, \MNW, \BWC,
\MCP, and \Hen\ models, respectively.
In comparison, the slope of the observed 
SMF of \citeauthor{Yang+09} 
is $\approx -1.65$ at $8.7 \leq \log (m/\msun) \leq 9.3$.
Hence, the observed low-end slope is best (worst) matched by the \BWC\ (\MCP)
model.
With the quiet galaxy merging scheme, the SMFs from the analytical
models
are similar, but
shifted down by 0.4 dex, thus matching less well the observed one.

\subsection{Age versus mass}
\label{sec:zvsm}

\begin{figure}
\centering
% SM: plotz50vsm_5models file 1 dmnew 1000
\includegraphics[width=\hsize,viewport=160 10 570 740]{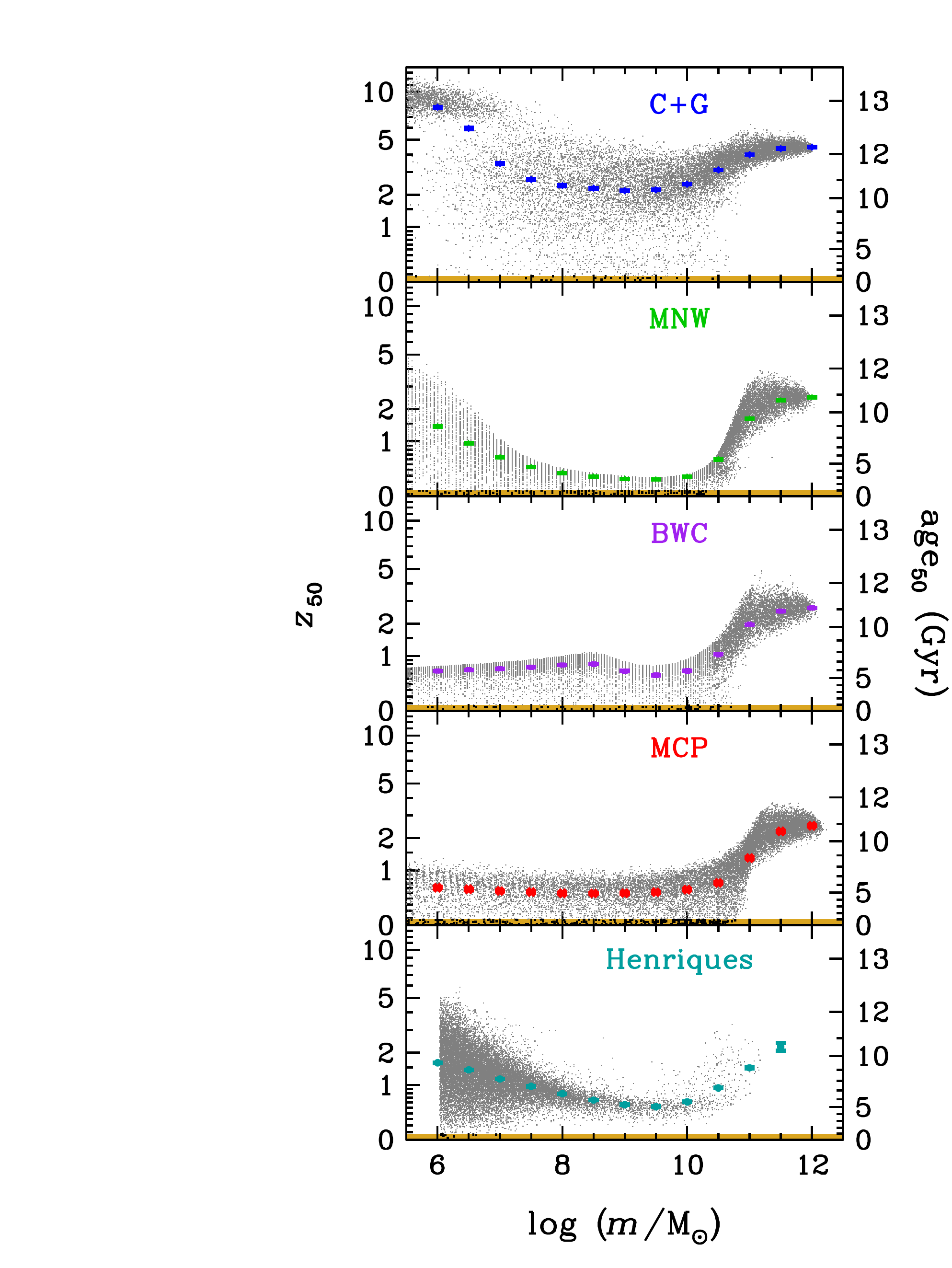}
\caption{Typical redshift of star formation versus galaxy stellar mass for
  the analytical models with bursty galaxy merging (median redshifts) and for
  the central galaxies in the Henriques semi-analytical model (arithmetic
  mean redshifts). The corresponding ages are shown on the right axis. 
The \emph{grey points} show a random subset of 20$\,$000 galaxies.
The \emph{large coloured symbols}  show the means (Henriques) or weighted
means (other models, eq.~[\ref{weight}])  of $\log 
(1\!+\!z_{50})$, while the error bars
display the uncertainties ($\sigma/\sqrt{N}$ for Henriques, and using 1000
bootstraps for the other models).
The \emph{gold shaded regions} indicate ages less than 1 Gyr, and the
very young galaxies (from the same random subset) are highlighted as
\emph{darker and thicker points}.
\label{fig:z50vsm}}
\end{figure}

We now study how the epoch, $z_{50}$,
when half the
$z$=0 stellar mass is formed in galaxies
varies with the $z$=0 stellar mass.
This is illustrated in
Figure~\ref{fig:z50vsm}
for the four analytical models of $F_{\rm SFE}(M,z)$,
as well as for the
\Hen\ SAM (for which the redshift corresponds to the arithmetic mean age of
the stellar population of the galaxy).

All models show downsizing at the high-mass
end, where $z_{50}$ increases with stellar mass. This is
 due to the negative  slope of the SFE - halo mass relations at high halo masses
(Fig.~\ref{fig:compare_sfe}): the galaxies with the highest stellar masses
 today live in high mass haloes that grow faster than low-mass ones
 (\citealp{vandenBosch02} and black curves of Fig.~\ref{fig:sfevsz}) and also
 faster than the mass of the peak SFE (whiter ridges in Fig.~\ref{fig:sfevsz}),  
leading to early quenching of star formation
(see \citealp{Cattaneo+06,Cattaneo+08}).

At low galaxy stellar mass, the behaviour of $z_{50}$ versus mass
relation differs between models.
While the
\BWC\   and  \MCP\   models show 
% instead
a fairly flat typical median age versus
mass for $\log (m/\msun) < 9.5$, 
the \CG\ and \MNW\  models lead to typically earlier star formation at increasingly lower galaxy
masses, i.e. \emph{upsizing}, for $\log (m/\msun)<9$. 
Finally, the \Hen\ SAM leads to an intermediate behaviour with weak upsizing for
 $\log (m/\msun)<10$.

In particular, the \CG\ model leads to very old low-mass ($m < 10^7\,\msun$)
galaxies,
most of which are formed by $z=8$ (see upper left panel
of Fig.~\ref{fig:z50vsm}). Since these low-mass galaxies have final halo
masses $\log (M/{\rm M_\odot}) < 10.2$ (upper panel of Fig.~\ref{fig:mvsM}), one can see that
 these old galaxies were quenched by the rising halo mass linked to the
 rising entropy and  temperature of the intergalactic medium during the epoch
 of reionization (see upper left panel of Fig.~\ref{fig:sfevsz}).
This is
qualitatively consistent with the star formation histories of low-mass
(ultra-faint) Local Group dwarf spheroidals \citep{Weisz+14}. However,
such low-mass
galaxies with very old stellar populations are currently difficult to observe
beyond
the Local Group, and may be limited to satellites that are ram
pressure-stripped by the fairly hot gas of their more massive centrals.

In the same mass range, Figure~\ref{fig:z50vsm} also displays a younger
galaxy population for the \CG\ model,
pretty similar in age to intermediate mass galaxies. These galaxies were formed
after the epoch of reionization. 
These two (very old and younger) galaxy populations for the \CG\ model are
related to 
our two $v_{\rm reion}$ values.  Their overlap in mass is caused by our adopted
stochasticity of $z_{\rm reion}$. This model is thus particularly well-suited 
in exploring the epoch and duration of reionization through the age distribution of dwarf galaxies.

The other source of discrepancy between the models is the median redshift of
star formation of
galaxies at intermediate stellar masses ($8 \leq \log(m/{\rm M}_\odot) \leq
10$):
roughly 2.5, 1, 0.7, 0.6, and 0.2 (the corresponding lookback times being roughly 
11, 7.5, 6.5, 5.5, and 2.5 Gyr) for the \CG, \Hen, \BWC, \MCP, and \MNW\ models,
respectively.

Finally, the scatter in $z_{50}$ is high in the \CG\ model (except at low
mass) and in the \Hen\ model (at low mass),
intermediate in the \BWC\ and \MCP\ models, and small in the \MNW\ model.

\section{Fraction of very young galaxies}
\label{sec:result}

\subsection{Estimation of (weighted) fractions of very young galaxies and their
  uncertainties} 
\label{sec:method2}
Given the data of Figure~\ref{fig:z50vsm}, we can derive the fraction of
VYGs  (i.e., $z_{50}<0.08$,
corresponding to median stellar ages of less than 1 Gyr).
We first need to normalize the raw fractions of VYGs by correcting
our flat  halo mass  function (in terms of $\log M$) to the cosmic halo mass
function, as  described in Sect.~\ref{sec:SMF}.
The fraction of VYGs in the $j$th bin of log stellar mass can
easily be obtained from equation~(\ref{mfunfromdata}) to yield (dropping all
the log terms for clarity)
\begin{equation}
f_{\rm VYG}(m_j) = 
{\displaystyle \sum_{i\in j,\rm VYG} w_i(M_k)\over 
\displaystyle  \sum_{i\in j} w_i(M_k)}
\label{fyoung}
\end{equation}
where $w_i(M_k)$ is, again, the weight for the galaxy $i$ in the $k$th bin among
the $N_k$ halo log mass bins, required to
normalize the measured SMF (eq.~[\ref{weight}]).

The uncertainties on $f_{\rm VYG}$ cannot be
 derived 
with the usual binomial formula
\begin{equation}
\sigma^2[f_{\rm VYG}(m_j)] = \sqrt{f_{\rm VYG}(m_j)\,
[1-f_{\rm VYG}(m_j)]\over N_{\rm eff}} \ ,
\label{errfyoung}
\end{equation}
because it is not clear what value to adopt for $N_{\rm eff}$, the effective
(given the weights) number of galaxies in bin $j$ of stellar mass. Instead, we
rely on bootstraps: for each bin, $j$, of stellar mass,
we bootstrap the sample 
to obtain $B$ values for 
 $f_{\rm VYG}$ using equation~(\ref{fyoung}), and 
then determine the uncertainty from the standard deviation of these $B$
values of $f_{\rm VYG}$. 
Unfortunately, the bootstrap method cannot provide errors when there are no VYGs.
In that case, we replace the numerator of $f_{\rm VYG}$ in
equation~(\ref{fyoung}) by $w_K/2$, where $K$ is the bin of weight (or halo
mass) that contains the youngest of all galaxies in the stellar mass bin
(by construction, the youngest galaxy will have a median age greater than 1 Gyr).

This weighting scheme is evidently not used for the \Hen\   model, since it
is applied to the Millennium simulations, which have realistic halo mass
functions.
In this case, we adopt the standard binomial errors (eq.~[\ref{errfyoung}],
replacing $N_{\rm eff}$ by the number $N$ of galaxies in the bin $j$ of
stellar mass). 
When there are no VYGs  in the bin $j$ of stellar mass (i.e.,
$f_{\rm VYG} = 0$), we adopt
the \cite{Wilson1927} upper limit:\footnote{See the\emph{ Wilson score} in the
  {\sc Wikipedia} entry on the {\sf Binomial proportion confidence
    interval}.}
\begin{equation}
f_{\rm VYG}^{\rm upper} = {z^2/N_{\rm eff} \over 1+z^2/N_{\rm eff}} 
\label{upperWilson}
\end{equation}
for $z =1.65$ standard deviations, corresponding to a 95 percent confidence
level for this upper limit.

\subsection{Fractions of very young galaxies for the different models}
\label{sec:fyoungmodels}
\begin{figure}
  \centering
% SM: fyoung_Dylan_new fyoung_Nov17 0 0 1 [1000 for read with 1000 bootstraps]    DO NOT ERASE THIS COMMENT!
\includegraphics[width=\hsize,viewport=0 0 560 540]{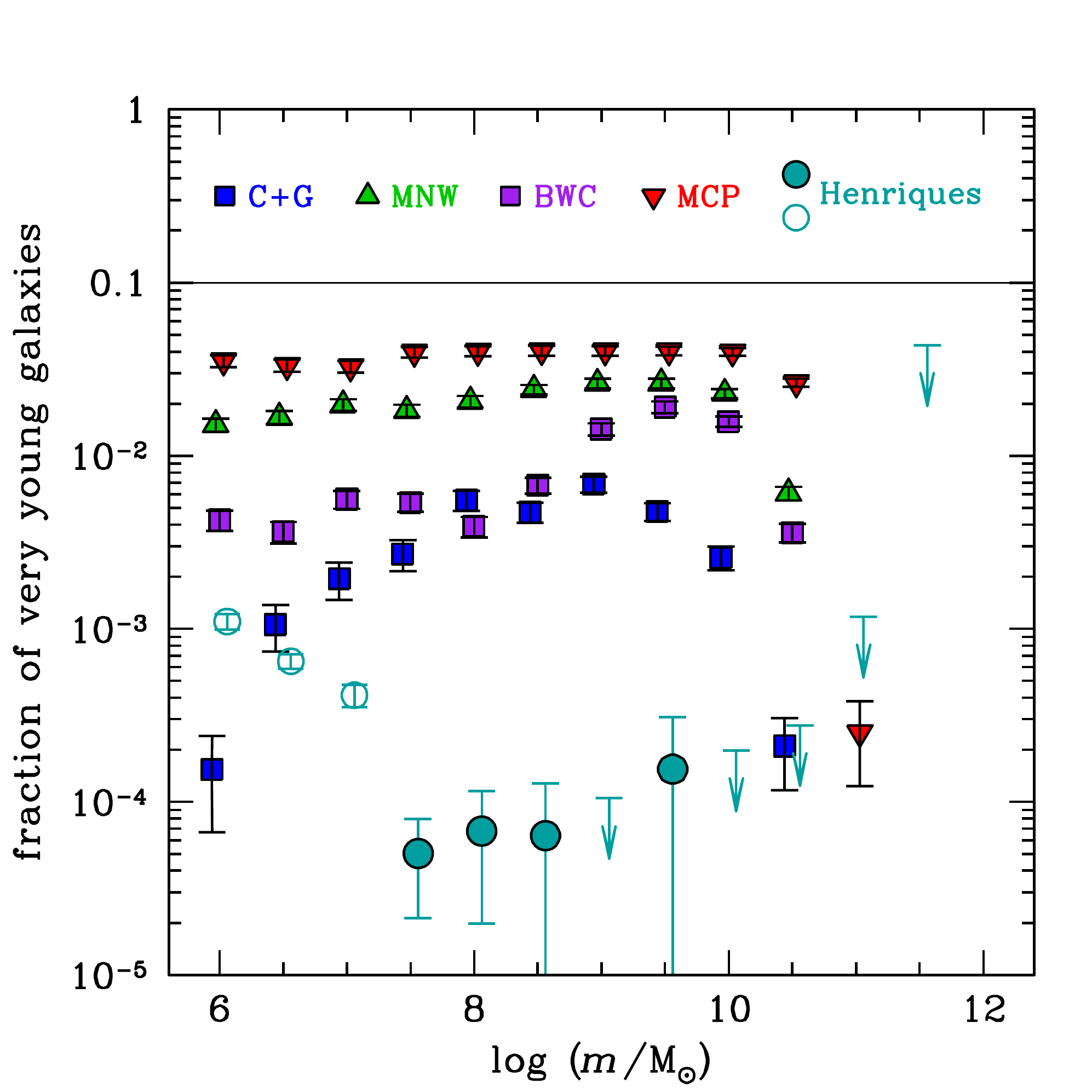} 
    \caption{Fraction
      of very young galaxies (most of the stellar mass formed within the last Gyr) at
      $z=0$ as a function of galaxy stellar mass for
      for
the analytical galaxy
formation models (see Fig.~\ref{fig:compare_sfe}) with the bursty galaxy
merging scheme.
Also shown are the fractions
 of central 
galaxies with mean stellar age less than 1 Gyr in the Henriques et
 al. (2015)
 semi-analytical model (\emph{blue-green circles}, \emph{filled} or
 \emph{empty} depending on whether the mass resolution is adequate or not).
For the analytical models, which need to be weighted by the halo mass
function, the fractions of very young galaxies are
taken from equation~(\ref{fyoung}), and the \emph{error bars} show the
uncertainties estimated through 1000
bootstraps (see end of Sect.~\ref{sec:method2}).
The abscissae are slightly shifted for clarity.
The uncertainties for the \Hen\ semi-analytical model are from binomial statistics,
with
\emph{upper limits} indicating 95
percent confidence using Wilson (1927) statistics, see 
eq.~[\ref{upperWilson}]). 
   \label{fig:fyoung}
}
\end{figure}

\nocite{Wilson1927}
Figure~\ref{fig:fyoung} shows our predicted fractions of VYGs
 as a function of galaxy stellar mass, for the analytical 
 galaxy formation models
with bursty galaxy merging
 and for the \Hen\ SAM.
The \MNW, \BWC\  and \MCP\ 
 models predict a fraction of VYGs in the range of 0.3 to 4
per cent for a
 stellar mass range of $10^6$ to $10^{10.5}\,\msun$, with maxima of 2 to
 4~per cent at $\log (m/\msun) \approx 9.5$. The 
 physically motivated \CG\ model  predicts the lowest fraction of VYGs, 
with a peak at 0.7~per cent at $\lm = 9.0$, and no plateau at lower masses.

Interestingly, for virtually all masses, the four models are ordered in the same way,
with increasingly higher fractions of VYGs at given
stellar mass for
the \CG, \BWC, \MNW, and \MCP\ models.
We will present a simple model in Sect.~\ref{sec:fyoungana}
to explain our results for the first three models.

In contrast, the \Hen\ SAM leads to very low fractions (less than 0.03 per
cent) of VYGs (defined here as mean mass-weighted age below 1 Gyr) at all
masses where it is well resolved ($\lm \geq 7.5$). The \Hen\ model thus
appears to be discrepant  
with the 4 analytical models, with 30 to 800 times
fewer VYGs predicted by the \Hen\ SAM at intermediate stellar masses (e.g.
$\log m = 10^9\,\msun$). 
While these predictions are for the central galaxies in the \Hen\ SAM, we
will see in Sect.~\ref{sec:limits} below that satellites and centrals have
similar fractions of VYGs, at these intermediate masses.
At masses below $10^{7.5}\,\msun$,
the \Hen\ satellites
are typically 5 times less likely to be VYGs than are centrals.
However, we do not trust the \Hen\ results below $10^7\,\msun$, because
the \Hen\ SMHM relation, displayed in Fig.~\ref{fig:mvsM}, shows a
suspicious flattening at $\log (m/{\rm M}_\odot) \approx 6-7$, which appears to
be caused by the limited mass resolution of the $N$ body simulation on which
the \Hen\ SAM was run.

\subsection{Dependence on critical age}

\begin{figure}
  \centering
  % SM cdfage50 cdfage_Nov17_dmnew dmnew 1
\includegraphics[width=0.99\hsize,viewport=1 160 560 730]{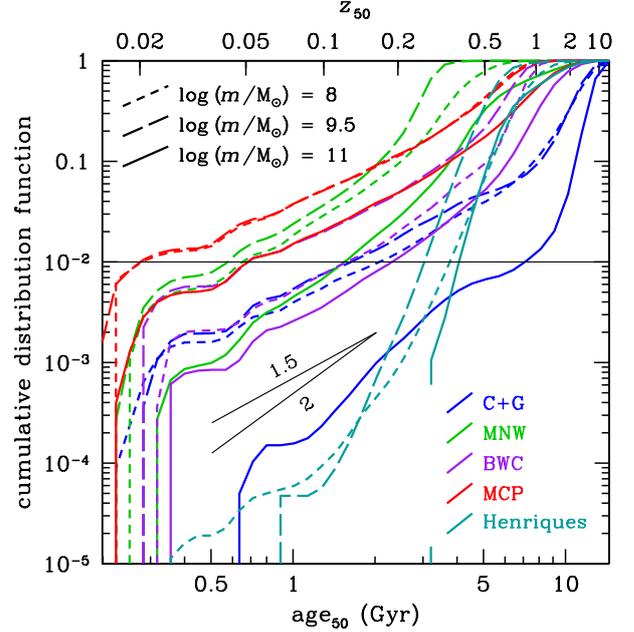} 
\caption{Cumulative distribution functions of galaxy ages
  for the different models (same colours as in
  Fig.~\ref{fig:fofm}) and for three different bins of stellar mass: $\log (m/{\rm M}_\odot)
  = 7.25-8.75$ (\emph{short dashed}), 8.75--10.25 (\emph{long-dashed}), 10.25-11.75
  (solid).
  The ages are medians (half the mass formed) for the analytical models
  (with the bursty galaxy merging scheme) and
  arithmetic means for the Henriques semi-analytical model.
The \emph{short segments} indicate slopes of 1.5 and 2, while the 
\emph{horizontal line} indicates the youngest 1~per cent of galaxies.
\label{fig:cdfage}} 
\end{figure}

One may ask how the fraction of VYGs depends on our choice of 1 Gyr for
the greatest allowed age for very young galaxies.
Figure~\ref{fig:cdfage} shows the cumulative distribution function (CDF) of
galaxy ages for the different models (using bursty galaxy merging for the
analytical ones) in three bins of stellar mass.
This allows us to measure how the VYG fraction scales with critical ages not
necessarily equal to 1 Gyr.
The slopes of the curves near 1 Gyr indicate that the fraction of VYGs
roughly scales
as $\rm age_{50}^{3/2}$ for most models, but is even more sensitive to the
critical age  (slope of
roughly 2) for models \CG\ and \Hen\ at high stellar mass.

Figure~\ref{fig:cdfage} also indicates wide variations between the models for
the ages of the youngest, say, 1~per cent of galaxies (horizontal line).
For the intermediate bin of stellar mass ($\log (m/{\rm M}_\odot)=9.5\pm0.75$,
long dashed lines), the ages of the 1~per cent youngest galaxies are 
0.25, 0.55, 0.65, 1.5, and 3 Gyr for the \MCP, \MNW, \BWC, \CG, and
\Hen\ models, respectively. The discrepancy is as pronounced for the high-mass
galaxies ($\log (m/{\rm M}_\odot)=11.0\pm0.75$, solid lines):
0.65, 1.5, 2.2, 4, and 7 Gyr, in the same order of models, except that
\CG\ switches with \Hen\ for the highest age.

\subsection{Dependence on the galaxy merging scheme}
\label{sec:mergscheme}

\begin{figure}
\centering
 % SM: fyoungvsm_2mergingschemes fyoungvsm_2mergschemes_Hen 1 1000 1
\includegraphics[width=\hsize,viewport=0 0 570 540]{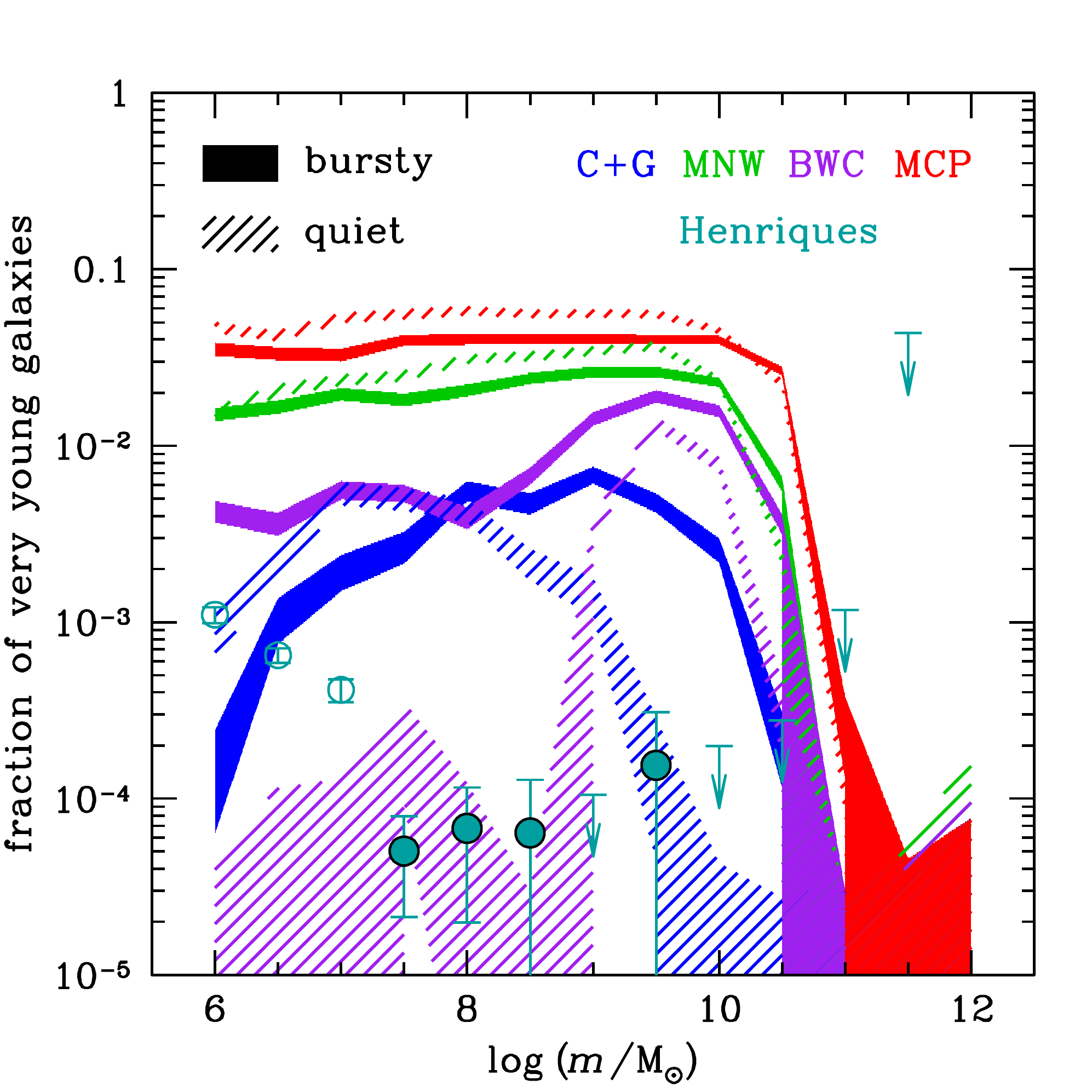}
\caption{Effect of the galaxy merging scheme used in the analytical models on the
  fraction of very young galaxies vs. stellar mass.
  The \emph{solid} and \emph{hatched shaded regions} denote the bursty and
  quiet merging schemes, respectively, where the weights are taken into
  account according to equation~(\ref{fyoung}), and the errors are based upon 1000 bootstraps. 
  The \emph{blue-green points}
  indicate the semi-analytical model Henriques for reference (open circles
  when the mass resolution is not adequate.
\label{fig:fyoung_quiet}}
\end{figure}

Figure~\ref{fig:fyoung_quiet} compares the VYG fractions versus mass for the two
galaxy merging schemes. It clearly shows the lack of sensitivity of the
VYG fractions predicted by the \MNW\ and \MCP\ models to the presence of a burst of star formation 
associated with the halo merger.
However, the merging scheme has an important effect
on the \CG\ and \BWC\ models.
For the \BWC\ model, the fraction of VYGs is above 0.1 per cent only in a
  narrow range of stellar masses around $\lm=9.5$
  for the quiet galaxy  merging scheme, with
  negligible fractions at masses lower than $\lm=8.5$.
  For the \CG\ model, the fraction of VYGs is mostly reduced at intermediate
  and high stellar masses for the quiet
  merging scheme.
  With the quiet merging scheme, only the \BWC\ and \CG\ analytical models reproduce the
very low VYG fractions obtained with the 
Henriques SAM for stellar masses $7.5 \leq \lm \leq 9$ and $9.5 \leq \lm \leq
10$, respectively,  while the other
analytical models predict
higher VYG fractions in these mass ranges. 
We will discuss the reasons for these behaviours in Sect.~\ref{sec:fyoungana}.  

\section{Discussion}
\label{sec:discuss}

\subsection{Limitations of the galaxy formation models}
\label{sec:limits}

The galaxy formation models that we considered to assess the frequency of
VYGs have several limitations.

The analytical and \Hen\ SAM were calibrated to observational data
in a limited range of redshifts and masses (highlighted in
Figures~\ref{fig:compare_sfe} and \ref{fig:sfevsz}, respectively, for the
analytical models). The fractions of VYGs at
  stellar log masses (in solar units) lower than 8.8 (\CG), 7.4 (\MNW\ and \BWC), 10.0 (\MCP)
  are obtained by extrapolating the models.  Similarly, the \Hen\ SAM is
  calibrated at $z=0$ to $\lm > 7.2$, and at lower stellar mass there is a
  suspicious flattening of the SMHM in Figure~\ref{fig:mvsM}, which may be
  due 
to insufficient mass resolution.

\begin{figure}
\centering
% SM: fyoungvsm_SAM_massresol file Henriques 1.5 [1-for-read]   !!! DO NOT REMOVE THIS LINE !!!
\includegraphics[width=\hsize,viewport=1 160 580 680]{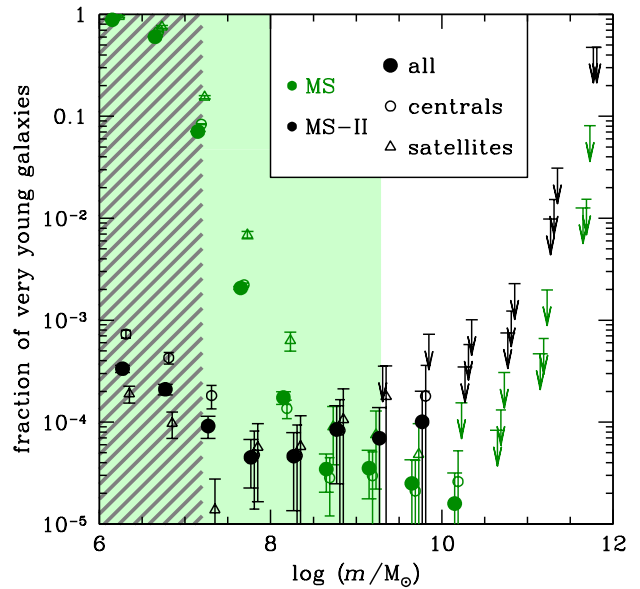}
\caption{Effect of the mass resolution of the parent dark matter simulation
  on the fractions of very young galaxies (mean stellar age less than 1 Gyr) 
obtained with the Henriques et
  al. (2015) semi-analytical model.
The \emph{green} and \emph{black symbols} respectively show the results for the Millennium
simulation and for the
  Millennium-II simulation (\emph{black}), which has 125 times better mass
  resolution. 
The \emph{light green} and \emph{hatched grey shaded regions} indicate the ranges of
stellar masses where the Millennium and Millennium-II simulations are not
correctly resolved.
The Figure highlights the differences between centrals (\emph{open circles}),
satellites (\emph{triangles}) and all galaxies (\emph{large filled circles}).
The abscissae are slightly shifted for clarity.
The \emph{error bars} are the uncertainties from binomial statistics (with 95
per cent Wilson upper limits  --- eq.~[\ref{upperWilson}]).
\label{fig:Henriques_massresol}}
\end{figure}
 
In fact, the mass and spatial resolution of the cosmological
simulation on which the \Hen\ SAM is run  has a strong effect on the fraction
of VYGs at low galaxy masses.
Figure~\ref{fig:Henriques_massresol} compares the fractions of VYGs versus
stellar mass found by the \Hen\ SAM, when run on the high resolution MS-II
simulation and
on the 125 times lower mass resolution MS simulation.
While the fractions of VYGs are similar at intermediate mass, the MS produces
increasingly higher VYG fractions at increasingly lower masses, where it is
poorly resolved, than does the
MS-II.
At $m=10^{7.25}\,\msun$, at the limit where the MS-II is still properly resolved,
the \Hen\ SAM run on the MS leads to VYG fractions up to 3 orders of magnitude
higher than found when it is run on the MS-II.  The \Hen\ SAM run on the poor mass
resolution MS forms galaxies at later times, which are thus more likely
to become VYGs.

The \Hen\ SAM
was run on dark matter
simulations that were rescaled in space and time to reproduce the
large-scale statistics for a more realistic cosmology (Planck
vs. 1st-year WMAP). Tests by \cite{Angulo&White10} suggest that the small-scale
quantities such as halo masses, bulk velocities, and the luminosities of their brightest galaxies
are changed by only 10, 5 and  25~per cent, respectively.
However, this rescaling may not
reproduce the astrophysical processes at small scales, i.e. in the nonlinear
regime, in particular the galaxy orbits including dynamical friction, as well as
tidal stripping of galaxies by their host groups.

Compared to the \Hen\ SAM, our 4 analytical models have the advantage of a much
superior halo mass resolution: our Monte-Carlo halo merger trees are built
for $z$=0 halo masses down to $10^7\,\msun$, corresponding to the mass of a
single particle in the MS-II simulation on which the \Hen\ SAM was run.  
Furthermore, the trees resolve branches down to $10^3\msun$.
Note that  our halo merger trees
were 
only tested by \cite{Jiang&vandenBosch14} down to $\log (h\,M_0/\msun) = 11$.

On
the other hand, our analytical models have also their own drawbacks.
The careful reader will have noted that some of the empirical models we use have
been calibrated assuming slightly different sets of cosmological parameters
(see Table~\ref{tab:models}). 
While models \MNW\ and \BWC\ have values of $\Omega_{\rm m}$, $h$ and $\sigma_8$
that are within 1~per cent of one another,
model \MCP\ has values of $\Omega_{\rm m}$, $h$, and $\sigma_8$ that are
respectively 8~per cent below, 4~per cent above, and 10~per cent above those
from the other two models.
Since the VYG
fractions of the \MNW\ model are in much better agreement with those of the
\MCP\ model than with those of the \BWC\ model  (Fig.~\ref{fig:fyoung}), it appears that the VYG
fractions are more sensitive to the baryonic physics than to the details of the
cosmological parameters. 

Our analytical models do not follow the satellite galaxies that survive by
$z=0$ against merging into the central galaxy of their halo.
However,  Fig.~\ref{fig:Henriques_massresol} shows that there are no large
differences in the VYG fractions of 
centrals (open circles) versus
satellites (triangles) in the \Hen\ SAM,
especially for the
SAM run on the higher resolution MS-II, except for
$\log( m / \msun) \leq 7.5$, where the satellites are typically 5 times less
likely to be VYGs. 

Moreover,
with the bursty galaxy merging scheme applied to our first three
analytical models, 
the stellar mass of the 
central galaxy is boosted at the time of the halo merger instead of when the
satellite mergers into the central, which is at least a third of a current Hubble time Gyr later (see
Sect~\ref{sec:delayedMergers}).
This boost is the consequence of the
larger halo mass of the central galaxy after the halo merger together with
the increasing stellar mass of the analytical models with halo mass (given
the shallower than --1 slopes in $F_{\rm SFE}$ versus $M$ 
in the bottom panel of
Fig.~\ref{fig:compare_sfe}). This will be discussed in detail in
Sect~\ref{sec:fyoungana}. In the fourth
model (MCP), the stellar mass growth is directly linked to the halo mass
growth. So, all four models naturally have star formation associated with
halo mergers.
This should be realistic for haloes that are rich in cold gas and
that merge in nearly head-on orbits. If the mergers are off-center, one ought
to delay the starburst by the dynamical friction time, and if the mergers are
gas-poor, there should be no burst from the halo merger.
Since our focus is on low and intermediate mass galaxies, which tend to be
gas rich (fig.~11 of \citealp{Baldry+08} and references therein), and whose progenitors
must have been even more  gas-rich, our galaxy merging scheme appears to be
sufficiently realistic. 

A final worry of the analytical models is that they may not properly consider
halo \emph{mass accretion histories} (MAHs).
Indeed,
galaxy properties appear to be not only be related to halo mass, but also to  halo
MAH.
In cosmological simulations, the halo MAH influences the
clustering of haloes \citep*{Gao+05} and their concentration
\citep{Wechsler+06}, a process known as \emph{assembly bias}.
Such \emph{galaxy assembly
  bias} is observed from mass modelling of low redshift galaxies traced by their
satellites, which  leads to red galaxies  (i.e., with older stellar
populations) having higher concentration haloes than blue (young stars)
galaxies of the same  stellar (or halo)
mass \citep{Wojtak&Mamon13}.
Galaxy assembly bias can be detected in the outputs of SAMs \citep{Wang+13}. 
Our Monte-Carlo halo merger trees incorporate assembly bias to a large extent,
since different haloes have different MAHs.
But it is not clear that the strong dependence of halo assembly bias with the
large scale environment (as predicted by \citealp{Yang+17}) is implicitly
incorporated
in our halo merger trees.

\subsection{Simple modelling of the fraction of very young galaxies in the analytical models}
\label{sec:fyoungana}
The relative importance of VYGs in the analytical models can be assessed
from first principles.
According to our definition, VYGs are produced when the stellar mass, summed
over all progenitors, increases 
by more than a factor of 2 in the last Gyr. This growth in stellar mass
occurs in two manners: 1) the quiescent growth
from the models in the absence of halo mergers (i.e., at fixed halo mass); 2)
the growth by halo mergers and corresponding galaxy mergers.
Below, we build a toy model of the stellar mass growth of galaxies that is
based on the simplifying assumption that this growth is independent of the
past history (in other words our simple model is Markovian).

\subsubsection{Model of quiescent growth during last Gyr}
\label{sec:natgrowth}
\begin{figure}
\centering
% DO NOT ERASE FOLLOWING COMMENTED LINE
% SM: mratpredvsM_new mgrowthnat_Jul17
\includegraphics[width=\hsize,viewport=1 160 580 690]{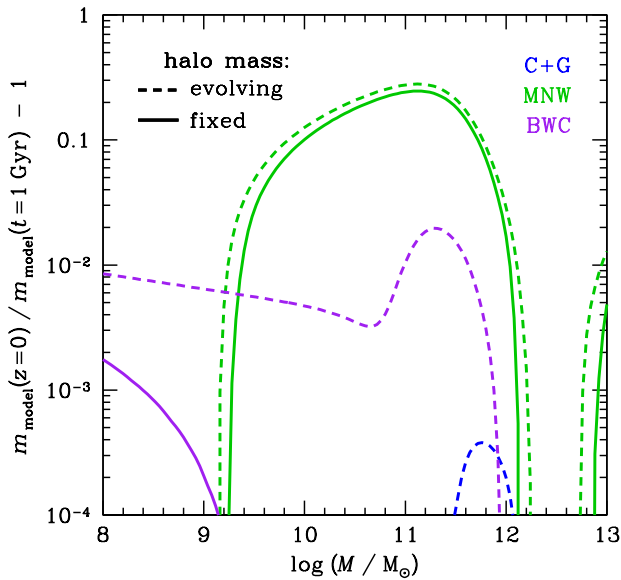}
\caption{
Quiescent growth of stellar mass for the first three analytical models
 (Cattaneo et al. (2011) with Gnedin (2000) [\CG] (\emph{blue}), 
Moster et al. (2013) [\MNW] (\emph{green}),
and
Behroozi et al. (2013) [\BWC] (\emph{purple},
see Fig.~\ref{fig:compare_sfe}) with the bursty galaxy merging scheme.
The growth is computed as the time integral of the maximum of the time derivative of the
stellar mass and zero (eq.~[\ref{naturalgrowthGyr}]). 
The \emph{solid curves} show the quiescent growth with fixed halo mass, while
the \emph{dashed curves} show a hybrid growth, combining the growth of the SFE
on top of the median evolution of halo mass.
The \CG\ model shows zero quiescent growth.
The time derivatives of the Mutch et al. (2013, \MCP, not shown) analytical model are zero at fixed
halo mass, hence the model leads to zero quiescent mass growth.
\label{fig:natgrowth}
}
\end{figure}
In the absence of mergers, VYGs
are produced if the quiescent growth in stellar mass from the analytical
model is a factor of 2 since 1 Gyr.
The stellar mass will vary during the last Gyr as
\begin{eqnarray} 
\Delta m_{\rm quiescent} &=& m(t_{\rm lookback}=1\,{\rm Gyr}) - m(t_{\rm lookback}=0)
\nonumber \\
&=& \int_{t_0-1}^{t_0} {\rm
  Max}\left(0, {{\rm d}m_{\rm model}\over {\rm d}t} \right)\,{\rm d}t \ ,
\label{naturalgrowthGyr}
\end{eqnarray}
where $t_0$ is the age of the Universe expressed in Gyr,
$m_{\rm model}(M_0,z)$ is given in equation~(\ref{mofMz_generic}), and the halo
mass is fixed to the  $z$=0 value.
If the quiescent growth is over 50~per cent, then \emph{all} galaxies should
be VYGs!

The gold band in Figure~\ref{fig:evolsfe} highlights the quiescent growth of
the analytical models during the last Gyr.
While model \MNW\ has
quiescent growth in the last few Gyr, models \CG\ and \BWC\ lead to negative
growth in the last Gyr. 
Solving the integral in equation~(\ref{naturalgrowthGyr}), we display in
Figure~\ref{fig:natgrowth} (solid lines) the relative jump in stellar mass during the last
Gyr in the absence of halo mergers.
Only the
\MNW\ model shows significant quiescent growth at some halo masses, with a peak growth of
25~per cent for $\lM=11.2$ (thick green curves in
Figure~\ref{fig:natgrowth}). 
In comparison, the \CG\ has no quiescent growth, 
while that of the \BWC\ model is small for $\lM<9.4$ and zero at higher halo mass.
The \MCP\ model leads to zero stellar mass growth when the halo masses are
fixed.

We can also consider the hybrid evolution of stellar mass combining the
analytical model $\widetilde m(M,z)$ using the median evolution of halo mass
instead of fixing the halo mass to its $z$=0 value. Although the median growth of halo
mass is small (typically 0.005 dex, increasing with final halo mass), this
halo evolution can make a difference in the stellar mass growth. 
Re-computing the integral of equation~(\ref{naturalgrowthGyr}) assuming the median
evolution of halo mass (black curves in Fig.~\ref{fig:sfevsz},
instead of fixed halo mass), we find 
 (dashed curves in Fig.~\ref{fig:natgrowth})
that
the relative growth of stellar mass is boosted slightly (to 28~per cent
for \MNW\ at $\lM=11.2$) or strongly (for the \CG\ and \BWC\ models).

The extremely low quiescent growth of  stellar mass in the analytical models
implies that mergers are necessary to produce VYGs in these 4 models.
But the \MNW\ model at $\lM=11$ requires less of a boost from mergers than it
does at other final halo masses, as well as compared to
the other models.

\subsubsection{Model of final Gyr stellar mass growth for bursty merging}
\label{sec:merggrowth}
We next consider the growth of stellar mass, summed over the progenitors,
during a halo merger
in the case of bursty merging.
To keep things simple, we consider (only) two haloes of
masses $M_1$ and $M_2$, with respective stellar masses $m_1$ and $m_2$,
merging
at timestep $s$ to reach a new mass of $M=M_1+M_2$.
This simple model does not account for multiple halo mergers in a given timestep.

Considering that the stellar mass of the satellites are frozen,
the stellar mass of the pair can be written
\begin{eqnarray} 
\Delta m_{\rm pair} &\!\!\!\!=\!\!\!\!&  {\rm Max} \left [ \widetilde m(M_1\!+\!M_2,s),
\widetilde m(M_1,s\!-\!1)
\right ]
- 
\widetilde m\left(M_1,s\!-\!1\right) \nonumber \\
&\!\!\!\!=\!\!\!\!& 
{\rm Max}\left[\widetilde m(M_1\!+\!M_2,s)\!-\!\widetilde m(M_1,s\!-\!1),
0\right]
\,.
\label{mjumpDM}
\end{eqnarray}
Since the
dynamical friction time (eq.~[\ref{tdf}]) is always greater than 6.7~Gyr
for lookback times less than 1 Gyr (Sect.~\ref{sec:delayedMergers}),
the satellites that were involved in halo mergers in the last Gyr do not have
time to merge with the main galaxy. 

\begin{figure}
\centering
% DO NOT ERASE FOLLOWING COMMENTED LINE
% SM: deltamovermpair_vsMratio 0.04 mgrowth_mergers_Nov17
\includegraphics[width=0.99\hsize,viewport=1 160 580 690]{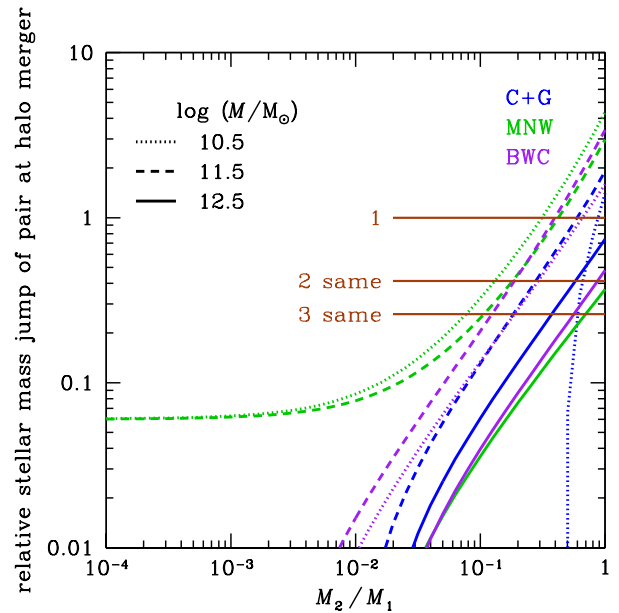} 
\caption{
Relative growth of stellar mass of merging pair (mass ratio minus unity) at
the time of the halo merger predicted from our simple model
(eq.~[\ref{mjumpDM}])
versus the halo mass ratio (above $10^{-4}$ in our
Monte Carlo simulations), 
for the three analytical models
shown in Figure~\ref{fig:natgrowth},
    with the bursty galaxy merging scheme,
    for three masses of the merged halo.
The halo merger is assumed to occur at $z=0.04$ (lookback time of 520 Myr).
For the \CG\ model, we used equation~(\ref{dlmCG}) to correct  the model
stellar mass at $\lM\leq11.2$ in equation~(\ref{mjumpDM}) to conform with the
mass found in Fig.~\ref{fig:mvsM}. 
The \emph{brown horizontal lines} indicate jumps in stellar mass that would 
double it in a single merger, or in 2 or 3 mergers of same halo mass ratios.
\label{fig:merggrowth}
}
\end{figure}

Figure~\ref{fig:merggrowth} shows how the stellar mass of the merging pair
evolves from the previous timestep to the halo merger one using
equation~(\ref{mjumpDM}), for the \CG, \MNW, and
\BWC\ models\footnote{We 
do not apply this test to the
       \MCP\ model, because we do not know how to determine the stellar mass
       of the pair in the normalization to relative variations.}.
Since we do
not follow satellites,
we thus normalize the jump in stellar mass by the stellar mass
$m_1$ of
the first galaxy.
While the $z$=0 SMHM matches quasi-perfectly the model prediction for the \MNW\ and
\BWC\ models, it predicts less stellar mass for the
\CG\ model at $\lM\leq11.2$ (Fig.~\ref{fig:mvsM}), because the stellar mass
is frozen when $F_{\rm SFE}$ reaches its maximum over 8 Gyr ago
(Fig.~\ref{fig:evolsfe}).  
The excess log stellar mass in
the \CG\ model is well fit (to 0.06 dex rms accuracy)  by 
\begin{eqnarray} 
\Delta \log \left ({m\over \msun}\right) &\!\!\!\!=\!\!\!\!& 1.100 - 1.678\,\log \left ({M\over
  10^{10}\,\msun}\right)
\nonumber \\
&\!\!\!\!\mbox{}\!\!\!\!& \qquad\quad + 0.633\,\log \left ({M\over 10^{10}
 \,\msun}\right)^2 \ . 
\label{dlmCG}
\end{eqnarray} 
We thus correct our \CG\ model stellar mass  using equation~(\ref{dlmCG}). 

Given that 1 Gyr corresponds to 3 timesteps, the galaxy has three
chances to boost its stellar mass by a factor 2 by mergers.
The brown horizontal lines in Figure~\ref{fig:merggrowth} indicate relative stellar
mass jumps amounting to a doubling of stellar mass in a single halo merger or
in 2 or 3 mergers of equal halo mass ratio.
The intersection of the curves with these lines provides the minimum halo mass
ratio of a single merger, or 2 or 3 halo mergers of the same mass ratio, 
to allow this doubling of
stellar mass.
Three halo masses are shown that, through the SMHM relation of
Fig.~\ref{fig:mvsM} correspond to stellar masses in the rough range $7.5 \la
\lm \la 10.6$ (depending on the model). 

For the \CG\ and \BWC\ models, the jumps in halo mass  for
$\lM=10.5$ (dotted curves in Fig.~\ref{fig:merggrowth})
are too small to reach the minimum necessary boost from 2 identical mergers, even for 1:1
mass ratios.
This means that it is virtually impossible to double
the  stellar mass in the last Gyr for these halo masses.
For the \MNW\ model, the plateau of the relative stellar mass jump at low
halo mass ratio, for low and intermediate halo masses,
suggests that this model may be able to produce low and intermediate stellar
mass VYGs through
minor halo mergers.

The expected numbers of halo mergers per 350 Myr timestep can be
obtained from theoretical analyses \citep{Neistein&Dekel08_mergers} or from
the analysis of cosmological simulations \citep*{Fakhouri+10}, both of which
are very similar.
We adopt the
differential halo merger rate per unit redshift of \citeauthor{Fakhouri+10}
\begin{eqnarray} 
{\cal R}(M,\mu,z) &\!\!\!\!\equiv\!\!\!\!& {{\rm d}^2N_{\rm merge}\over {\rm d}\mu\,{\rm d}z}
\nonumber \\
&\!\!\!\!=\!\!\!\!& 
 A \left({M\over M_{\rm ref}}\right)^b
 \mu^c\,\exp\left[\left({\mu\over \mu_{\rm ref}}\right)^d\right]\,(1\!+\!z)^e \ ,
\label{rateFakhouri}
\end{eqnarray}
where $\mu=M_2/M_1$ is the halo mass ratio (with $M_1+M_2=M$) and where
$A=0.010$, $M_{\rm ref}=10^{12}\,\msun$, 
$b=0.133$,
$c=-1.995$,
$\mu_{\rm ref}=0.00972$,
$d=0.263$,
and
$e=0.010$.
Integrating over halo mass ratios, we first infer that in the 350 Myr
timestep, 
a halo of mass  $M = 10^{10}$, $10^{11}$, or $10^{12}\,\msun$ is expected to be
involved in respectively $\langle N_{\rm m}\rangle = 2$, 3, or 4 mergers
with  halos  of mass ranging from $10^{-4}\,M$ (the extreme mass ratio allowed in our
Monte-Carlo halo merger tree, see Table~\ref{tab:models}) to $M$.
Hence, halo mergers are ubiquitous in our halo merger tree, although there
are non-negligible probabilities ($100\,\exp(\langle-N_{\rm m}\rangle) = 14$,
5, and 2~per cent respectively for 
$\lM=10$, 11, and 12) that a halo suffers no merger in a given timestep.
We note here that according to the tests of \cite{Jiang&vandenBosch14}, the
\cite{Parkinson+08} code reproduces the halo merger rates of \citeauthor{Fakhouri+10} to better than 20
per cent for all mass ratios for low- and intermediate-mass haloes 
($M=10^{11}\,h^{-1}\,\msun$
and $10^{13}\,h^{-1},\msun$ at $z=0$).
 
We can go one step further and integrate the halo merger rate. 
Let $g_{\rm q}(M)$ and $g_{\rm m}(M,\mu)$ respectively be the quiescent
relative growth over 1 Gyr and the relative growth through a halo merger.
The predicted fractions of VYGs obtained in two mergers can then be written
(dropping the dependence of halo mass and redshift in the expression for the
halo merger rate ${\cal R}$ for clarity)
\begin{equation}
f_{\rm VYG} = \left({0.08\over 3}\right)^2\,
\int_{0.0001}^1 {\cal R}(\mu_1) \,{\rm d}\mu_1 \,\int_{\mu_2^{\rm min}}^1 {\cal
    R}(\mu_2)\,{\rm d}\mu_2 \ ,
\label{ratedouble}
\end{equation}
where the factor in front of the integral in equation~(\ref{ratedouble})  is
for the conversion from $\Delta z = 1$ to 
one-third of a Gyr,
while
the lower limit of the first integral is the resolution of the Monte-Carlo
halo merger tree that we have used (Table~\ref{tab:models}).
In equation~(\ref{ratedouble}),  
 $\mu_2^{\rm min}$ is the solution of the equation 
\begin{equation}
(1+\eta\,g_{\rm q})\,\left\{1+ g_{\rm m}\left[\left({M\over 1+\mu_2}\right),\mu_1\right]\right\}\,\left[1  + g_{\rm m}(M,\mu_2)\right]=2 \ ,
\label{mu2min}
\end{equation}
for $\mu_2$,
with $\eta=1/3$ to account for the fact that only one-third of the Gyr
interval is available for quiescent growth (since we do not bother to perform
the triple integral required for the three timesteps).

\begin{figure}
\centering
% SM (first) predVYGvsM 1000000 fixed Nov17 (5 min CPU)
% SM plotpredVYGvsm device Nov17
\includegraphics[width=\hsize,viewport=1 160 580 690]{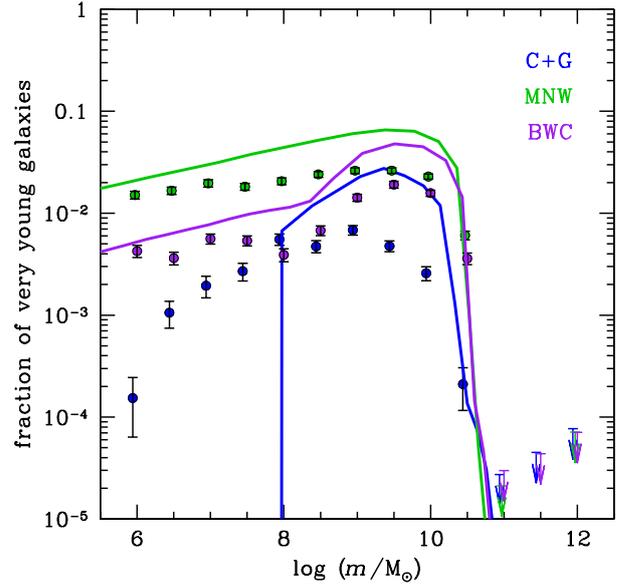} 
\caption{
  Predicted fractions of very young galaxies for the first three analytical
  models (with the bursty galaxy merging scheme) as a function of $z$=0 stellar
  mass,
  from our simple Markovian model (\emph{curves})
combining quiescent growth and growth during halo mergers
(eqs.~[\ref{naturalgrowthGyr}--\ref{ratedouble}]), compared to those 
  from the Monte Carlo halo merger trees (\emph{symbols} as in
  Fig.~\ref{fig:fyoung}), where
the abscissae are slightly shifted for clarity.
\label{fig:fyoung_pred} 
}
\end{figure}

The curves in Figure~\ref{fig:fyoung_pred} show the fractions of VYGs predicted by our
simple model (which involves no Monte-Carlo halo merger trees), as given by
equation~(\ref{ratedouble}).
The figure captures the basic features of VYG fractions we found with the analytical
models applied to the Monte-Carlo
halo merger trees
(Fig.~\ref{fig:fyoung} and shown as symbols in Fig.~\ref{fig:fyoung_pred}).
In particular, the \MNW\ and \BWC\ models are both remarkably well reproduced:
the \MNW\ model shows a few per cent VYGs at low and intermediate 
stellar masses and a sharp drop at higher masses
(with the simple model overpredicting the Monte Carlo halo merger tree
calculations by a factor of 2).
Similarly, our simple model catches the details of the \BWC\ model.
However, our model is less successful in predicting  the VYG fractions with the \CG\ model,
as it misses the 0.001 fraction of VYGs at $\lm < 8$.
(Had we not corrected for the excess stellar mass of
the \CG\ model galaxies compared to their model predictions,
eq.~[\ref{dlmCG}], the predicted \CG\ VYG fraction would be a factor 3 above
that of the \MNW\ model at low mass.)
These discrepancies may be attributable to the simplicity of
our model and to the large scatter in the SMHM of the \CG\ model at low mass.
Nevertheless, the agreement for the \MNW\ and \BWC\ models of our simple VYG model with the VYG fraction
measured by the full model on the Monte-Carlo halo merger trees suggests that
the galaxy histories before 1 Gyr matter little in predicting the \MNW\ and
\BWC\ VYG
fractions, while the \CG\ VYG fractions, on the contrary, appear to depend on past history.

\subsubsection{Model of final Gyr stellar mass growth for quiet merging}
\label{sec:mgrowthquiet}
For the quiet merging scheme, we can better understand the jump in stellar
mass with a toy model based on equation~(\ref{mnew_quiet}).

Consider a halo of mass $M_0$ at
lookback time $t=t_0$ (corresponding to $z=0$) and assume that its stellar
mass $m_0$ is given by its model stellar mass $\widetilde m_0=\widetilde
m(M,0)$.
This corresponds to the first
term within the brackets of 
equation~(\ref{mnew_quiet}) dominating the second one.
Let ${\cal F}_i(M)$ be the fraction of  halo mass $M$ at lookback
time $t_i$ (corresponding to redshift $z_i$)
that came from branches that merged at lookback time $\geq t_i$ and whose host
satellites are not expected to 
merge (after dynamical friction) before lookback time $t$.
In other words, ${\cal F}$ measures the fraction of halo mass coming from
branches with surviving satellites. 
The stellar mass at $z=0$ can then be written
\begin{equation}
    m_0 \equiv \widetilde m_0 = \widetilde m
  \left (\left[1\!-\!{\cal F}_0(M_0)\right] M_0, 0\right) \ .
  \label{m0}
\end{equation}
At lookback time $t_1=1\,\rm Gyr$, corresponding to redshift $z_1$,
we write the stellar mass summed over the
progenitors of the $z=0$ galaxy as
\begin{equation}
  m_1 = \widetilde m\left (\left[1\!-\!{\cal F}_1(M_1)\right] M_1, z_1\right)
  + \sum m_{\rm sats} \ ,
  \label{m1}
\end{equation}
where $M_1$ is the mass of its main progenitor at lookback time $t_1$.
The last term in equation~(\ref{m1})  is the total stellar mass in satellites that merge with
the central between lookback times $t_1$ and 0
(recall that the stellar masses of satellites are frozen).

Let  $\overline{\cal F}_i(M)$ be the median value expected for the values of ${\cal F}_i(M)$
and $\eta_i(M) ={\cal F}_i(M)/\overline{\cal F}_i(M)$, for $i=1$ and 0.
Thus, $\eta_i(M)$ represents the excess, relative to the median, of the mass fraction  of halo of mass $M$ at
lookback time $t_i$ that originates from branches whose galaxies would survive as
satellites until a lookback time $t < t_i$. A halo with $\eta > 1$ is one
that could be
considered fortunate to have a higher fraction of its mass than usual from branches with surviving satellites.

The condition for obtaining a VYG is $m_0 > 2\,m_1$ using
equations~(\ref{m0}) and (\ref{m1}). 
Solving this inequality is difficult, because of the range of redshifts
when haloes merged with the main progenitor of the current halo, yet their
satellites survived. So, for simplicity, we assume that the satellite term in
equation~(\ref{m1}) contributes to a fraction $f_{\rm sats}$ of the mass
$m_1$.
Moreover, we adopt for $M_1$ the median mass $\overline M_1$ expected for the
$z$=0 halo mass $M_0$. The black curves of Figure~\ref{fig:sfevsz} indicate a
very slow typical growth of the main halo progenitor in the last Gyr, of order of
0.005 dex (Sect.~\ref{sec:natgrowth}). We shall see (Sect.~\ref{sec:statgrowth} below) that contrary to the case of
bursty galaxy merging, VYGs are not related to rapid jumps in halo mass with the quiet merging scheme.  
With these assumptions and
equations~(\ref{m0}) and (\ref{m1}),
the condition for a VYG is
\begin{equation}
  \left.
  \begin{aligned}
 &   \widetilde m\left(\left[1-\eta_0(M_0)\, \overline
    {\cal F}_0\left(M_0\right)\right] M_0,0\right)
    \\
& \quad >  {2\over 1-f_{\rm sats}}\,
  \widetilde m\left(\left[1-\eta_1(\overline M_1)\,\overline
    {\cal F}_1\left(\overline M_1\right)\right] \overline M_1,z_1\right)
  \ .
\end{aligned}
\right.
\label{condVYGquiet}    
\end{equation}

Of course, since VYGs are predicted to be rare at all masses,
typical values of ${\cal F}(M)$ (i.e. $\eta_0 = \eta_1 = 1$)
will not lead to VYGs according to equation~(\ref{condVYGquiet}). 
Instead, the requirement for obtaining a VYG is to have
$\eta_0 < 1$ and/or $\eta_1 > 1$, i.e. a final halo with a low mass fraction
from branches with surviving satellites and/or progenitors at 1 Gyr lookback time
with high mass fraction from  branches with surviving satellites (to 1 Gyr
lookback time).
Here, we assume
$\eta_1=1$ and determine the maximum value $\eta_0^{\rm VYG}$ of $\eta_0$ for obtaining a
VYG. If  $\eta_0^{\rm VYG}$ is only slightly below unity, VYGs will be easier to
form than if it is much lower than unity.

\begin{figure}
\centering
\includegraphics[width=\hsize,viewport=0 0 560 550]{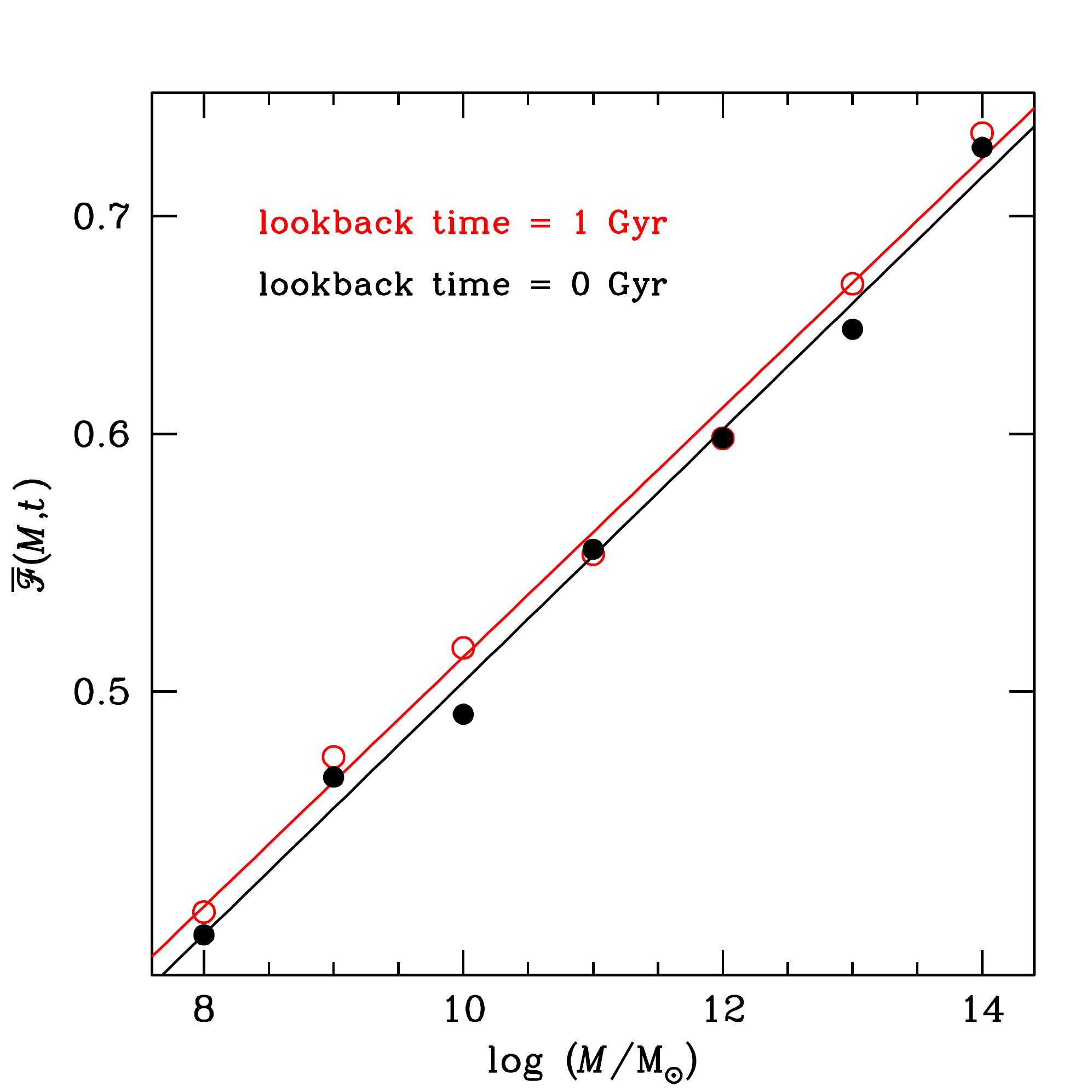} 
\caption{Median fraction of halo mass from branches
    hosting surviving satellites (from eq.~[\ref{F2}]),  as function of halo mass
  for two lookback times. The
  \emph{lines}
  are linear fits of $\log \overline {\cal F}$ vs log halo mass.
\label{fig:frac_nomerge}}
\end{figure}
The median value $\overline{\cal F}(M)$
is not straightforward to extract, e.g. from extended Press Schechter theory
(e.g., \citealp{Neistein&Dekel08_mergers,Parkinson+08}).

Instead, we estimate  $\overline{\cal F}(M)$ by considering the median halo growth:
\begin{equation}
  {\cal F}_i(M) = {1\over M}\,
  \sum_{\substack{t'>t_i \\ \tau_{\rm df} > t'-t_i}}
\left [ \overline M(t'\!-\!\Delta t_i) - \overline M(t') \right ]
\ ,
\label{F2}
\end{equation}
where $\overline M(t')$ is the median mass expected for a halo at lookback
time $t'$,  whose mass at
lookback time $t_i$ is $M$,
$\Delta t$ is 
the previous timestep,\footnote{In equation~(\ref{F2}), $\Delta t$ is subtracted
  (not added) to $t'$ because
$t$ is a lookback time, which decreases with time.} 
and where the dynamical friction time is
$\tau_{\rm df}\equiv\tau_{\rm df}\left[\overline M(t'),\overline
  M(t'-\Delta t) - \overline  M(t'), t'\right]$.
The expression for ${\cal F}$ given in equation~(\ref{F2}) 
corresponds to summing up all the masses of secondary progenitors
merging with the primary ones.
It is an underestimate, because it
assumes that halo merging is binary. If, instead, many branches merge at
once, each branch will correspond to a lower mass progenitor and its
dynamical friction time would be longer. We thus assume that the lower mass
branches bring negligible mass to the final halo.

Figure~\ref{fig:frac_nomerge} shows that $\overline {\cal F}(M)$ is of
order one-half, weakly increasing with halo mass, 
and with very little
dependence on the final considered lookback time.
The linear fit in log-log space yields
\[
  \log \overline {\cal F}(M) \simeq a + b\,\log \left ({M\over \msun}\right)
  \ ,
\]
with $a = -0.686$ and $b=0.0389$ for $z=0$,
and $a = -0.674$ and $b=0.0383$ for lookback time of 1 Gyr.
This median surviving fraction $\overline {\cal F}(M)$
increases with halo mass, because large halo masses grow faster,
hence the values of $M_2$ are relatively more important (this offsets the
shorter dynamical friction times, since they are always greater than one-third
of the current Hubble time).

\begin{figure}
% SM: eta0VYGvsmstars device ! DO NOT ERASE THIS LINE!
\centering
\includegraphics[width=\hsize,viewport=0 0 560 580]{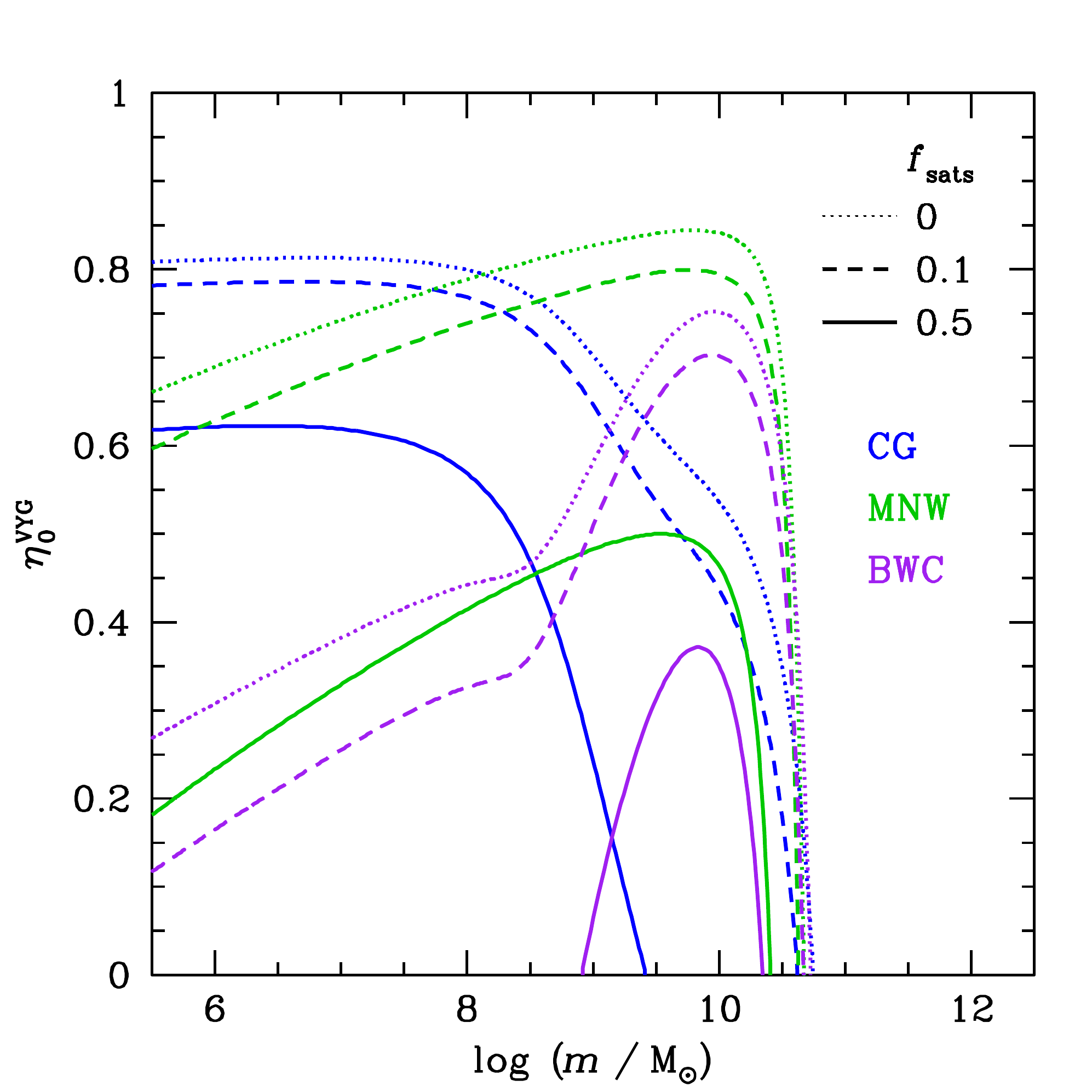}
\caption{Maximum excess fraction $\eta_0^{\rm VYG}$ of final halo mass coming from branches that
    merged recently enough that the galaxies survive as satellites at $z=0$
    to yield very young galaxies
    versus final stellar mass. The parameter $\eta_0^{\rm VYG}$ is derived by solving
    equation~(\ref{condVYGquiet}), using Fig.~\ref{fig:frac_nomerge} for
    $\overline {\cal F}(M)$, assuming $\eta_1 = 1$ and the $z$=0 stellar - halo
    mass relation predicted by the model.
    The \emph{dotted}, \emph{dashed}, and \emph{solid} curves display the
    results when assuming that $f_{\rm sats}=0$, 0.1, and 0.5, respectively.
    \label{fig:eta0VYG}
  }
\end{figure}

We can combine equations~(\ref{condVYGquiet}) and (\ref{F2}) to estimate the
maximum value of $\eta_0^{\rm VYG}$ to obtain VYGs.
Figure~\ref{fig:eta0VYG} shows the maximum values of $\eta_0$ to obtain a
VYG, for the \CG, \MNW, and \BWC\ models. We assumed $\eta_1 = 1$ and converted the final halo masses to stellar
masses using the SMHM predicted by the model.
Figure~\ref{fig:eta0VYG} can be compared to the hatched shaded regions of 
Figure~\ref{fig:fyoung_quiet}, corresponding to the VYG fractions obtained
with the quiet merging scheme applied to the Monte Carlo halo merger tree. For all choices of $f_{\rm sats}$, we recover
the sharp drop of VYG fractions at $\lm \simeq 10.7$ for all 3 models.
We also recover most of the hierarchy between the 3 galaxy formation models:
\CG\ should produce a much higher fraction of VYGs than \BWC\ at low mass,
\BWC\ should overtake \CG\ at intermediate mass, and
the \MNW\ model should dominate the others at intermediate masses.
The only weakness of our simple model is  that it fails to recover the higher
fraction of VYGs expected with the \MNW\ model at low masses in comparison
to the \CG\ model. This might be explained by a higher value of $f_{\rm
  sats}$ for the \CG\ model, whose quiescent
evolution is frozen earlier than the other models by our prevention of
decreasing stellar masses, as seen in Figure~\ref{fig:evolsfe}.

\subsection{Evolutionary histories}
\begin{figure*} 
\centering
% SM: define land 1
% SM: evolMhalos evolMhalos 3 3 1                   DO NOT ERASE!
\includegraphics[width=0.3\hsize,angle=-90,viewport=0 0 600 770]{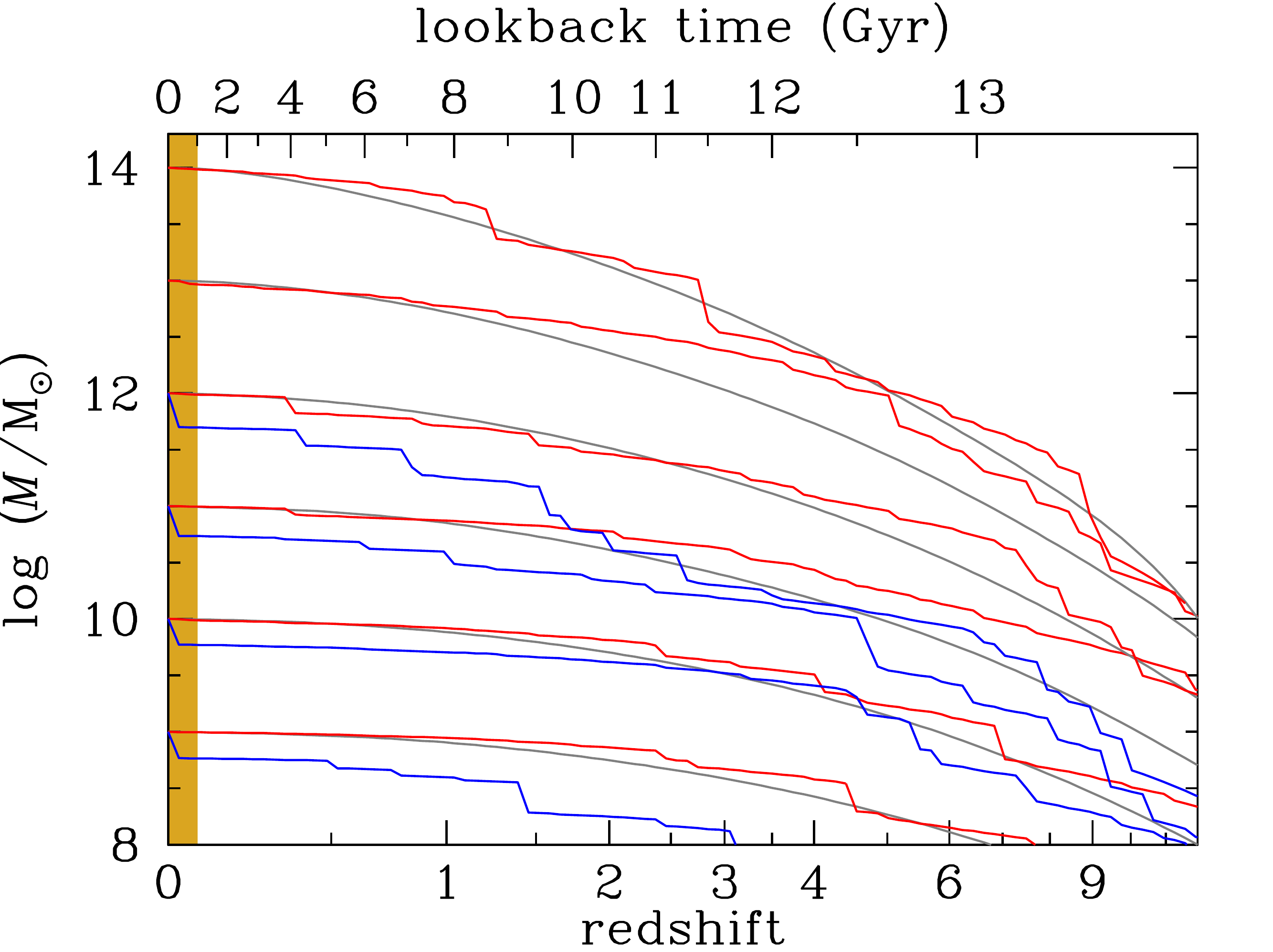} 
% SM: evolvchaloes evolvchalos 3 3 1                DO NOT ERASE!
\includegraphics[width=0.3\hsize,angle=-90,viewport=0 0 600 770]{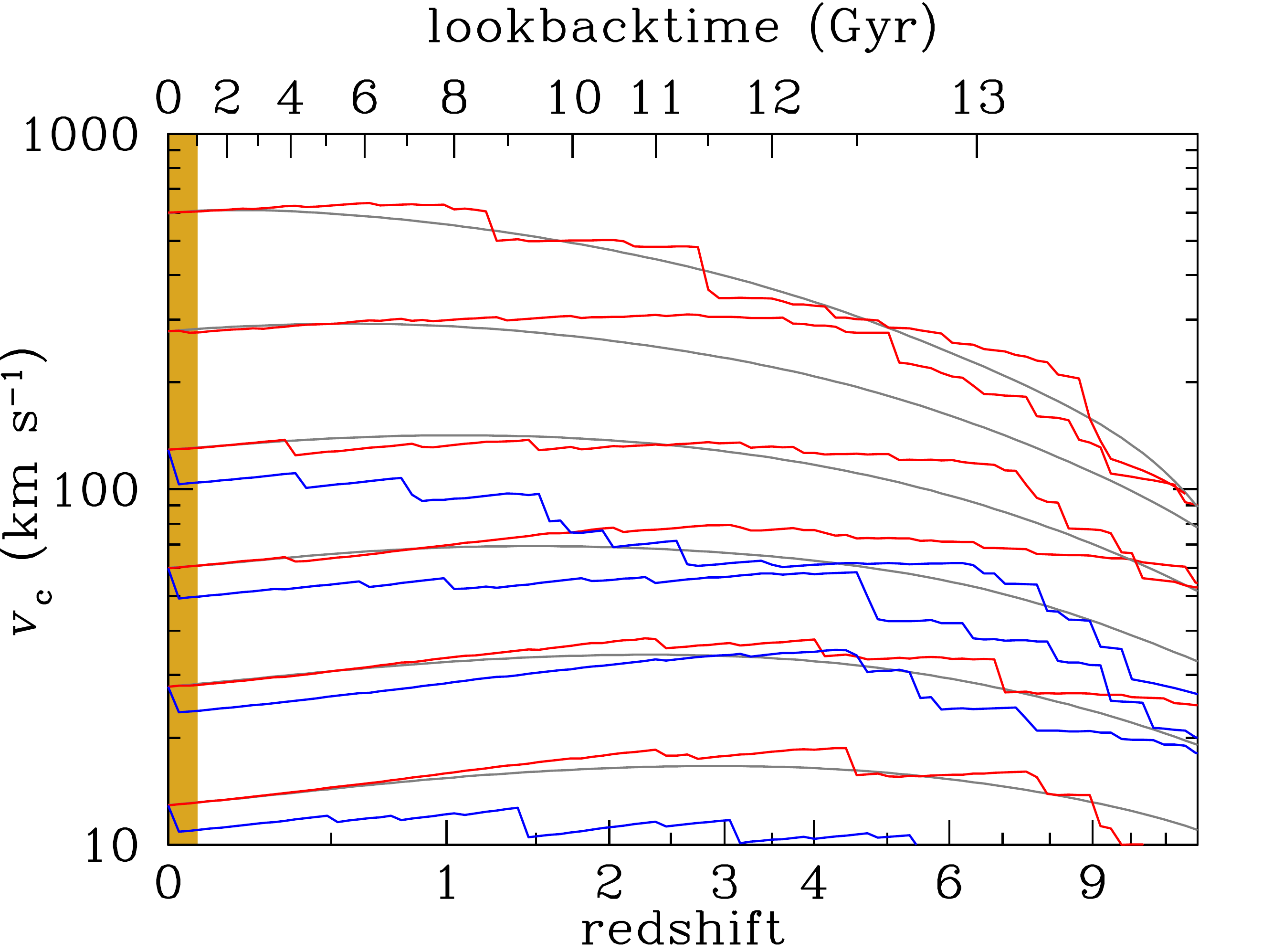} 
% SM: evolgals2 CG MNW evolgals_CG_MNW 3 3 1        DO NOT ERASE!
\includegraphics[width=0.3\hsize,angle=-90,viewport=240 35 570 760]{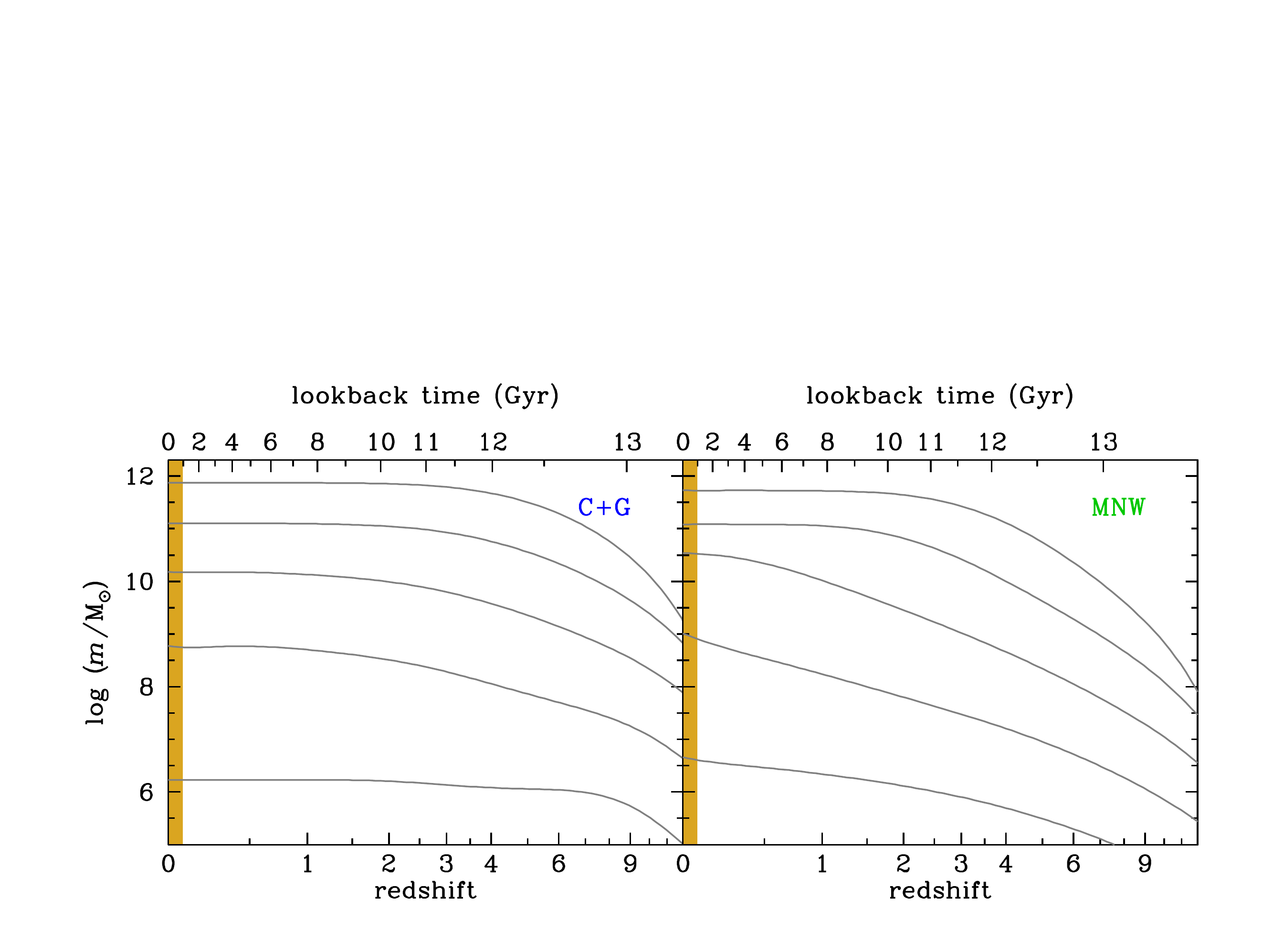}
% SM: evolgals2 BWC MCP evolgals_BWC_MCP 3 3 1      DO NOT ERASE!
% SM: define land delete
\includegraphics[width=0.3\hsize,angle=-90,viewport=240 35 570 760]{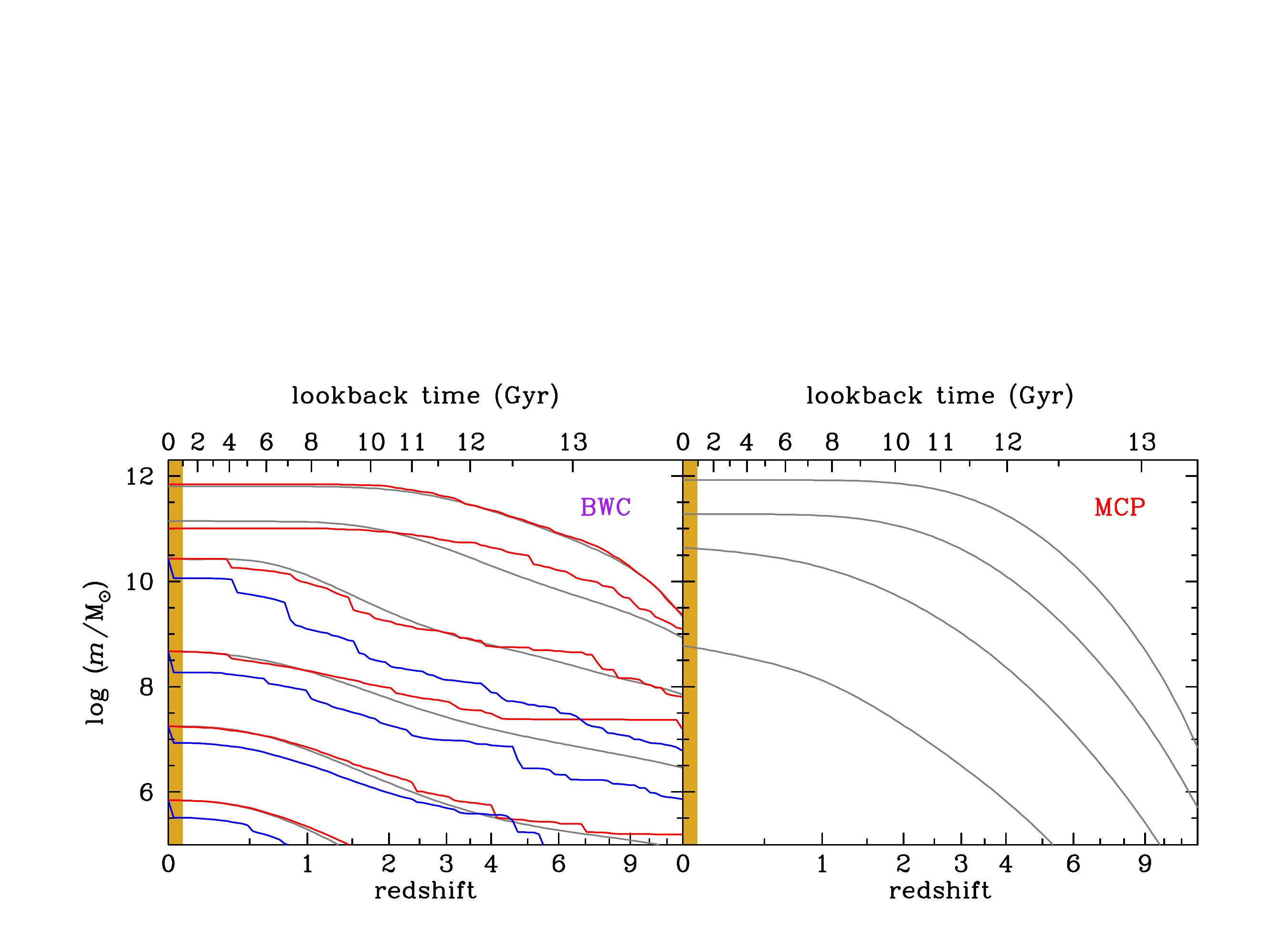}
\caption{Evolutionary histories of haloes and galaxies for haloes with final
  log masses from 9 to 14 in steps of 1 (going upwards).  \emph{Top}: Evolution
  of the main progenitor of the $z$=0 haloes in mass (\emph{top left}) and
  circular velocity (\emph{top right}).  \emph{Middle \& bottom}: Evolution
  of the total stellar mass (summed over all progenitors) of galaxies for the
  four models with bursty galaxy merging, with the same $z$=0 halo log
  masses.
The \emph{smooth grey} curves represent the median evolutionary histories.
For the \BWC\ model, the \emph{broken curves} represent particular histories of
galaxies, which at $z=0$ are very young (\emph{blue}) or normal (\emph{red}).
The evolution of the corresponding haloes are displayed in the same colours
(\emph{top panels}).  
Note the
different cosmologies used for each model to link lookback time and redshift.
\label{fig:evolhist}
}
\end{figure*}

Figure~\ref{fig:evolhist} shows (grey curves) the median evolution of the main halo
progenitor in mass and circular velocity (top panels) and the median
evolution  of
the stellar mass, summed over all progenitors, for the 4 analytical models.
One notices that for intermediate halo masses ($\lM=13$), the median
evolution of haloes since $z=1$ is at constant circular
velocity (see \citealp{Mamon+12}),
while at low masses halo mass growth is so weak that halo mass is
nearly constant.

As seen in the lower 4 panels of Fig.~\ref{fig:evolhist},
the median stellar mass evolution (whose time derivative corresponds to the
star formation rate) can be growing or nearly frozen depending on the
analytical model and the final halo mass.

The bottom left panel of Figure~\ref{fig:evolhist} displays the
evolution of 
stellar mass for 
representative very young (blue) and old (red) galaxies for the \BWC\ model
with delayed
galaxy merging
(which produces no VYGs for massive haloes, $\lM \geq 13$).
The VYGs can be easily spotted by a sharp growth in the
stellar mass at low redshift, 
and are
caused by the rapid halo mass growth 
(top left panel).
The bottom left panel of Figure~\ref{fig:evolhist} shows examples of VYGs
produced with unusual long-term histories (the example ending at $\lm=10.5$) and
others with very typical histories (the examples ending at $\lm=8.7$, 7.3,
and 5.9).
This illustrates the Markovian nature of the \BWC\ model
(Sect.~\ref{sec:fyoungana}): 
the production of VYGs is not determined by the long-term histories of the
galaxies.

\subsection{Statistics of halo growth in last Gyr}
\label{sec:statgrowth}
\begin{figure}
% SM: Mratpdfsall4 Mratpdfsall4_bursty 0 1 dmnew   DO NOT ERASE THIS COMMENT!
\includegraphics[width=\hsize,viewport=1 0 560 540]{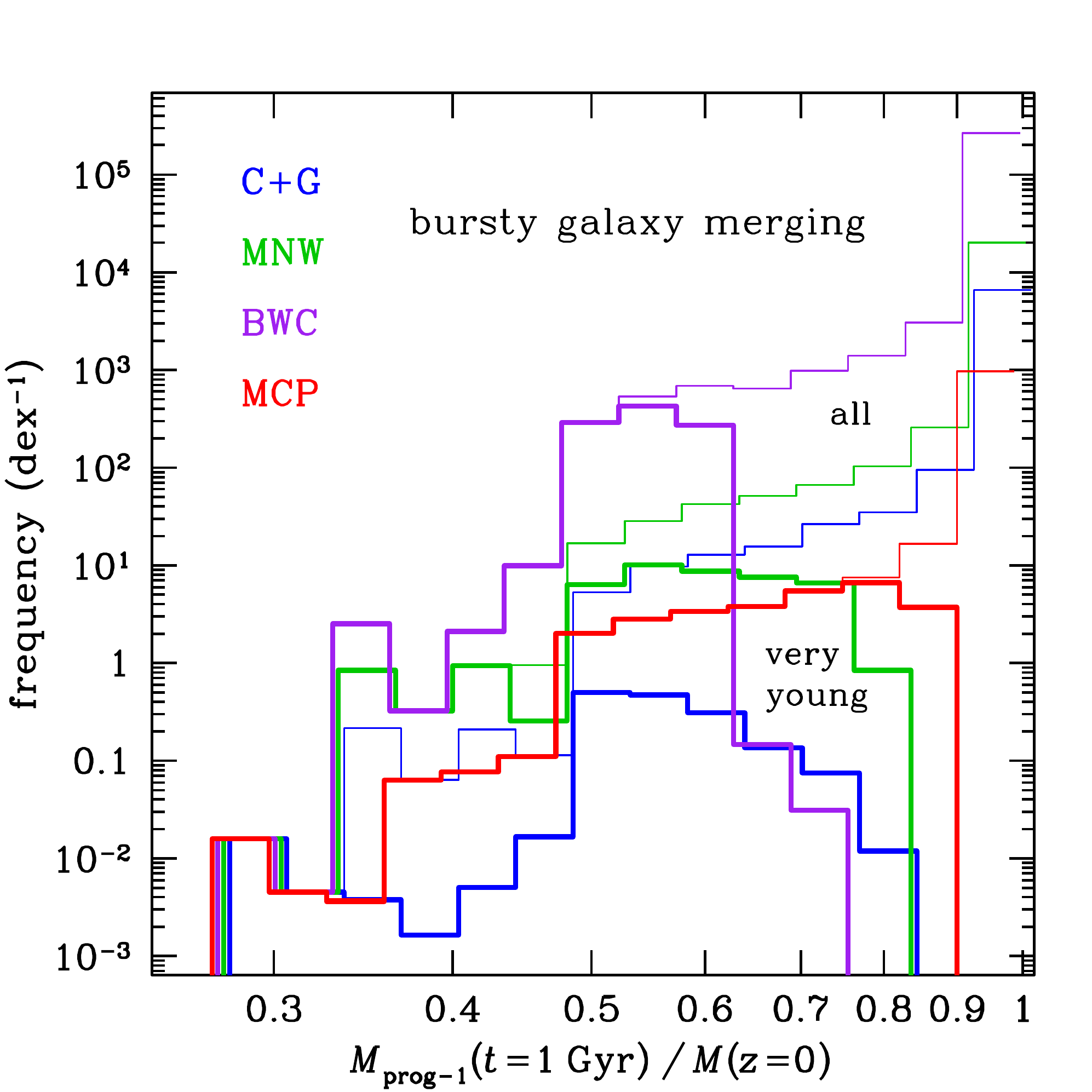} 
\caption{Frequencies of the inverse growth of the most massive
  halo progenitor in the last Gyr,
for all  (\emph{thin}) and very young (\emph{thick histograms}) galaxies,  for
the 4 analytical models
 (see Fig.~\ref{fig:compare_sfe}) for bursty galaxy merging.
The abscissae
above 3/4 indicate no mergers   with mass ratios
between 1:1 and 3:1, while those below 1/2 indicate more halo growth than a
single 1:1 major merger. 
The frequencies of halo mass ratios are weighted according to equation~(\ref{weight}). 
\label{fig:pdfMrat}}
\end{figure}

One may ask whether, in the analytical models, the final rapid growth of
stellar mass (summed over all progenitors) is indeed linked with a corresponding rapid growth in
the most massive progenitor of a galaxy's halo.
Figure~\ref{fig:pdfMrat} 
shows the statistics of halo mass growth in the last Gyr.
The abscissae (slightly shifted for clarity) above 3/4 indicate no mergers   with mass ratios
between 1:1 and 3:1, while those below 1/2 indicate more halo growth than a
single 1:1 major merger. 
In the \CG\ and \BWC\ models, the rapid recent growth in total stellar mass
for the VYGs indeed comes
hand-in-hand with rapid mass growth of the most massive progenitor of the halo.
In contrast, the \MNW\ and, especially, \MCP\ models  show fairly frequent cases of rapid
final growth in stellar mass that occurs with slow halo mass growth.
To summarize, the median halo relative mass growth from 1 Gyr to the present is of order
0.015 for all galaxies, but for VYGs it is as high as 0.4 (\MCP), 0.6 (\MNW),
0.8 (\BWC), and 0.9 (\CG).
This suggests the important role of major halo mergers. However, 
in nearly one-quarter of the VYGs in the \MCP\ model, the haloes grow by less than 25~per
cent, indicating that more minor halo mergers suffice. %% None of the haloes

\begin{figure}
  \centering
% SM: Mratpdfsall4 Mratfreqs_Nov17_quiet 0 1 freeze2   DO NOT ERASE THIS COMMENT!
\includegraphics[width=\hsize,viewport=0 0 570 540]{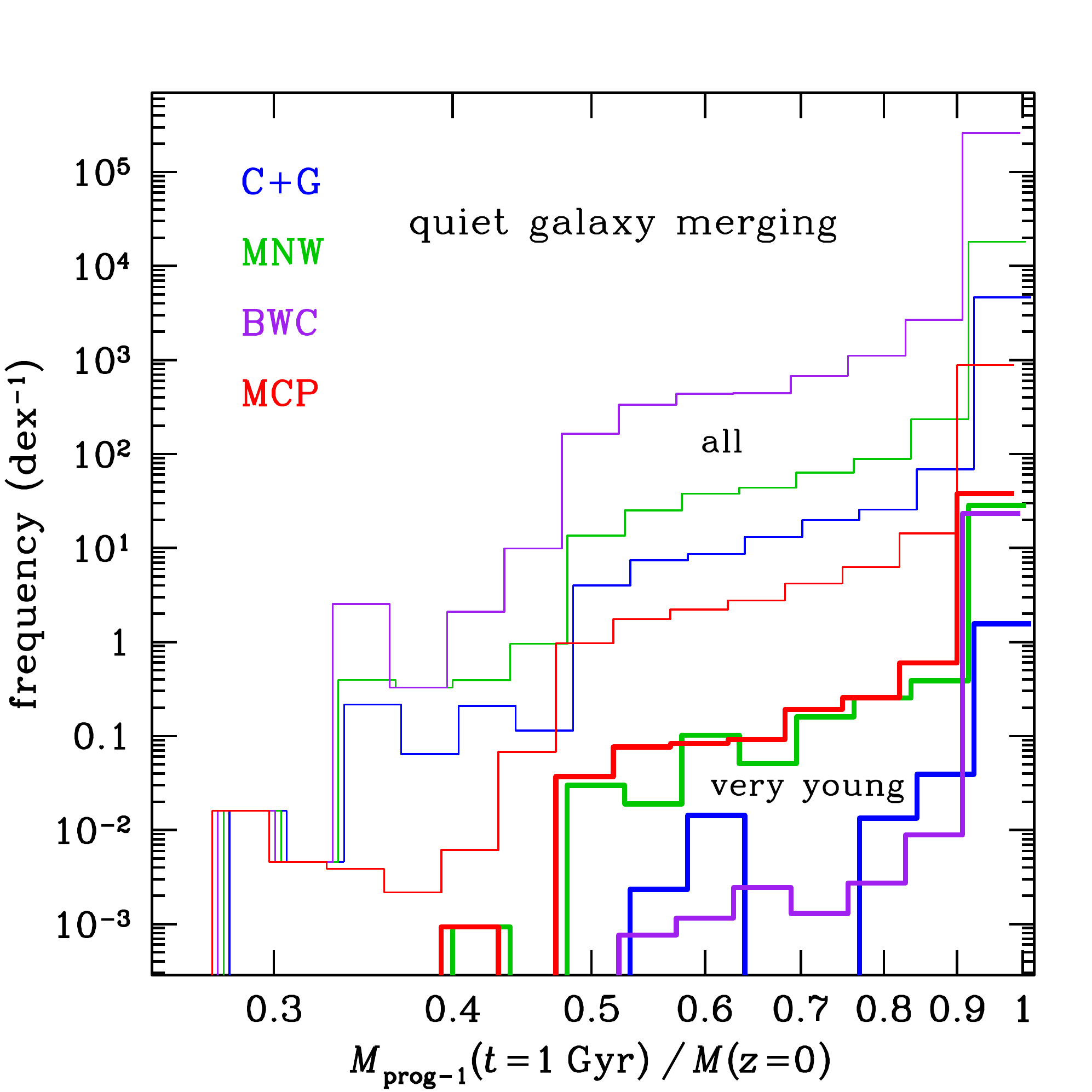} 
\caption{Same as Fig.~\ref{fig:pdfMrat}, but for the quiet galaxy merging scheme.
\label{fig:pdfMrat_quiet}}
\end{figure}
Figure~\ref{fig:pdfMrat_quiet} shows a very different halo merging history
for the VYGs obtained with quiet galaxy  merging: most of the VYGs are
associated with haloes that only slowly increase their mass in the last Gyr.
The distribution of the values of halo mass growth in the last Gyr are
indistinguishable between the VYGs and all galaxies.
This is the consequence of the stellar mass no longer being a function of
the current halo mass, but instead of the current halo mass minus the mass in
the  subhaloes that host
surviving satellites (eq.~[\ref{mnew_quiet}]).

\subsection{The effect of the galaxy environment}

Our analytical models completely miss the effects
of a galaxy's environment. Analyses of large observational samples of
galaxies by \cite{Weinmann+06} and \cite{vonderLinden+10} indicate that the
fraction of star forming galaxies is reduced in both massive global environments
(high-mass clusters) and in the inner local environments (the inner regions
of groups and clusters).  Since most galaxies belong to
groups or clusters (e.g., \citealp{Yang+07}), one may expect that the lack of
environmental effects in our analytical models may cause substantial
modifications to the fraction of VYGs. 

The abundance matching analytical models mix the different environments to
yield a global SMHM relation. However, moving from high to low-mass groups, 
their central galaxies become increasingly likely to be of 
 spiral morphology
\citep{Weinmann+06} and with efficient star formation \citep{Woo+13}. 
One needs to go beyond standard abundance matching and also incorporate the
link between stellar ages and halo formation times (as indirectly observed by
\citealp{Wojtak&Mamon13}, see Sect.~\ref{sec:limits}, above), as this allows 
 reproducing correctly several environmental
trends \citep{Hearin&Watson13}, in particular the fraction of star forming
central galaxies as a 
function of their luminosity (which is known to correlate strongly with
stellar and group mass).

Therefore, the quenching of star formation by the environment may explain why
the \Hen\ SAM leads to typically 100  times lower fractions of VYGs at
intermediate mass ($8 \leq \lm \leq 10$) in comparison with the analytical
models with bursty galaxy merging.
  However, one cannot unreservedly believe the results of the \Hen\ SAM,
because of its limited mass resolution and also because the treatment of the
physics of quenching in SAMs is still very approximate.
Moreover, Figure~\ref{fig:z50vsm} indicates that the typical ages of galaxies
in the \Hen\ SAM are of the same order as those found by \MNW, \BWC, and
\MCP\ 
models, but their spread is very narrow at intermediate stellar mass, leading
to much lower VYG fractions in this mass range.

\subsection{Effects of the primordial density fluctuation spectrum}

Recently, there has been much interest in allowing for a warmer primordial
density fluctuation spectrum than allows the $\Lambda$CDM model.
Moving from  $\Lambda$CDM to  $\Lambda$WDM means that high
wavenumber primordial density fluctuations are suppressed. This decreases
(but does not suppress) the variance of low-mass fluctuations, making them
rarer at given epochs. This in turn means that haloes of given $z$=0 mass
collapse later. 
This effectively reduces the numbers of low-mass haloes at all epochs, 
potentially solving (\citealp*{Bode+01}; \citealp{Avila-Reese+01}) 
the problem of the overabundance of
such low mass haloes in cosmological $N$-body simulations
\citep{Moore+99}. 
Since present-day stellar mass varies monotonically with halo
mass, one concludes that low-mass galaxy formation is delayed in the
$\Lambda$WDM model.
Therefore,
the fraction of VYGs should be enhanced in the $\Lambda$WDM
cosmology.

\begin{figure}
\centering
% SM: fyoungvsm_cosmo Menci fyoungvsm_cosmo_Jul17 1000 1 1   DO NOT ERASE THIS COMMENT!
\includegraphics[width=\hsize,viewport=1 160 580 680]{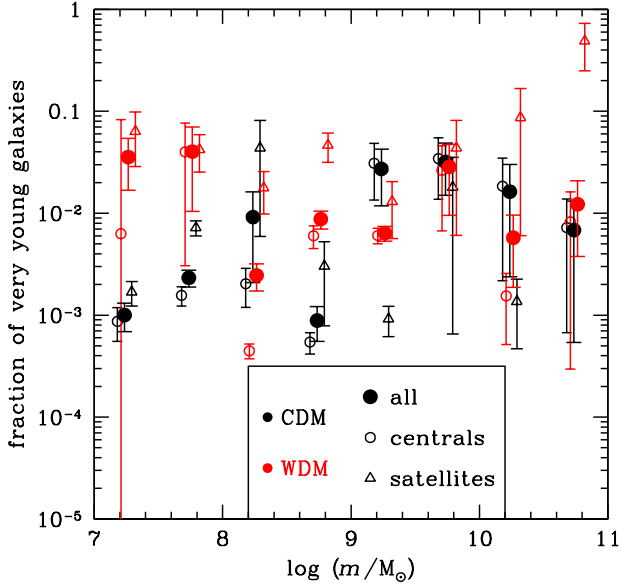} 
\caption{Effect of the primordial density fluctuation spectrum on the
  fractions of very young galaxies (half the stellar mass formed in the last
  1 Gyr), for the Menci et al. (2014)
  semi-analytical model, run on Monte-Carlo halo merger trees based on two
  cosmologies: Cold Dark Matter (\emph{black}) and Warm Dark Matter (with
  particle mass $m_\chi = 0.75\,\rm keV$, \emph{red}).
The figure highlights the differences between centrals (\emph{open circles}),
satellites (\emph{triangles}) and all galaxies (\emph{large filled circles}).
The abscissae are slightly shifted for clarity.
The \emph{error bars} are the uncertainties estimated from 1000 bootstraps.
\label{fig:fyoung_menci}}
\end{figure}

We have extracted the fractions of VYGs from the \cite{Menci+14}
SAM (first presented in \citealp{Menci+08}, see Sect.~\ref{sec:menci}), 
run  on two forests of Monte-Carlo halo merger trees, one
for $\Lambda$CDM, the other for $\Lambda$WDM.
The Menci et al. SAM was run with the exact same parameters on both forests.
Figure~\ref{fig:fyoung_menci} shows that, for the CDM model, the fraction of VYGs among
intermediate mass galaxies is typically 50 times greater than predicted by
the \Hen\ SAM run on the high-resolution MS-II simulation.
The VYG fraction  among
low-mass satellites is significantly greater in the \citeauthor{Menci+08} SAM (run on both
CDM and WDM halo merger trees) than among centrals in the same mass range,
which is opposite to the trend found in the \Hen\ SAM (Fig.~\ref{fig:Henriques_massresol}).

Figure~\ref{fig:fyoung_menci} also illustrates the effects of the primordial density
fluctuation spectrum on the fractions of VYGs.
We see that the power spectrum has little effect on the fraction of VYGs  at
intermediate masses, while the passage from CDM to WDM boosts the fraction of
VYGs among high-mass satellites. On the other hand,  moving from CDM to WDM leads to 10 to 30
times more VYGs at low masses ($m < 10^9 \,\msun$, except for one of the four
mass bins -- at $\lm = 8.25$), both for centrals and
satellites, although this increase is not statistically significant for the
centrals (except in the bin at $\lm=8.75$).
This basically confirms the basic trend for younger galaxies with $\Lambda$WDM 
found by \cite{Calura+14}
for this SAM. The boost in the fraction of VYGs among low-mass galaxies but
not higher mass ones is an illustration of the WDM power
spectrum that suppresses high wavenumbers. Star formation in low-mass
galaxies is thus delayed, 
making a higher fraction of low-mass galaxies
appear very young at $z=0$.
With the WDM particle mass of 0.75 keV, the boost in the VYG fraction occurs at masses $m
< 10^8\,\msun$. Had we considered a higher WDM particle mass, the boost in
the VYG fraction would
have occurred at even lower galaxy mass.

These conclusions must be somewhat tempered by the fact that the Menci Monte Carlo
halo merger trees are not well resolved at low masses.
At stellar masses $\lm = 7\pm0.25$ and $8\pm0.25$,
the corresponding halo masses have a median of $\lM = 9.5$ and 10,
respectively (see also the bottom panel of Fig.~\ref{fig:mvsM}), so given the 
minimum branch mass of $\lM=7.7$ (Table~\ref{tab:models}), the mass resolution of
the trees is only respectively 60 and 200, compared to $10^4$ for our four analytical models. 

Note also that the \cite{Menci+14} SAM, based upon Monte-Carlo halo merger trees,
includes a rough treatment of the environmental effects on galaxies.
It leads to intermediate fractions of VYGs at
intermediate mass (i.e. for their CDM run) in comparison with the high VYG fractions of the analytical
models and the low VYG fractions of the \Hen\ SAM (compare
Figs.~\ref{fig:fyoung} and \ref{fig:fyoung_menci}).

\section{Conclusions}
\label{sec:concl}

\subsection{Method and models}
In this work, we modelled the frequency of VYGs (defined to be those
with over half the stellar
mass having formed within the last 1 Gyr, corresponding to $z=0.08$), among
central galaxies, as a
function of present-day stellar mass. For this, we first  produced
 over a quarter million Monte-Carlo  halo
merger trees derived from the code of \cite{Parkinson+08}. We then
ran  four simple models of galaxy formation on these halo merger trees.
The first three models give the galaxy stellar mass as a function of
halo mass and redshift. These are the physically-motivated model (\citealp{Cattaneo+11}
with the refined reionization feedback of \citealp{Gnedin00}), and
the empirical models of \cite{Moster+13} and \cite{Behroozi+13}
obtained by abundance matching of the halo mass function with the observed
SMF (with, for the latter, additional constraints on the cosmic star
formation rate and the specific star formation rate as a function of mass).
We also considered a fourth analytical model by \cite{Mutch+13}, where
the stellar mass
growth rate is proportional to halo mass growth rate and a simple function of
halo mass and redshift.
 
These analytical models
present differences in the  star formation
efficiency as a function of halo mass and redshift
(Figs.~\ref{fig:compare_sfe} and \ref{fig:sfevsz}), because
i) one (\CG) is physically motivated while the other 3 are empirical;
ii) only one
(\BWC) is calibrated to data extending to the epoch of reionization;
iii)
only one (\MCP) is based on mass growth rates instead of masses.
We also considered two SAMs, which incorporate much more complex physics: one
(Henriques) with the advantage of being based on realistic galaxy positions (based on
subhaloes from a dark matter cosmological simulation), the other (Menci) that
was run on both CDM and WDM cosmologies.
Our analytical models run on Monte-Carlo halo merger trees have
the advantage of having more statistics at the high halo mass end and higher
mass resolution.

We updated the stellar masses of our analytical models using
two
galaxy
merging
schemes,
one
involving a starburst at the time of the halo merger,
and one without it.
These two schemes should bracket the evolution of stellar masses.

The galaxy formation models applied to the halo merger trees produce
somewhat different halo to stellar mass relations (Fig.~\ref{fig:mvsM}). Yet, they
generally match well (especially the \MNW\ and \BWC\ models) 
the analytical stellar to halo mass relations that they predict at the redshift where half
the mass in stars is formed.
These models predict stellar mass functions that generally agree with the
observations, but with noticeable differences in the low-end slopes.

The dependence on stellar mass of the epoch when half the final stellar
mass is formed (measuring the stellar mass evolution by considering all the
progenitors, not just the main one) shows differences between the galaxy
formation models ran with the bursty merging scheme.
All four analytical models and the SAM show downsizing at the high mass end: 
the median stellar ages decrease with decreasing stellar masses, reaching
a minimum around $\log (m/\msun) \simeq 9.5$ (see Fig.~\ref{fig:z50vsm}). 
The \CG\ model shows strong
 \emph{upsizing} (stellar age increasing with decreasing stellar mass) at the
 low end, while the \MNW, \BWC, and \Hen\ models show weak upsizing.
The typical median
ages at $m = 10^{8-10}\, \msun$ range from 2.5 to 11 Gyr (Fig.~\ref{fig:z50vsm}), according to the
model. 
These differences between models can be explained by differences in their
star formation efficiencies as a function of halo mass and redshift
(Fig.~\ref{fig:sfevsz}) combined with our preventing stellar masses to
decrease in time.

\subsection{Frequency of very young galaxies}

At $z=0$, the fractions of galaxies that are very young  depend on the galaxy stellar mass, as well
as on the model and the scheme for galaxy mergers.
For bursty galaxy merging,
the predicted  fraction of VYGs is roughly flat up
to $10^{10}\,\msun$ for the 
analytical models,  with  peaks at typically a few percent
at stellar mass near $10^9\, \msun$
(Fig.~\ref{fig:fyoung}). 
For quiet galaxy merging, the VYG fractions are similar to those with
bursty merging for two models
(\MNW\ and \MCP), but suppressed for two others, at respectively intermediate
(\CG) and low (\BWC) stellar masses.

With bursty galaxy merging,
VYGs are always  (\CG\ and \BWC) or often
(\MNW\ and \MCP) associated with rapid late halo
growth
(Fig.~\ref{fig:pdfMrat}),
indicating that most
VYGs
are associated with recent major halo mergers.
However, the \MNW, and
especially \MCP\ models
also allow for stellar mass buildup with halo mergers of intermediate mass
ratios
(Fig.~\ref{fig:pdfMrat}).
With quiet galaxy merging, we found, instead, no relation between halo mass
growth and galaxy stellar mass growth
(Fig.~\ref{fig:pdfMrat_quiet}).

The VYG fractions
with bursty galaxy merging
are very well reproduced for the \MNW\ and \BWC\ galaxy
formation models 
(Fig.~\ref{fig:fyoung_pred}) 
by a simple, quasi-analytical, 
Markovian model (Sect.~\ref{sec:merggrowth})
computing the growth of stellar mass in the last
Gyr by combining the growth during halo mergers with the quiescent growth.
This simple model reproduces less well the VYG fractions for the \CG\ model,
for which the past
history appears to play a role in the stellar mass buildup in the last Gyr,
hence the VYG fraction.
The trends of VYG fractions with stellar mass for the case of quiet galaxy merging are
qualitatively reproduced with another simple model
(Sect.~\ref{sec:mgrowthquiet}),
which involves a lower
fraction of $z$=0 halo mass in subhaloes hosting surviving satellites than haloes 1
Gyr ago (Fig.~\ref{fig:eta0VYG}).
The lower fractions of VYGs with the \CG\ and \BWC\ models, regardless of the
merging scheme, thus appear to be
linked to the slow recent growth of their star formation efficiency (Fig.~\ref{fig:evolsfe}).

The semi-analytical model of galaxy formation
run by \cite{Henriques+15} on the MS-II produces 30 to 800 times lower
fractions of VYGs in comparison to the
analytical models run on Monte-Carlo halo merger trees with bursty galaxy
merging,
predicting less than 0.03 percent of
VYGs for $10^8\,\msun \leq m \leq 10^{10}\,\msun$ (lacking mass resolution at
the low end and statistics at the high end). These very low VYG fractions are
consistent with one (\BWC) or another (\CG) analytical model with quiet
galaxy merging, but not both at once (Fig.~\ref{fig:fyoung_quiet}).
These discrepancies may highlight the importance of more accurate
modeling such as in the SAM. Conversely, they may indicate that the \Hen\ SAM
underestimates the effects of starbursts during galaxy mergers or
quenches too much galaxies within groups.

Finally, only at low masses ($m < 10^8 \,\msun$) is the fraction of VYGs 
significantly boosted with  Warm Dark Matter compared to Cold Dark Matter,
but the mass resolution of the tree code used for the SAM of \cite{Menci+14} 
comparing CDM and
WDM may not be sufficient for accurate results.

\subsection{Final remarks}

The  semi-analytical models used here may not have the necessary
mass resolution to probe the growth of stellar mass to form dwarf galaxies such as
I~Zw~18,  while the physical analytical model is probably too simple and
missing some important astrophysical processes that may be present in the
semi-analytical model. The empirical analytical models should be more accurate,
but the 3 such models tested here differ in their predictions, because of
subtle differences in their stellar to halo mass relations, as confirmed with
the simple models we introduced in Sects.~\ref{sec:merggrowth} and \ref{sec:mgrowthquiet}.

We conclude that modelling the fraction of VYGs  is a promising sensitive
test of galaxy formation models,
given the wide range of predictions among the analytical models and with the state-of-the-art SAM. 
In Paper~II, we confront these models with observations, 
measuring the fractions of VYGs in the local Universe  as a function of
stellar mass, using the spectral database of the SDSS.

\section*{Acknowledgments}
We thank the 
anonymous
 referee for many useful comments and additional
 references, which considerably improved our manuscript.
We acknowledge Hannah Parkinson for making her halo merger tree
code publicly available and to Shaun Cole for helpful comments on it.
We thank Steve Murray for making his {\sc HMFCalc} code publicly available.
Thanks also to Gerry Williger for a critical reading.
DT and TXT are grateful to the Institut d'Astrophysique de Paris for
its hospitality, where a large part of
this work was performed, while GAM thanks the University of Virginia for
its hospitality.
This work was supported by the French \emph{Minist\`ere de l'Enseignement
Sup\'erieur et de la Recherche}, the \emph{Minist\`ere de Affaires Europ\'eennes
et Etrang\`eres} and the Israeli \emph{Ministry Of Science and Technology}
within a France--Israel research program  for the project
\emph{Central
  Issues on Galaxy Formation}, awarded to AD and GAM.
FC acknowledges funding from
the INAF PRIN-SKA 2017 program 1.05.01.88.04.
AD acknowledges support from a \emph{Lagrange Fellowship}.
The Millennium Simulation databases used in this paper and the web
application providing online access to them were constructed as part of the
activities of the German Astrophysical Virtual Observatory (GAVO).

\bibliography{master}

% Don't change these lines
%\bsp	% typesetting comment
\label{lastpage}

\end{document}